\documentclass[superscriptaddress,twocolumn,pre,nofootinbib]{revtex4-1}

\usepackage{amsmath}
\usepackage{graphicx,color}

\newcommand{\be}{\begin{equation}}
\newcommand{\ee}{\end{equation}}
\newcommand{\bea}{\begin{eqnarray}}
\newcommand{\eea}{\end{eqnarray}}

\begin{document}
\sloppy

%-title page-%

\title{Models of universe with a polytropic equation of state: \\
II. The late universe}

\author{Pierre-Henri Chavanis}
%\email{chavanis@irsamc.ups-tlse.fr}
\affiliation{Laboratoire de Physique Th\'eorique (IRSAMC), CNRS and UPS, Universit\'e de Toulouse, France}

\begin{abstract}

We construct models of universe with a generalized equation of state
$p=(\alpha \rho+k\rho^{1+1/n})c^2$ having a linear component and a
polytropic component. The linear equation of state $p=\alpha\rho c^2$
describes radiation ($\alpha=1/3$), pressureless matter ($\alpha=0$),
stiff matter ($\alpha=1$), and vacuum energy ($\alpha=-1$). The
polytropic equation of state $p=k\rho^{1+1/n} c^2$ may be due to
Bose-Einstein condensates with repulsive ($k>0$) or attractive ($k<0$)
self-interaction, or have another origin. In this paper, we consider
negative indices $n<0$. In that case, the polytropic component
dominates in the late universe where the density is low. For
$\alpha=0$, $n=-1$ and $k=-\rho_{\Lambda}$, we obtain a model of late
universe describing the transition from the matter era to the dark
energy era. Coincidentally, we live close to the transition between
these two periods, corresponding to $a_2=8.95\, 10^{25}\, {\rm m}$ and
$t_2=2.97\, 10^{17}\, {\rm s}$. The universe exists eternally in the
future and undergoes an inflationary expansion with the cosmological
density $\rho_{\Lambda}=7.02\, 10^{-24}\, {\rm g}/{\rm m}^3$ on a
timescale $t_{\Lambda}=1.46\, 10^{18} {\rm s}$. For $\alpha=0$, $n=-1$
and $k=\rho_{\Lambda}$, we obtain a model of cyclic universe appearing
and disappearing periodically. If we were living in this universe, it
would disappear in about $2.38$ billion years. We make the connection
between the early and the late universe and propose a simple equation
describing the whole evolution of the universe. This leads to a model
of universe that is eternal in past and future without singularity
(aioniotic universe). It generalizes the $\Lambda$CDM model by
removing the primordial singularity (Big Bang). This model exhibits a
nice ``symmetry'' between an early and late phase of inflation, the
cosmological constant in the late universe playing the same role as
the Planck constant in the early universe. We interpret the
cosmological constant as a fundamental constant of nature describing
the ``cosmophysics'' just like the Planck constant describes the
microphysics. The Planck density and the cosmological density
represent fundamental upper and lower bounds differing by ${122}$
orders of magnitude. The cosmological constant ``problem'' may be a
false problem.  We determine the potential of the scalar field
(quintessence, tachyon field) corresponding to the generalized
equation of state $p=(\alpha
\rho+k\rho^{1+1/n})c^2$. We also propose a unification
of pre-radiation, radiation and dark energy through the quadratic
equation of state $p/c^2=-4\rho^2/3\rho_P+\rho/3-4\rho_{\Lambda}/3$.

\end{abstract}

\maketitle

\section{Introduction}

According to contemporary cosmology, the present energy content of the
universe is composed of approximately $5\%$ ordinary matter, $20\%$
dark matter and $75\%$ dark energy \cite{bt}.  The expansion of the
universe began in a tremendous inflationary burst driven by the vacuum
energy with the Planck density $\rho_P=5.16 \, 10^{99}\, {\rm g}/{\rm
m}^3$. Between $10^{-35}$ and $10^{-33}$ seconds after the beginning
(Big Bang), the universe expanded by a factor $10^{30}$
\cite{guth,linde}. Inflation does not offer any explanation for the
time before that ``beginning''. The universe then entered in the
radiation era and, when the temperature cooled down below
approximately $10^3\, {\rm K}$, in the matter era \cite{weinberg}. At
present, it undergoes an accelerated expansion \cite{novae} presumably
due to the cosmological constant or to some form of dark energy with
negative pressure violating the strong energy condition
\cite{cst}. This corresponds to a second period of inflation, which is
different from the first since it is driven by the cosmological energy
density $\rho_{\Lambda}=7.02\, 10^{-24}\, {\rm g}/{\rm m}^3$ instead
of the Planck density.  Despite the success of this model, the nature
of dark matter, dark energy, and of the very early universe
(pre-radiation era) remains very mysterious and leads to many
speculations.

The phase of inflation in the early universe is usually described by some hypothetical scalar field $\phi$ with its origin in quantum fluctuations of vacuum \cite{linde}. This leads to an equation of state $p=-\rho c^2$, implying a constant energy density,  called the vacuum energy. This energy density is usually identified with the Planck density $\rho_P$. As a result of the vacuum energy, the universe expands exponentially rapidly on a timescale of the order of the Planck time $t_P=5.39\, 10^{-44}{\rm s}$ (early inflation). This phase of inflation is followed by the radiation era described by an equation of state $p=\rho c^2/3$. In our previous paper (Paper I), we showed that  a unified description of the pre-radiation ($\rho=\rho_P$) and radiation ($\rho\propto a^{-4}$) eras, connected by  a phase of exponential inflation, could be obtained from a single equation of state of the form $p=(1/3)\rho(1-4\rho/\rho_P)c^2$. This provides a non-singular model of the early universe.

The phase of acceleration in the present universe is usually ascribed to the cosmological constant $\Lambda$  which is equivalent to a constant energy density $\rho_{\Lambda}=\Lambda/(8\pi G)$ called dark energy. This can be modeled by an equation of state $p=-\rho c^2$, implying a constant energy density identified with the cosmological density  $\rho_{\Lambda}$. As a result of the dark energy, the universe expands exponentially rapidly on a  timescale of the order of the cosmological time $t_{\Lambda}=1.46\, 10^{18} {\rm s}$ (de Sitter solution). This leads to a second phase of inflation (late inflation). Inspired by the analogy with the early inflation, some authors have represented the dark energy by a scalar field called quintessence \cite{quintessence}. As an alternative to the quintessence, other authors  \cite{chaplygin} have proposed to model the acceleration of the universe by an exotic fluid with an equation of state of the form $p=-A/\rho$  called the Chaplygin gas (see \cite{cst} for a complete list of references). At late times, this equation of state leads to a constant energy density implying  an exponential growth of the scale factor that is similar to the effect of the cosmological constant. At earlier times, this equation of state returns the results of the cold dark matter model. Therefore, it provides a unification of dark matter ($\rho\propto a^{-3}$) and dark energy ($\rho=\rho_{\Lambda}$) in the late universe. Furthermore, it gives a real velocity of sound which is non-trivial for fluids with negative pressure.  Some generalizations of this equation of state have been considered in the form $p=-A/\rho^{a}$ with $a\ge -1$ \cite{chaplygin,cst}. As mentioned in \cite{chaplygin}, the Chaplygin gas has some connection with string theory and can be obtained from the Nambu-Goto action for $d$-branes moving in a $(d+2$)-dimensional spacetime in the light-cone parametrization. Furthermore, it is the only fluid which, up to now, admits a supersymmetric generalization.

From a theoretical point of view, it is desirable to study models of universe with a generalized equation of state $p=(\alpha \rho+k\rho^{1+1/n}) c^2$ having a standard linear component and a polytropic component. The linear equation of state $p=\alpha\rho c^2$ describes radiation ($\alpha=1/3$), pressureless matter ($\alpha=0$), stiff matter ($\alpha=1$), and vacuum energy ($\alpha=-1$). The polytropic equation of state $p=k\rho^{\gamma} c^2$ with $\gamma=1+1/n$ may be due to Bose-Einstein condensates with repulsive ($k>0$) or  attractive ($k<0$) self-interaction \cite{c4}, or have another origin. When $n>0$, the polytropic component dominates the linear component when the density is high. These models, studied in Paper I, describe the early universe. Conversely, when $n<0$, the polytropic component dominates  the linear component when the density is low. These models, studied in the present paper, describe the late universe. As we shall see, these two studies are strikingly symmetric.  Interestingly, this symmetry seems to reflect the true evolution of the universe.

In this paper, we propose an exhaustive study of the equation of state $p=(\alpha \rho+k\rho^{1+1/n}) c^2$ with $n<0$. When $k>0$, the universe exhibits a future peculiarity: Its density vanishes in infinite time ($n\le -2$), or periodically in time ($n>-2$), while its radius tends to a constant value (the universe is singular when $n>-1$ since the pressure becomes infinite in finite time).  For $\alpha=0$, $n=-1$ and $k=\rho_{\Lambda}$, we obtain a model of cyclic universe, equivalent  to the anti-$\Lambda$CDM model, in which the universe ``disappears'' (it becomes empty) and ``reappears'' periodically. According to this model, the universe would disappear in about $2.38$ billion years.   When $k<0$, the universe exists for all times in the future and there is no singularity. For $\alpha=0$, $n=-1$ and $k=-\rho_{\Lambda}$, we obtain a model of late universe, equivalent to the $\Lambda$CDM model, describing in a unified manner the transition from the matter era to the dark energy era. Coincidentally, we live close to the transition between these two periods, corresponding to $a_2=8.95\, 10^{25}\, {\rm m}$ and $t_2=2.97\, 10^{17}\, {\rm s}$. This universe exists eternally in the future and undergoes an inflationary expansion with the cosmological density $\rho_{\Lambda}=7.02\, 10^{-24}\, {\rm g}/{\rm m}^3$ on a timescale $t_{\Lambda}=1.46\, 10^{18} {\rm s}$.

The paper is organized as follows. In Sec. \ref{sec_basic}, we recall
the basic equations of cosmology. In Secs. \ref{sec_ges} and
\ref{sec_dark}, we study the generalized equation of state $p=(\alpha
\rho+k\rho^{1+1/n}) c^2$ for any value of the parameters $\alpha$, $k$
and $n<0$. In Sec. \ref{sec_solid}, we consider the case $\alpha=0$,
$n=-1$, and $k=- \rho_{\Lambda}$, providing a model of non-singular
inflationary universe describing the transition from the matter era to
the dark energy era ($\Lambda$CDM model). In Sec. \ref{sec_cyclic}, we
consider the case $\alpha=0$, $n=-1$, and $k= \rho_{\Lambda}$, leading
to a model of peculiar cyclic universe (anti-$\Lambda$CDM model).  In
Sec. \ref{sec_simple}, we discuss the connection between the early and
the late universe, and propose a simple equation [see Eq. (\ref{pr7})]
describing the whole evolution of the universe. This leads to a model
of universe that is eternal in past and future without singularity
(aioniotic universe). This model exhibits a nice ``symmetry'' between
an early and late phase of inflation, the cosmological constant in the
late universe playing the same role as the Planck constant in the
early universe. This model refines the standard $\Lambda$CDM model by
removing the primordial singularity. In Sec. \ref{sec_scalar}, we
determine the potential of the effective scalar field (quintessence,
tachyon field) corresponding to the generalized equation of state
$p=(\alpha \rho+k\rho^{1+1/n}) c^2$.

\section{Basic equations of cosmology}
\label{sec_basic}

In a space with uniform curvature, the line element is given by the
Friedmann-Lema\^itre-Roberston-Walker (FLRW) metric
\begin{eqnarray}
\label{b1}
ds^2=c^2 dt^2-a(t)^2\left\lbrace \frac{dr^2}{1-kr^2}+r^2\,  (d\theta^2+ \sin^2\theta\,  d\phi^2)\right \rbrace,\nonumber\\
\end{eqnarray}
where $a(t)$ represents the radius of curvature of the $3$-dimensional space, or the scale factor. By an abuse of language, we shall call it the ``radius of the universe''. On the other hand, $k$ determines the curvature of space. The universe can be closed ($k>0$), flat ($k=0$), or open ($k<0$).

If the universe is isotropic and homogeneous at all points in conformity with the line element (\ref{b1}), and contains a uniform perfect fluid of energy density $\epsilon(t)=\rho(t) c^2$ and isotropic pressure $p(t)$, the Einstein equations can be written as
\begin{equation}
\label{b2}
\frac{d\rho}{dt}+3\frac{\dot a}{a}\left (\rho+\frac{p}{c^2}\right )=0,
\end{equation}
\begin{equation}
\label{b3}
\frac{\ddot a}{a}=-\frac{4\pi G}{3} \left (\rho+\frac{3p}{c^2}\right )+\frac{\Lambda}{3},
\end{equation}
\begin{equation}
\label{b4}
H^2=\left (\frac{\dot a}{a}\right )^2=\frac{8\pi G}{3}\rho-\frac{kc^2}{a^2}+\frac{\Lambda}{3},
\end{equation}
where we have introduced the Hubble parameter $H=\dot a/a$ and accounted for a possibly non-zero cosmological constant $\Lambda$. These are the well-known Friedmann equations describing a non-static universe \cite{weinberg}. Among these three equations, only two are independent. The first equation can be viewed as an ``equation of continuity''. For a given barotropic equation of state $p=p(\rho)$, it determines the relation between the density and the scale factor. Then, the temporal evolution of the scale factor is given by Eq. (\ref{b4}).

We will also need the thermodynamical equation
\begin{equation}
\label{t2}
\frac{dp}{dT}=\frac{1}{T}(\rho c^2+p),
\end{equation}
which can be derived from the first principle of thermodynamics
\cite{weinberg}. For a given barotropic equation of state $p=p(\rho)$,
this equation can be integrated to obtain the relation $T=T(\rho)$
between the temperature and the density. It can be shown
\cite{weinberg} that the Friedmann equations conserve the entropy of
the universe
\begin{equation}
\label{t4}
S=\frac{a^3}{T}(p+\rho c^2).
\end{equation}

In this paper, we consider a flat universe ($k=0$) in agreement with the observations of the cosmic microwave background (CMB) \cite{bt}. On the other hand, we set $\Lambda=0$ because the effect of the cosmological constant will be taken into account in the equation of state. The Friedmann equations reduce to
\begin{equation}
\label{b7}
\frac{d\rho}{dt}+3\frac{\dot a}{a}\left (\rho+\frac{p}{c^2}\right )=0,
\end{equation}
\begin{equation}
\label{b8}
\frac{\ddot a}{a}=-\frac{4\pi G}{3} \left (\rho+\frac{3p}{c^2}\right ),
\end{equation}
\begin{equation}
\label{b9}
H^2=\left (\frac{\dot a}{a}\right )^2=\frac{8\pi G}{3}\rho.
\end{equation}
Introducing the equation of state parameter $w=p/\rho c^2$, we see from Eq. (\ref{b8}) that the universe is decelerating if $w>-1/3$ (strong energy condition) and accelerating if $w<-1/3$. On the other hand, according to Eq. (\ref{b7}), the density decreases with the scale factor if $w>-1$ (null dominant energy condition) and increases with the scale factor if $w<-1$.

\section{Generalized equation of state with negative index}
\label{sec_ges}

We consider a generalized equation of state of the form
\begin{equation}
\label{b10}
p=(\alpha \rho+k\rho^{1+1/n}) c^2.
\end{equation}
This is the sum of a standard linear equation of state $p=\alpha\rho c^2$ and a polytropic equation of state $p=k\rho^{\gamma} c^2$, where $k$ is the polytropic constant and $\gamma=1+1/n$ is the polytropic index. Concerning the linear equation of state, we assume $-1\le \alpha\le 1$  (the case $\alpha=-1$ is treated specifically in Appendix \ref{sec_eosgm}). This equation of state describes radiation ($\alpha=1/3$), pressureless matter ($\alpha=0$) and vacuum energy ($\alpha=-1$). Concerning the polytropic equation of state, we remain very general. Such an equation of state may correspond to self-gravitating BECs with repulsive ($k>0$) or attractive ($k<0$) self-interaction, but it may have another origin.

For an equation of state of the form (\ref{b10}) with positive index $n>0$, the polytropic component dominates the linear component when the density is high. These models, studied in paper I,  describe the early universe. Conversely, when $n<0$, the polytropic component dominates  the linear component when the density is low. These models, studied in the present paper, describe the late universe. We shall see that these two situations are strikingly similar. They reveal a form of ``symmetry'' between the early and the late universe. In this paper, as in Paper I, we assume that $\alpha+1+k\rho^{1/n}>0$. This corresponds to the ``normal'' case where the density decreases with the radius ($w>-1$). The opposite case $\alpha+1+k\rho^{1/n}<0$, leading to a ``phantom universe'' where the density increases with the radius ($w<-1$), is treated in Paper III.

The purpose of our series of papers  is to make an exhaustive study of all the possible cases $(\alpha,n,k)$, even if some of them seem to be unphysical or in conflict with the known  properties of our universe. Actually, by rejecting the cases leading to past or future singularities (or peculiarities), we will be led naturally to the ``good'' model (see Sec. \ref{sec_aioniotic}).

Some of the equations established in Paper I remain valid when $n<0$. To avoid repetitions, we shall directly refer to these formulae. For example, Eq. (I-60) refers to Eq. (60) of Paper I.

\subsection{The density}
\label{sec_gesdensity}

For the equation of state (\ref{b7}), assuming $w>-1$, the Friedmann equation (\ref{b2}) can be integrated into
\begin{equation}
\label{ges2}
\rho=\frac{\rho_*}{\left\lbrack (a/a_*)^{3(1+\alpha)/n}\mp 1\right\rbrack^n},
\end{equation}
where  $\rho_*=\lbrack (\alpha+1)/|k|\rbrack^n$ and $a_*$ is a constant of integration. The upper sign
corresponds to $k>0$  and the lower sign
corresponds to $k<0$.

\begin{figure}[!h]
\begin{center}
\includegraphics[clip,scale=0.3]{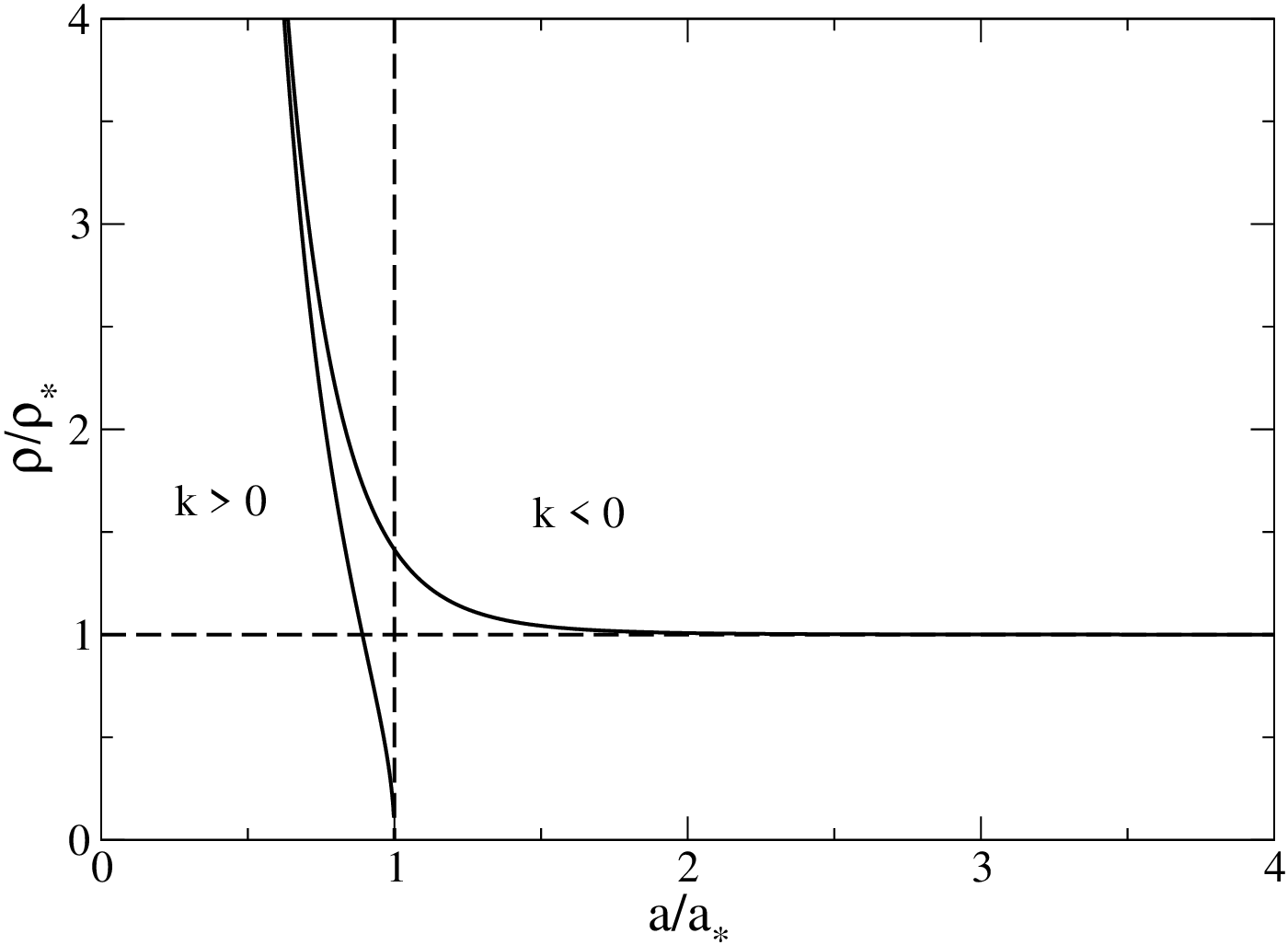}
\caption{Density as a function of the scale factor for  $k>0$ and $k<0$. We have taken $n=-1/2$ and $\alpha=0$.}
\label{densiteNneg}
\end{center}
\end{figure}

For $k>0$, the density is defined only for $a<a_*$. When $a\rightarrow 0$, $\rho/\rho_*\sim (a_*/a)^{3(1+\alpha)}\rightarrow +\infty$ corresponding to the linear equation of state. When $a\rightarrow a_*$,
\begin{equation}
\label{ges3}
\frac{\rho}{\rho_*}\sim \left\lbrack \frac{3(1+\alpha)}{|n|}\right \rbrack^{|n|} (1-a/a_*)^{|n|}\rightarrow 0.
\end{equation}
For the behavior of the pressure, we must consider different cases. When $a\rightarrow 0$, $p\rightarrow +\infty$ for $\alpha>0$, $p\rightarrow -\infty$ for $\alpha<0$, $p\rightarrow +\infty$ for $(\alpha=0,n<-1)$,  $p\rightarrow 0$ for $(\alpha=0,n>-1)$, and $p=kc^2$ for $(\alpha=0,n=-1)$.
When $a\rightarrow a_*$,  $p \rightarrow +\infty$ for $n>-1$, $p$ tends to a finite value for $n=-1$, and $p\rightarrow 0$ for $n<-1$.

For $k<0$, the density is defined for all $a$. When
$a\rightarrow 0$,  $\rho\sim \rho_* (a_*/a)^{3(1+\alpha)}\rightarrow +\infty$ corresponding to the linear equation of state. When $a\rightarrow +\infty$, the density tends to a finite value  $\rho_*$.
For the behavior of the pressure, we must consider different cases. When $a\rightarrow 0$, $p\rightarrow +\infty$ for $\alpha>0$, $p\rightarrow -\infty$ for $\alpha<0$, $p\rightarrow -\infty$ for $(\alpha=0,n<-1)$,  $p\rightarrow 0$ for $(\alpha=0,n>-1)$, and $p=kc^2$ for $(\alpha=0,n=-1)$.
When $a\rightarrow +\infty$,   $p \rightarrow -\rho_* c^2$.

Some curves giving the evolution of the density $\rho$  as a function of the scale factor $a$ are plotted in Fig.  \ref{densiteNneg} for $k>0$ and $k<0$.

\subsection{The temperature}
\label{sec_gestemperature}

For the equation of state (\ref{b10}), the evolution of the temperature is given by Eqs. (I-60) and (I-61). Let us consider some asymptotic limits.

When $a\ll a_*$, Eq. (I-61) reduces to Eq. (I-62) returning the relation valid for a linear equation of state. When $a\rightarrow 0$, $T\rightarrow +\infty$ for $\alpha>0$, $T\rightarrow 0$ for $\alpha<0$, and $T\rightarrow T_*$ for $\alpha=0$.

When $k>0$ and $a\rightarrow a_*$,
\begin{equation}
\label{ges8}
T/T_*\sim \left\lbrack \frac{|n|}{3(1+\alpha)}\right \rbrack^{n+1} \frac{1}{(1-a/a_*)^{n+1}}.
\end{equation}
Therefore, $T\rightarrow 0$ for $n<-1$, $T\rightarrow T_*$ for $n=-1$ and  $T\rightarrow +\infty$ for $n>-1$.

When $k<0$ and $a\rightarrow +\infty$, $T/T_*$ is given by Eq. (I-64). Therefore, $T\rightarrow 0$ for $\alpha+n+1>0$, $T\rightarrow T_*$ for $\alpha+n+1=0$ and  $T\rightarrow +\infty$ for $\alpha+n+1<0$.

The extremum of temperature (when it exists) is located at the point defined by Eqs. (I-65)-(I-67). Finally, the entropy (\ref{t4}) is given by Eq. (I-68).

\begin{figure}[!h]
\begin{center}
\includegraphics[clip,scale=0.3]{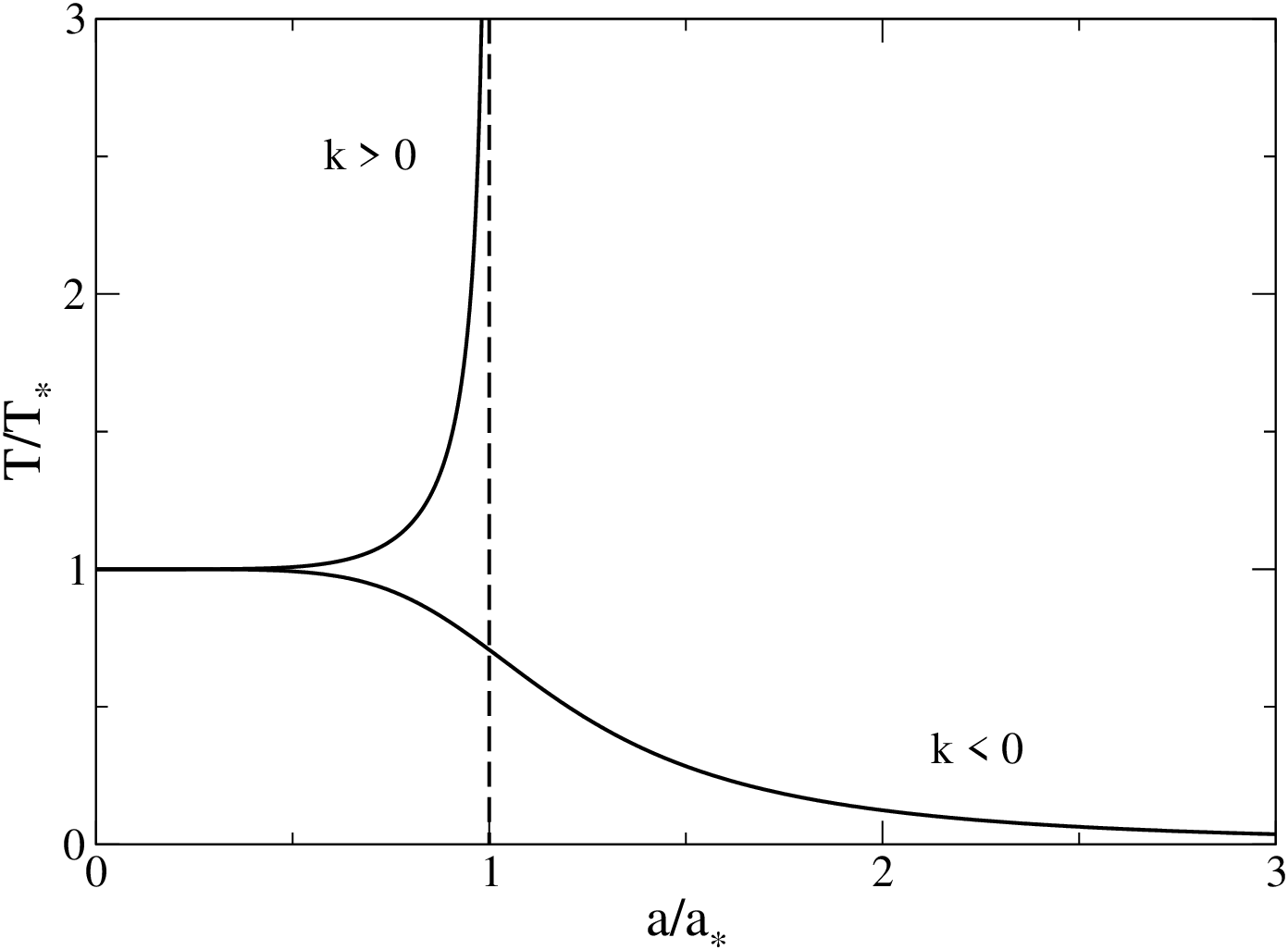}
\caption{Temperature as a function of the scale factor for  $k>0$ and $k<0$. We have taken $n=-1/2$ and $\alpha=0$.}
\label{temp}
\end{center}
\end{figure}

Some curves giving the evolution of the temperature $T$  as a function of the scale factor $a$ are plotted in Fig.  \ref{temp} for $k>0$ and $k<0$.

\subsection{The equation of state parameter $w(t)$}
\label{sec_gesw}

We can rewrite the equation of state (\ref{b10}) as $p=w(t) \rho c^2$ where $w(t)$ is given by Eq. (I-69). The point $(a_w,\rho_w,T_w)$ where the pressure vanishes is defined by Eqs. (I-70) and (I-71).

For $k>0$,  $w\rightarrow \alpha$ when $a\rightarrow 0$, and $w\rightarrow +\infty$ when $a\rightarrow a_*$. For $\alpha>0$, the pressure is always positive ($w>0$).  For $\alpha<0$, the pressure is negative when $a<a_w$ and positive when $a_w<a<a_*$.

For $k<0$,  $w\rightarrow \alpha$ when $a\rightarrow 0$ and  $w\rightarrow -1$ when $a\rightarrow +\infty$. For $\alpha<0$, the pressure is always negative ($w<0$). For $\alpha>0$, the pressure is positive when $a<a_w$ and negative when $a>a_w$.

\begin{figure}[!h]
\begin{center}
\includegraphics[clip,scale=0.3]{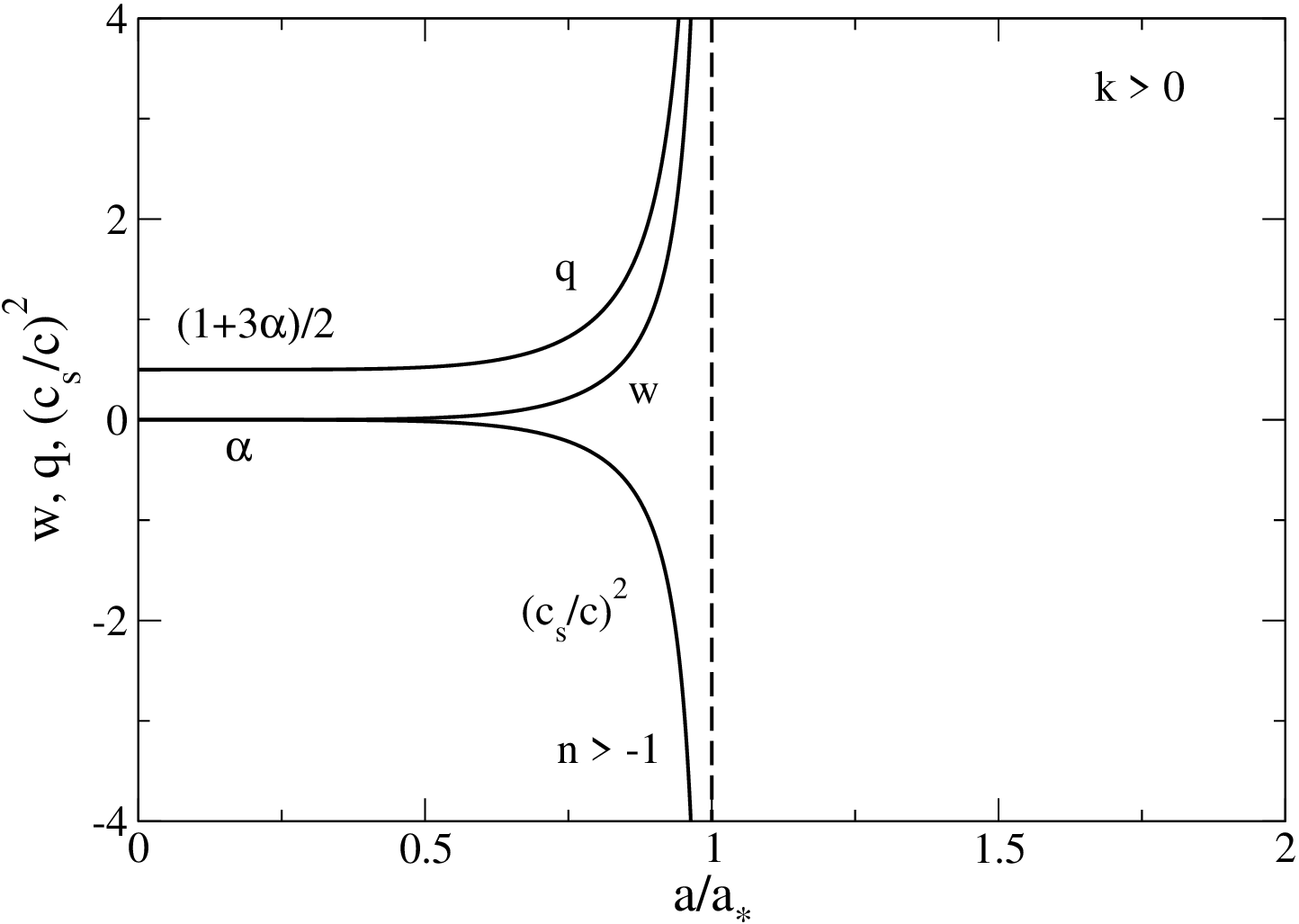}
\caption{The parameters $w$, $q$ and $c_s^2/c^2$ as a function of the scale factor $a$ for $k>0$. We have taken $n=-1/2$ and $\alpha=0$.}
\label{wKposNneg}
\end{center}
\end{figure}

\begin{figure}[!h]
\begin{center}
\includegraphics[clip,scale=0.3]{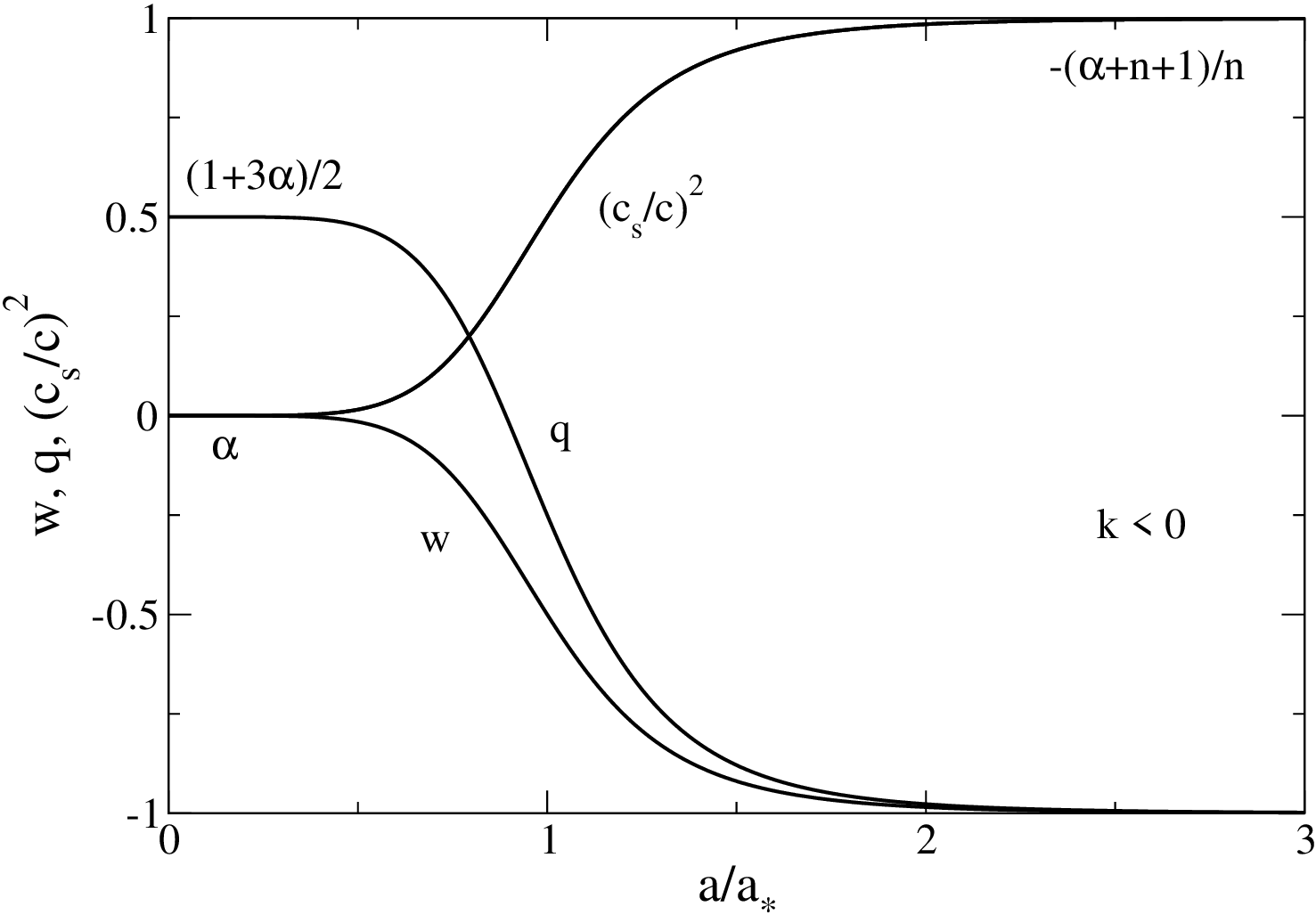}
\caption{The parameters $w$, $q$ and $c_s^2/c^2$ as a function of the scale factor $a$ for $k<0$. We have taken $n=-1/2$ and $\alpha=0$.}
\label{wKnegNneg}
\end{center}
\end{figure}

Some curves giving the evolution of $w$ as a function of
the scale factor $a$ are plotted in Figs. \ref{wKposNneg} and \ref{wKnegNneg}
for $k>0$ and $k<0$.

\subsection{The velocity of sound}
\label{sec_gessound}

For the equation of state (\ref{b10}), the velocity of sound is given by Eq. (I-72).
The velocity of sound vanishes at the point $(a_e,\rho_e,T_e)$ defined by Eqs.  (I-65)-(I-67) where the temperature is extremum. At that point the pressure is extremum with value $p_e$ given by Eq. (I-73). The case $c_s^2<0$ corresponds to an imaginary velocity of sound. The point $(a_s,\rho_s,T_s)$ where
the velocity of sound is equal to the speed of light is defined by Eqs. (I-74)-(I-76). Different cases have to be considered.

We first assume $k>0$. When $a\rightarrow 0$, $(c_s/c)^2\rightarrow \alpha$; when $a\rightarrow a_*$, $(c_s/c)^2\rightarrow +\infty$ for $n<-1$ and $(c_s/c)^2\rightarrow -\infty$ for $n>-1$. For $n<-1$ and $\alpha>0$,  $c_s^2$  is always positive. For $n<-1$ and $\alpha<0$, $c_s^2$  is negative when $a<a_e$ and positive when $a_e<a<a_*$. The velocity of sound is smaller than the speed of light when $a<a_s$ and larger when $a_s<a<a_*$. For $n>-1$ and $\alpha<0$,  $c_s^2$  in always negative. For $n>-1$ and $\alpha>0$,  $c_s^2$  is positive when $a<a_e$ and negative when $a_e<a<a_*$. For $n>-1$ and $\alpha\le 1$, the velocity of sound is always smaller than the speed of light.

We now assume $k<0$.  When $a\rightarrow 0$, $(c_s/c)^2\rightarrow \alpha$; when $a\rightarrow +\infty$, $(c_s/c)^2\rightarrow -(\alpha+n+1)/n$. For $\alpha+n+1>0$ and $\alpha>0$, $c_s^2$  is always positive. For $\alpha+n+1>0$ and $\alpha<0$,  $c_s^2$  is negative when $a<a_e$ and positive when $a>a_e$. For $\alpha+n+1<0$ and $\alpha<0$,  $c_s^2$  in always negative. For $\alpha+n+1<0$ and $\alpha>0$,  $c_s^2$  is positive when $a<a_e$ and negative when $a>a_e$. For $\alpha\le 1$ and $\alpha+2n+1<0$, the velocity of sound is always smaller than the speed of light. For $\alpha\le 1$ and $\alpha+2n+1>0$,  the velocity of sound is smaller than the speed of light when $a<a_s$ and larger when $a>a_s$.

Some curves giving the evolution of  $(c_s/c)^2$  as a function of
the scale factor $a$ are plotted in Figs. \ref{wKposNneg} and \ref{wKnegNneg}
for $k>0$ and $k<0$.

\section{Evolution of the scale factor}
\label{sec_dark}

\subsection{The deceleration parameter}
\label{sec_dec}

The deceleration parameter is defined by Eq. (I-77). In a flat universe, we obtain the relation of Eq. (I-78). The universe is decelerating when $q>0$ and accelerating when $q<0$.  For the equation of state (\ref{b10}), we get Eq. (I-79). The point $(a_c,\rho_c,T_c)$ defined by Eqs. (I-80) and (I-81) corresponds to a possible inflexion point ($q=\ddot a=0$) in the curve
$a(t)$. Different cases have to be considered.

We first assume $k>0$. When $a\rightarrow 0$, $q\rightarrow  (1+3\alpha)/2$; when $a\rightarrow a_*$, $q\rightarrow +\infty$. For $\alpha>-1/3$, the universe is always decelerating
($q>0$).  For $\alpha<-1/3$, the universe is accelerating when $a<a_c$ and decelerating when $a_c<a<a_*$.

We now assume $k<0$. When $a\rightarrow 0$, $q\rightarrow (1+3\alpha)/2$; when $a\rightarrow +\infty$, $q\rightarrow -1$. For $\alpha<-1/3$, the universe is always accelerating ($q<0$). For $\alpha>-1/3$, the universe is decelerating when $a<a_c$ and accelerating when $a>a_c$.

Some curves giving the evolution of  $q$  as a function of
the scale factor are plotted in Figs. \ref{wKposNneg} and \ref{wKnegNneg}
for $k>0$ and $k<0$.

\subsection{The differential equation}
\label{sec_darkdiff}

The temporal  evolution of the scale factor $a(t)$ is determined by the Friedmann equation (\ref{b4}).
Introducing the normalized radius $R=a/a_*$, the density (\ref{ges2}) can be written as Eq. (I-82).
Substituting this expression in Eq. (\ref{b4}), we obtain the differential equation
\begin{equation}
\label{dark6}
\dot R=\frac{\epsilon K R}{\lbrack R^{3(1+\alpha)/n}\mp 1\rbrack^{n/2}},
\end{equation}
where $K=({8\pi G\rho_*}/{3})^{1/2}$ and $\epsilon=\pm 1$. In general, we shall select the sign $\epsilon=+1$ corresponding to an expanding universe ($\dot R>0$), except in the case of a cyclic universe where both signs of $\epsilon$ must be considered. The solution of Eq. (\ref{dark6}) can be written as in Eq. (I-84) or, after a change of variables $x=R^{3(1+\alpha)/n}$, as in Eq. (I-85).

The solution of Eq. (\ref{dark6}) can be expressed in terms of hypergeometric functions. Some simple analytical expressions can be obtained for specific values of $n$. Actually, we can have a good idea of the behavior of the solution of Eq. (\ref{dark6}) by considering asymptotic limits (see below). The complete solution is represented in the figures by solving Eq. (\ref{dark6}) numerically.

\subsection{The case $k>0$}
\label{sec_darkreppos}

For $k>0$, the universe starts at $t=0$ with a vanishing radius $R=0$, an infinite density and an infinite pressure (Big Bang singularity). When $t\rightarrow 0$, we recover the solution corresponding to a  linear equation of state
\begin{eqnarray}
\label{dark9b}
R\sim \left \lbrack \frac{3(\alpha+1)}{2}Kt\right \rbrack^{2/\lbrack 3(1+\alpha)\rbrack},
\end{eqnarray}
\begin{eqnarray}
\label{toto1}
\frac{\rho}{\rho_*}\sim \frac{1}{\left \lbrack \frac{3}{2}(\alpha+1)Kt\right \rbrack^{2}}.
\end{eqnarray}
The final fate of the universe depends on the value of $n$. Different cases must be considered.

For $n<-2$, the radius tends to a finite value $R=1$ in infinite time, while the density and the pressure decrease to zero. When $t\rightarrow +\infty$,
\begin{eqnarray}
\label{dark11}
1-R\sim \left \lbrace \frac{2}{|n|-2}\left\lbrack \frac{|n|}{3(\alpha+1)}\right \rbrack^{|n|/2} \frac{1}{Kt}\right \rbrace^{2/(|n|-2)},\qquad
\end{eqnarray}
\begin{eqnarray}
\label{toto3}
\frac{\rho}{\rho_*}\sim  \left \lbrack \frac{2}{|n|-2} \frac{|n|}{3(\alpha+1)} \frac{1}{Kt}\right \rbrack^{2|n|/(|n|-2)}.
\end{eqnarray}
The density decreases algebraically rapidly.

For $n=-2$, the radius tends to a finite value $R=1$ in infinite time, while the density and the pressure decrease to zero. When $t\rightarrow +\infty$,
\begin{eqnarray}
\label{dark12}
1-R\sim Ae^{-\frac{3(1+\alpha)}{2}Kt},
\end{eqnarray}
\begin{eqnarray}
\label{toto4}
\frac{\rho}{\rho_*}\sim  \left\lbrack \frac{3}{2}(1+\alpha)A\right\rbrack^2e^{-3(1+\alpha)Kt}.
\end{eqnarray}
The density decreases exponentially rapidly.

For $n>-2$, the evolution of the universe is cyclic. Starting from $R=0$ at $t=0$ (Big Bang), the radius increases, reaches its maximum value $R=1$ at $t=t_*$, then decreases and reaches zero at $t=2t_*$ (Big Crunch), increases again and so on (since $\dot a=\infty$ at $t=2t_*$, it may not be allowed to extend the solution after $2t_*$). The density is infinite when $R=0$ and vanishes when $R=1$. When $t\rightarrow t_*$,
\begin{equation}
\label{dark10}
1-R\sim  \left \lbrace \frac{2-|n|}{2}\left\lbrack \frac{3(\alpha+1)}{|n|}\right \rbrack^{|n|/2} K(t_{*}-t)\right \rbrace^{2/(2-|n|)}.
\end{equation}
\begin{equation}
\label{toto2}
\frac{\rho}{\rho_*}\sim  \left \lbrack \frac{2-|n|}{2}\frac{3(\alpha+1)}{|n|} K(t_{*}-t)\right \rbrack^{2|n|/(2-|n|)}.
\end{equation}
The universe displays a future peculiarity at $t=t_{*}$ since its density vanishes (the universe ``disappears'') while its radius has a finite value $R=1$. At $t=t_*$,  the pressure tends to zero for $n<-1$, is constant  for $n=-1$, and is infinite for $n>-1$. Therefore, for $n>-1$ there is a future singularity (since $\ddot a=\infty$ at $t=t_*$, it may not be allowed to extend the solution after $t_*$ in that case).

\begin{figure}[!h]
\begin{center}
\includegraphics[clip,scale=0.3]{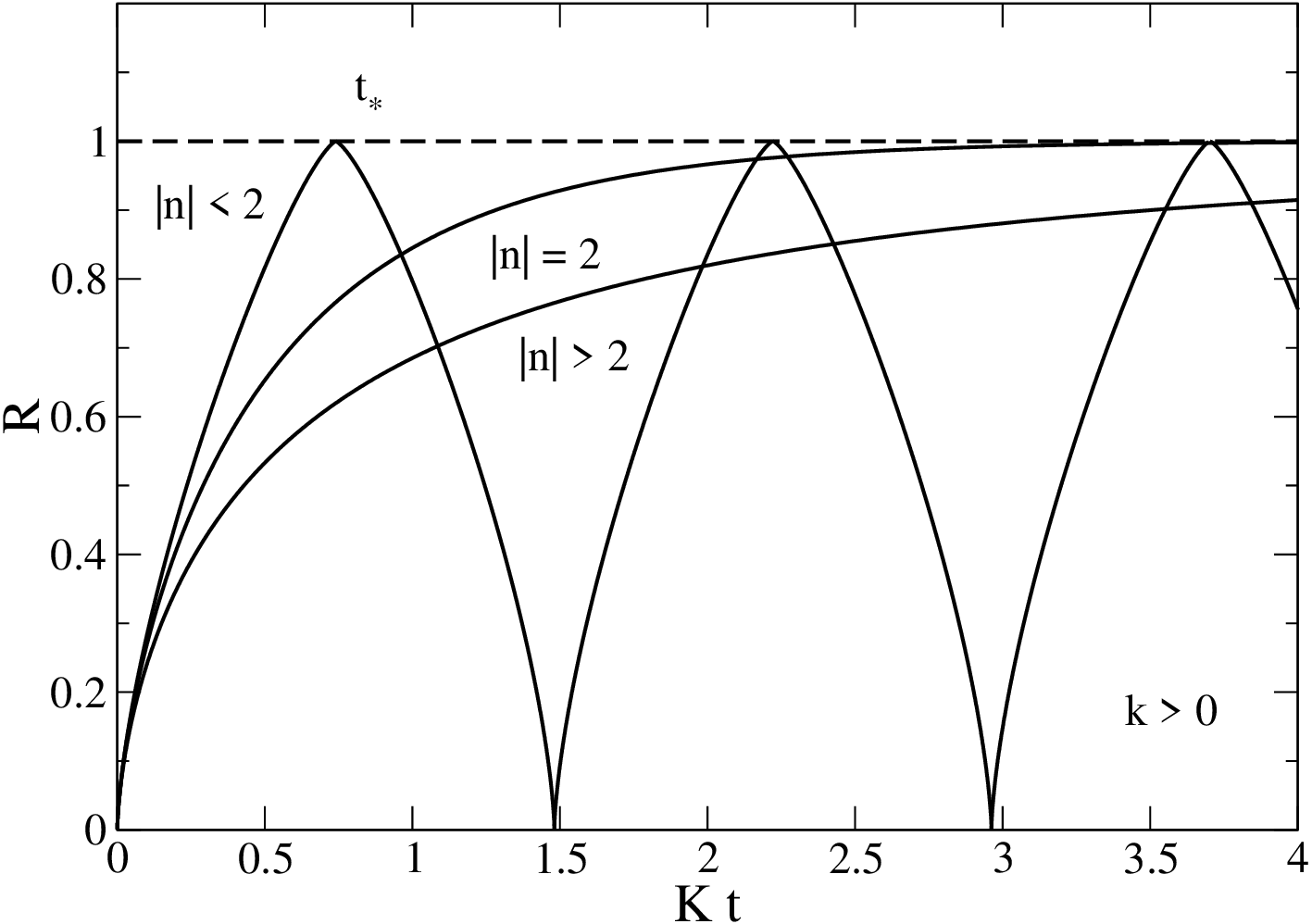}
\caption{Evolution of the radius $R$ as a function of time for $k>0$. We have represented the cases $n<-2$, $n=-2$, and $n>-2$ (specifically $n=-3$, $n=-2$, and $n=-1/2$). We have taken $\alpha=0$. The universe evolves towards a future peculiarity in which the density tends to zero while the radius tends to a finite value (see text for details). For $n>-1$, the pressure is infinite at $t=t_*$. For $n<-1$, the velocity of sound becomes larger than the speed of light when $R>R_s=\lbrace (\alpha+2n+1)/\lbrack n(1-\alpha)\rbrack \rbrace^{{n}/\lbrack {3(1+\alpha)}\rbrack}$.}
\label{rayonKposNneg}
\end{center}
\end{figure}

The evolution of the scale factor is represented in Fig. \ref{rayonKposNneg} in the different cases described above. Some simple analytical results can be obtained in particular cases.

For $n=-1$, using the identity
\begin{eqnarray}
\label{dark13}
\int \frac{1}{\sqrt{x-1}}\, \frac{dx}{x}=2 \arctan \left (\sqrt{x-1}\right ),
\end{eqnarray}
we obtain
\begin{eqnarray}
\label{dark14}
R=\sin^{2/\lbrack 3(1+\alpha)\rbrack}\left\lbrack \frac{3}{2}(1+\alpha)Kt\right\rbrack,
\end{eqnarray}
\begin{eqnarray}
\label{tata1}
\frac{\rho}{\rho_*}=\frac{1}{\tan^{2}\left\lbrack \frac{3}{2}(1+\alpha)Kt\right\rbrack}.
\end{eqnarray}
This describes a peculiar cyclic universe (see Sec. \ref{sec_cyclic}). The radius reaches its maximum value $R=1$ with a vanishing density $\rho=0$ for the first time at $Kt_*=\pi/\lbrack 3(1+\alpha)\rbrack$. We can explicitly check that Eq. (\ref{dark14}) has the asymptotic forms (\ref{dark9b}) and (\ref{dark10}).

For $n=-2$, using the identity
\begin{eqnarray}
\label{dark15}
\int \frac{1}{x-1}\, \frac{dx}{x}=\ln \left (\frac{x-1}{x}\right ),
\end{eqnarray}
we obtain
\begin{eqnarray}
\label{dark16}
R=\left\lbrack 1-e^{-\frac{3}{2}(1+\alpha)K t}\right\rbrack^{2/\lbrack 3(1+\alpha)\rbrack},
\end{eqnarray}
\begin{eqnarray}
\label{tata2}
\frac{\rho}{\rho_*}=\frac{1}{\left\lbrack e^{\frac{3}{2}(1+\alpha)K t}-1\right\rbrack^{2}}.
\end{eqnarray}
This describes a universe reaching a peculiarity $R=1$ with $\rho=0$ in infinite time. Equation (\ref{dark16}) has the asymptotic forms (\ref{dark9b}) and (\ref{dark12}) with $A=2/\lbrack 3(1+\alpha)\rbrack$.

For $n=-1/2$, we have  the identity
\begin{eqnarray}
\label{tutu2}
\int \frac{1}{(x-1)^{1/4}}\, \frac{dx}{x}=\sqrt{2}\arctan\left\lbrack 1+\sqrt{2}(x-1)^{1/4}\right\rbrack\nonumber\\
-\sqrt{2}\arctan\left\lbrack 1-\sqrt{2}(x-1)^{1/4}\right\rbrack\nonumber\\
+\ln\left\lbrack \frac{-1+\sqrt{2}(x-1)^{1/4}-\sqrt{x-1}}{1+\sqrt{2}(x-1)^{1/4}+\sqrt{x-1}}\right\rbrack,\qquad
\end{eqnarray}
which determines $t(R)$ using Eq. (I-85). For $\alpha=0$, this solution corresponds to the anti-Chaplygin gas $p=A/\rho$.

\subsection{The case $k<0$}
\label{sec_darkrepneg}

\begin{figure}[!h]
\begin{center}
\includegraphics[clip,scale=0.3]{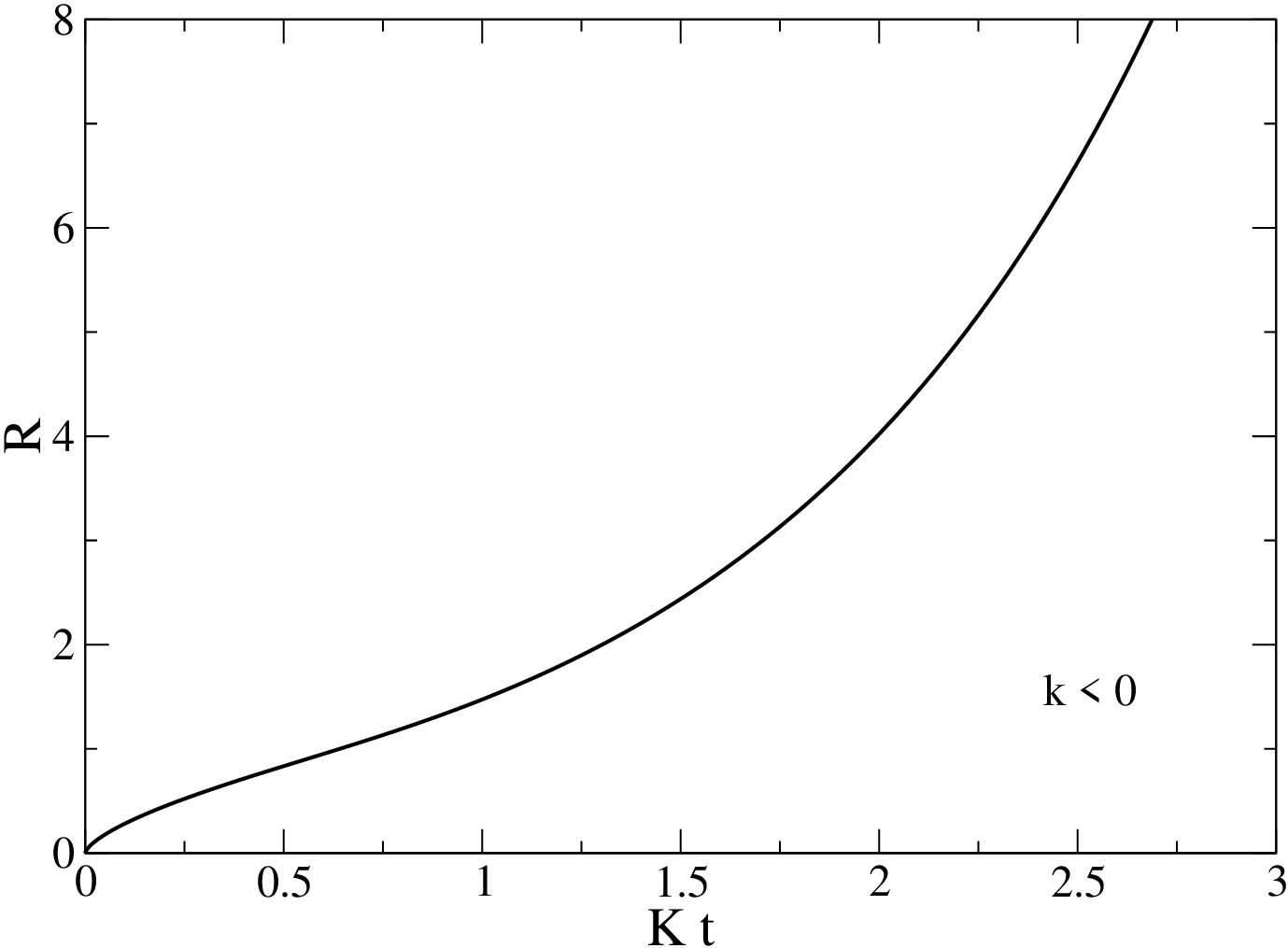}
\caption{Evolution of the radius $R$ as a function of time for $k<0$. We have taken $n=-1/2$ and $\alpha=0$. This corresponds to the original Chaplygin gas \cite{chaplygin}.}
\label{rayonKnegNneg}
\end{center}
\end{figure}

For $k<0$, the universe starts at $t=0$ with a vanishing radius $R=0$, an infinite density, and an infinite pressure (Big Bang singularity). When $t\rightarrow 0$, we recover the solution (\ref{dark9b})-(\ref{toto1}) corresponding to a  linear equation of state. When $t\rightarrow +\infty$, the density tends to a constant value $\rho_*$ leading to an exponential growth of the radius
\begin{eqnarray}
\label{dark18}
R\sim A e^{Kt}.
\end{eqnarray}
This late inflationary phase can mimic the effect of the cosmological constant. This was the original motivation of the Chaplygin gas model corresponding to $\alpha=0$ and  $n=-1/2$ \cite{chaplygin}. The evolution of the scale factor is represented in Fig. \ref{rayonKnegNneg}. Some simple analytical results can be obtained in particular cases.

For $n=-1$, using the identity
\begin{eqnarray}
\label{dark19}
\int \frac{1}{\sqrt{x+1}}\, \frac{dx}{x}=\ln \left (\frac{\sqrt{1+x}-1}{\sqrt{1+x}+1}\right ),
\end{eqnarray}
we obtain
\begin{eqnarray}
\label{dark20}
R=\sinh^{2/\lbrack 3(1+\alpha)\rbrack}\left\lbrack \frac{3}{2}(1+\alpha)Kt\right\rbrack,
\end{eqnarray}
\begin{eqnarray}
\label{tata3}
\frac{\rho}{\rho_*}=\frac{1}{\tanh^{2}\left\lbrack \frac{3}{2}(1+\alpha)Kt\right\rbrack}.
\end{eqnarray}
We can explicitly check that Eq. (\ref{dark20}) has the asymptotic forms (\ref{dark9b}) and (\ref{dark18}) with $A=1/2^{2/\lbrack 3(1+\alpha)\rbrack}$.

For $n=-2$, using the identity
\begin{eqnarray}
\label{dark21}
\int \frac{1}{x+1}\, \frac{dx}{x}=\ln \left (\frac{x}{1+x}\right ),
\end{eqnarray}
we obtain
\begin{eqnarray}
\label{dark22}
R=\left\lbrack e^{\frac{3}{2}(1+\alpha)K t}-1\right\rbrack^{2/\lbrack 3(1+\alpha)\rbrack},
\end{eqnarray}
\begin{eqnarray}
\label{tata4}
\frac{\rho}{\rho_*}=\frac{1}{\left\lbrack 1-e^{-\frac{3}{2}(1+\alpha)K t}\right\rbrack^{2}}.
\end{eqnarray}
Eq. (\ref{dark22}) has the asymptotic forms (\ref{dark9b}) and (\ref{dark18}) with $A=1$.

For $n=-1/2$, we have  the identity
\begin{eqnarray}
\label{tutu1}
\int \frac{1}{(x+1)^{1/4}}\, \frac{dx}{x}=2\arctan\left\lbrack (1+x)^{1/4}\right\rbrack\nonumber\\
+\ln\left\lbrack \frac{(1+x)^{1/4}-1}{(1+x)^{1/4}+1}\right\rbrack,
\end{eqnarray}
which determines  $t(R)$ using Eq. (I-85). For $\alpha=0$, this returns the result of \cite{chaplygin} for the Chaplygin gas $p=-A/\rho$.

\section{A model of late inflationary  universe}
\label{sec_solid}

The generalized equation of state (\ref{b10}) with $n<0$ and $k<0$ can be used to describe the transition between a phase where the universe has a linear equation of state ($p=\alpha\rho c^2$) and a phase of late inflation where its density is constant ($\rho=\rho_*$). Therefore, if we take $\alpha=0$, it describes the transition between the matter era and the dark energy era. This provides a model of late inflationary universe. For simplicity, we shall take $n=-1$. In this way, we obtain a model of late universe that is ``symmetric''  to the model of early universe studied in Paper I, corresponding to an index $n=1$.

\subsection{The basic equations}
\label{sec_sb}

For $\alpha=0$, $n=-1$, and $k<0$, Eqs. (\ref{b10}) and (\ref{ges2}) become
\begin{equation}
\label{s1}
p=k c^2,
\end{equation}
\begin{equation}
\label{s2}
\rho=\rho_{*}\left\lbrack \left (\frac{a_2}{a}\right )^{3}+ 1\right\rbrack,
\end{equation}
where $\rho_*=|k|$ and we have written $a_2$ for $a_*$.

When $a\gg a_2$, the density is approximately constant: $\rho\simeq \rho_*$. This gives rise to a phase of late inflation. The value $\rho_*$ defines a fundamental lower bound $\rho_{min}$ for the density. We shall identify it with the dark energy density $\rho_{\Lambda}=7.02\, 10^{-24}\, {\rm g}/{\rm
m}^3$ (see Appendix \ref{sec_pcscales}). Hence, we take
\begin{equation}
\label{s3}
\rho_*=\rho_{min}=\rho_{\Lambda}.
\end{equation}
This fixes $k=-\rho_{\Lambda}$. This is how the dark energy (cosmological constant) is taken into account in our model.

When $a\ll a_2$, Eq. (\ref{s2}) reduces to $\rho\simeq \rho_{\Lambda}(a_2/a)^3$ corresponding to a pressureless universe ($p=0$). This describes the matter era. The conservation of $\rho_{m} a^3$ implies $\rho_{\Lambda} a_2^3=\rho_{m,0}a_0^3$. Writing $\rho_{m,0}=\Omega_{m,0}\rho_0$ and  $\rho_{\Lambda}=\Omega_{\Lambda,0}\rho_0$ where $\Omega_{m,0}=\Omega_{B,0}+\Omega_{DM,0}$ is the present fraction of matter (including baryons and dark matter) and $\Omega_{\Lambda,0}$ is the present fraction of dark energy, we obtain
\begin{equation}
\label{s5a}
\frac{a_2}{a_0}=\left (\frac{\Omega_{m,0}}{\Omega_{\Lambda,0}}\right )^{1/3}=0.677,
\end{equation}
where we have used $\Omega_{B,0}=0.0455$, $\Omega_{DM,0}=0.1915$,  $\Omega_{m,0}=0.237$, and $\Omega_{\Lambda,0}=0.763$ \cite{bt}.  Introducing the cosmological length  $l_{\Lambda}=4.38\, 10^{26}{\rm m}$ (see Appendix \ref{sec_pcscales}), we obtain
\begin{equation}
\label{s5b}
\frac{a_2}{l_{\Lambda}}=\left (\frac{\Omega_{m,0}}{\Omega_{\Lambda,0}}\right )^{1/3}\left (\frac{3\Omega_{\Lambda,0}}{8\pi}\right )^{1/2}=0.204.
\end{equation}
The value of the  characteristic length $a_2$ is
\begin{equation}
\label{s5}
a_2=8.95\, 10^{25}\, {\rm m}.
\end{equation}
It corresponds to the typical radius of the universe at the transition between the matter era and the dark energy era.

Regrouping these results, we obtain the basic equations of the model
\begin{equation}
\label{s6}
p=-\rho_{\Lambda} c^2,
\end{equation}
\begin{equation}
\label{s6b}
w=-\frac{\rho_{\Lambda}}{\rho}, \qquad q=\frac{1}{2}\left (1-3\frac{\rho_{\Lambda}}{\rho}\right ),
\end{equation}
\begin{equation}
\label{s7}
\rho=\rho_{\Lambda}\left\lbrack \left (\frac{a_2}{a}\right )^{3}+ 1\right\rbrack.
\end{equation}
The temperature $T=T_*$ is constant. Finally, we note that $K=({8\pi}/{3})^{1/2}t_{\Lambda}^{-1}$ where we have introduced the cosmological time $t_{\Lambda}=1.46\, 10^{18} {\rm s}$ (see Appendix \ref{sec_pcscales}).

As the universe expands from $a=0$ to $a=+\infty$, the density decreases from $+\infty$ to $\rho_{\Lambda}$, the equation of state parameter $w$ decreases from $0$ to $-1$, and the deceleration parameter decreases from $1/2$ to $-1$.

{\it Remark:} The equation of state (\ref{s6}) used to describe the late universe dominated by dark matter and dark energy corresponds to a  constant negative pressure (the velocity of sound $c_s=0$).  This equation of state is related to, but different from, the linear equation of state  $p=-\rho c^2$. This linear equation of state describes only the dark energy era while the constant equation of state (\ref{s6}) describes {\it both} the matter era and the dark energy era in a unified manner. Furthermore, it can be given a scalar field representation (see Sec. \ref{sec_scalar}). It turns out that the equation of state (\ref{s6}) is equivalent to the standard $\Lambda$CDM model (see Sec. \ref{sec_late}), although this was not obvious {\it a priori}. The Chaplygin gas model \cite{chaplygin} is often presented as a unification of dark matter and dark energy. We note that a constant (not linear) negative pressure (\ref{s6}) does the same job. In our opinion, this gives a new light to the $\Lambda$CDM model.

\subsection{The inflationary phase}
\label{sec_ip}

During the inflationary phase ($a\gg a_2$),  the density has an approximately constant value $\rho\simeq \rho_{\Lambda}$. The corresponding Hubble parameter is
\begin{equation}
\label{s9}
H=\frac{\dot a}{a}\simeq \left (\frac{8\pi G}{3}\rho_{\Lambda}\right )^{1/2}\simeq \left (\frac{8\pi}{3}\right )^{1/2}\frac{1}{t_{\Lambda}}.
\end{equation}
Numerically, $\rho=7.02\, 10^{-24}\, {\rm g}/{\rm
m}^3$ and $H=1.98\, 10^{-18}\, {\rm s}^{-1}$. Integrating Eq. (\ref{s9}), we find that the scale factor increases with time as
\begin{equation}
\label{s10}
a(t)\propto e^{(8\pi/3)^{1/2}t/t_{\Lambda}}.
\end{equation}
This corresponds to de Sitter's solution.  The radius of the universe increases  exponentially rapidly on a timescale of the order of the cosmological time $t_{\Lambda}=1.46\, 10^{18} {\rm s}$. The universe exists at any time in the future and there is no singularity.

\subsection{The matter era}
\label{sec_rp}

Before the late inflation ($a\ll a_2$), the universe is in the matter era. In that case, due to the smallness of $\rho_{\Lambda}$, we can take $p=0$ and we recover the standard Einstein-de Sitter (EdS)  model. The density is related to the scale factor by $\rho\sim \rho_{\Lambda} a_2^3/a^3$.  The Friedmann equation (\ref{b9}) becomes
\begin{equation}
\label{s11}
H=\frac{\dot a}{a}\sim \left (\frac{8\pi G}{3}\frac{\rho_{\Lambda} a_2^3}{a^3}\right )^{1/2}\sim \left (\frac{8\pi}{3}\right )^{1/2}\frac{1}{t_{\Lambda}}\left (\frac{a_2}{a}\right )^{3/2},
\end{equation}
yielding
\begin{equation}
\label{s12}
\frac{a}{a_2}\sim (6\pi)^{1/3}\left (\frac{t}{t_{\Lambda}}\right )^{2/3},
\end{equation}
where we have determined the constant of integration such that $a=0$ at $t=0$. This is not quite correct because we should match the matter era to the radiation era (see Sec. \ref{sec_simple}), but this is a good approximation. Using the approximate expression (\ref{s12}) of the scale factor in the matter era, we find that the time at which $a=a_2$ is
\begin{equation}
\label{s13}
t_2=\frac{t_{\Lambda}}{\sqrt{6\pi}}.
\end{equation}
Numerically, $t_2=0.230 t_{\Lambda}=3.37\, 10^{17}{\rm s}$. This marks the end of the matter era and the beginning of the dark energy era. This numerical value will be revised in Sec. \ref{sec_gp} using the exact solution (\ref{s15}) of the Friedmann equations.  We also have
\begin{equation}
\label{s14}
\rho\sim \frac{\rho_{\Lambda}}{(a/a_2)^3}\sim \frac{\rho_{\Lambda}}{6\pi}\left (\frac{t_{\Lambda}}{t}\right )^2.
\end{equation}
During the matter era, the density decreases algebraically as the universe expands.

\subsection{The general solution}
\label{sec_gp}

The equation of state (\ref{s6}) interpolates smoothly between the matter era described by a density $\rho\propto a^{-3}$ and the dark energy era described by a constant density $\rho=\rho_{\Lambda}$ (cosmological density). It provides therefore a unified description of the late universe. For the equation of state (\ref{s6}), the general solution of the Friedmann equation (\ref{b9}) is
\begin{eqnarray}
\label{s15}
\frac{a}{a_2}=\sinh^{2/3}\left ( \sqrt{6\pi}\frac{t}{t_{\Lambda}}\right ).
\end{eqnarray}
The density evolves as
\begin{eqnarray}
\label{s15b}
\rho=\frac{\rho_{\Lambda}}{\tanh^{2}\left (\sqrt{6\pi}\frac{t}{t_{\Lambda}}\right )}.
\end{eqnarray}
The time at which $a=a_2$ is
\begin{equation}
\label{s16}
t_2=\frac{1}{\sqrt{6\pi}}{\rm argsinh}(1)t_{\Lambda}.
\end{equation}
Numerically $t_2=0.203 t_{\Lambda}=2.97\, 10^{17}{\rm s}$. The corresponding density is $\rho_2=2\rho_{\Lambda}=1.40\, 10^{-23}\, {\rm g}/{\rm
m}^3$. For $t\ll t_2$, the universe is in the matter era. Its radius increases algebraically according to Eq. (\ref{s12}) while its density decreases algebraically according to Eq. (\ref{s14}). For $t\gg t_2$, the universe is in the dark energy era. Its radius increases exponentially according to Eq. (\ref{s10}) while its density remain approximately constant $\rho\sim \rho_{\Lambda}$. From Eq. (\ref{s15}) we obtain the asymptotic behavior
\begin{equation}
\label{s17}
a(t)\sim l_{\Lambda} e^{(8\pi/3)^{1/2}(t-t_f)/t_{\Lambda}},
\end{equation}
with
\begin{equation}
\label{s18}
t_f=\left (\frac{3}{8\pi}\right )^{1/2}\left\lbrack \frac{2}{3}\ln 2+\ln\left (\frac{l_{\Lambda}}{a_2}\right )\right \rbrack t_{\Lambda}.
\end{equation}
Numerically $t_f=0.708 t_{\Lambda}=1.03\, 10^{18}{\rm s}$. For $t>t_f$, the radius of the universe is larger than the cosmological length $l_{\Lambda}=4.38\, 10^{26}{\rm m}$.

\begin{figure}[!h]
\begin{center}
\includegraphics[clip,scale=0.3]{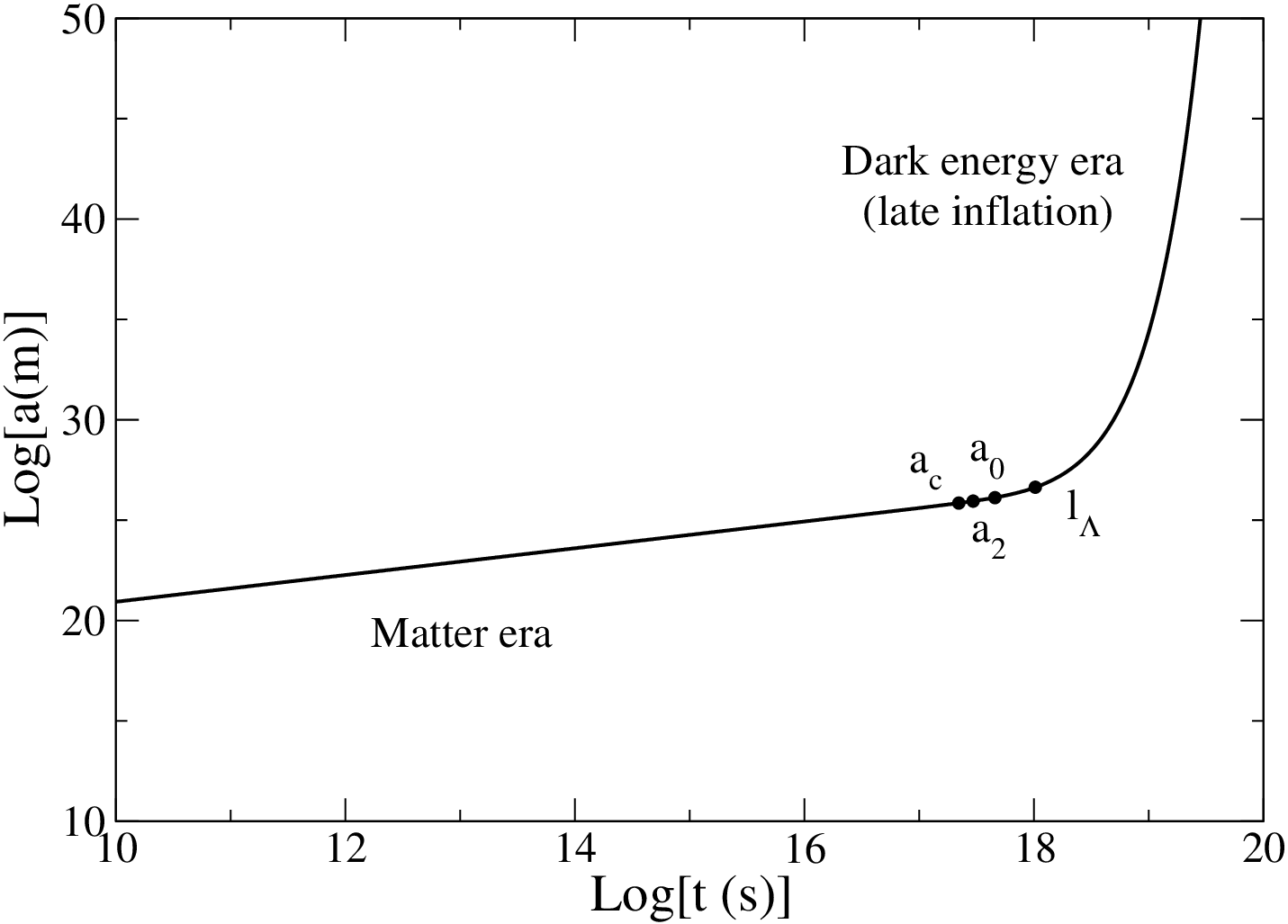}
\caption{Evolution of the scale factor $a$ with the time $t$ in logarithmic scales. This figure clearly shows the transition between the matter era (EdS) and the dark energy era (de Sitter). In the dark energy era, the radius increases exponentially rapidly on a timescale of the order of the cosmological time $t_{\Lambda}=1.46\, 10^{18} {\rm s}$. This corresponds to a phase of late inflation. The universe is decelerating for $a<a_c$ and accelerating for $a>a_c$ with $a_c=7.11\, 10^{25} {\rm m}$. The transition between the matter era and the dark energy era takes place at $a_2=8.95\, 10^{25} {\rm m}$. Coincidentally, the present universe turns out to be close to the transition point ($a_0\sim a_2$).  }
\label{tasansradiationLOGLOG}
\end{center}
\end{figure}

\begin{figure}[!h]
\begin{center}
\includegraphics[clip,scale=0.3]{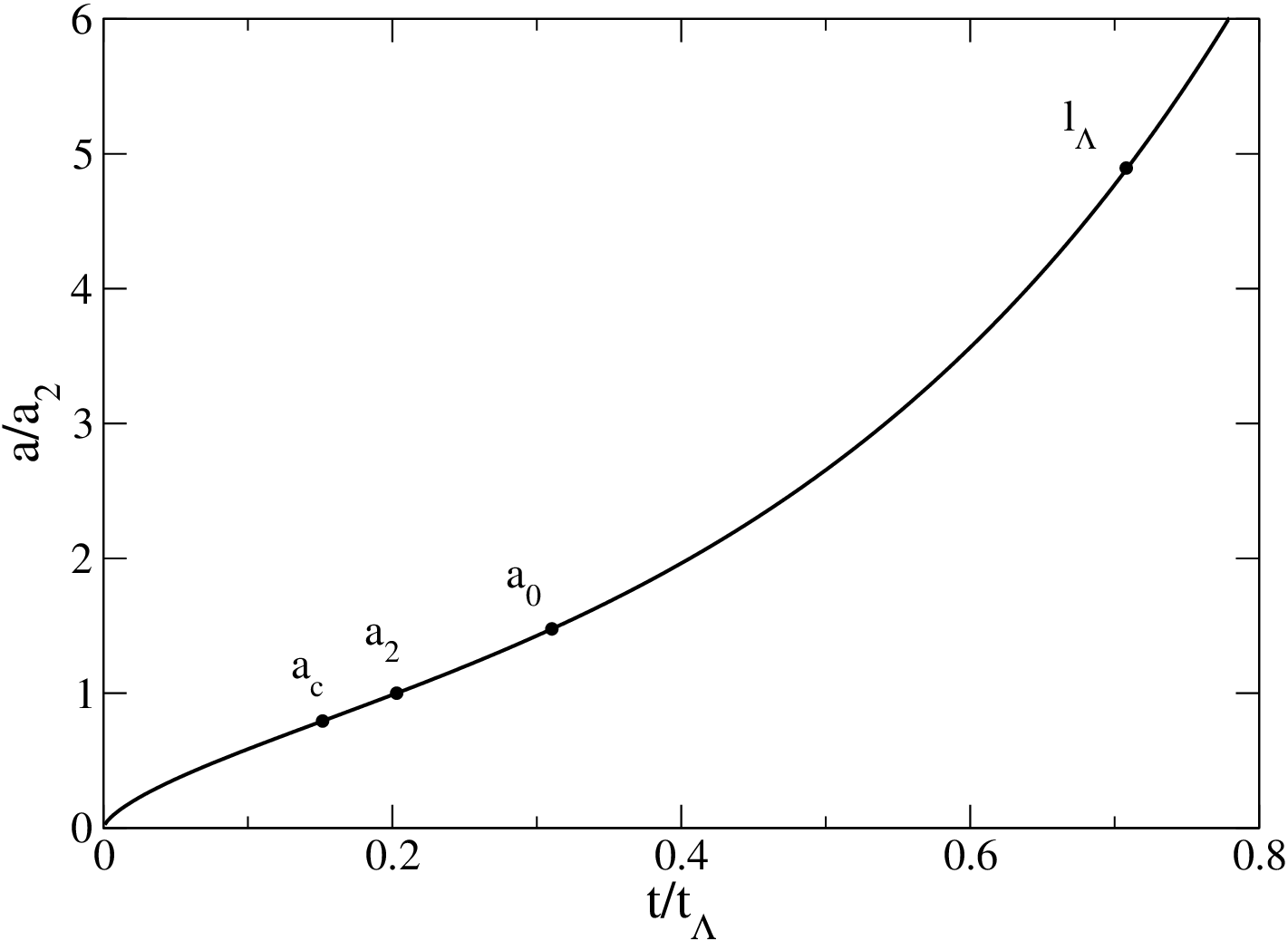}
\caption{Evolution of the scale factor $a$ with the time $t$ in linear scales.}
\label{tasansradiationLINLIN}
\end{center}
\end{figure}

\begin{figure}[!h]
\begin{center}
\includegraphics[clip,scale=0.3]{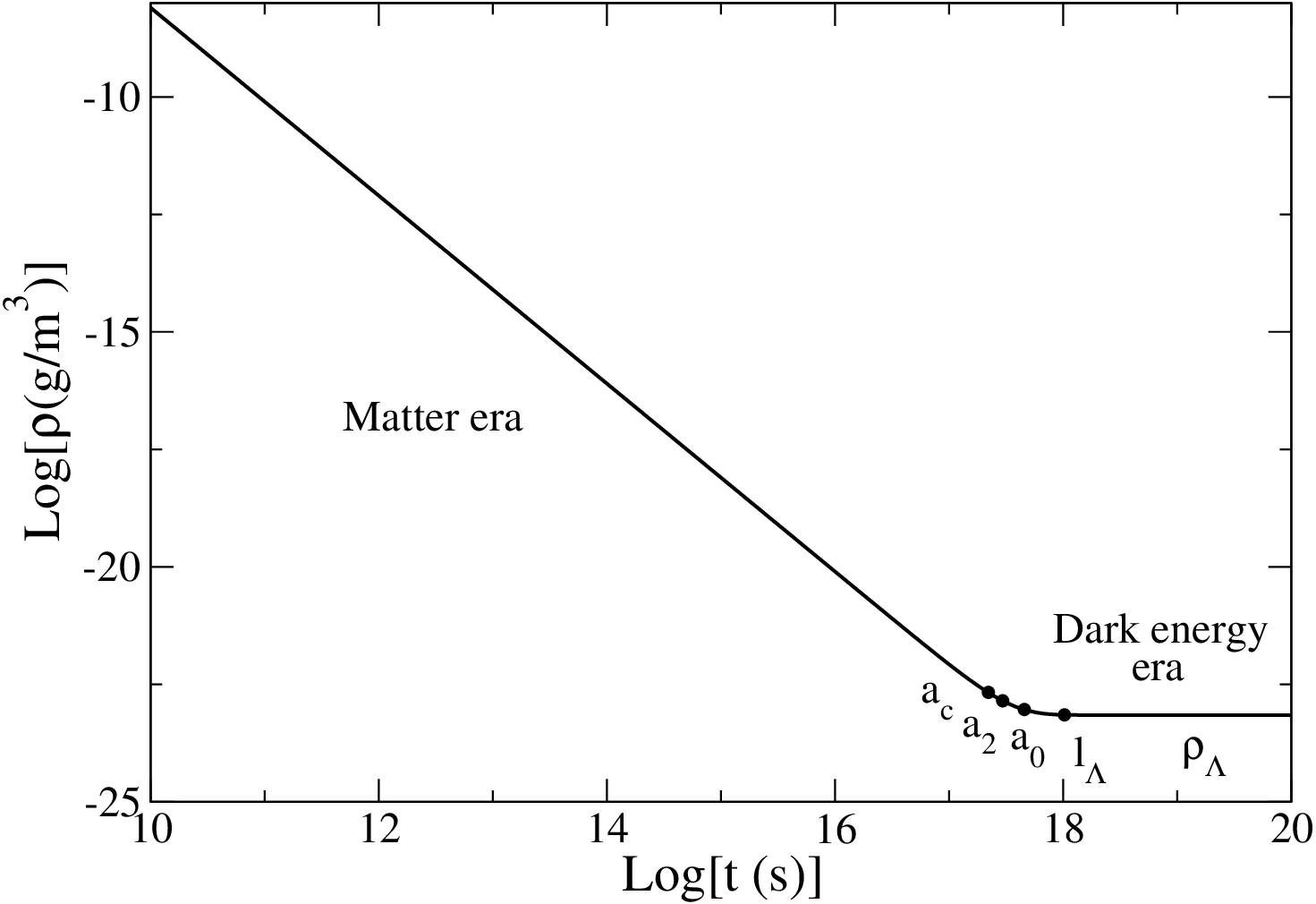}
\caption{Evolution of the density $\rho$ with the time $t$ in logarithmic scales. During the late inflation, the density remains approximately constant, with the cosmological value $\rho_{min}=\rho_{\Lambda}$ representing a lower bound.}
\label{trhosansradiationLOGLOG}
\end{center}
\end{figure}

\begin{figure}[!h]
\begin{center}
\includegraphics[clip,scale=0.3]{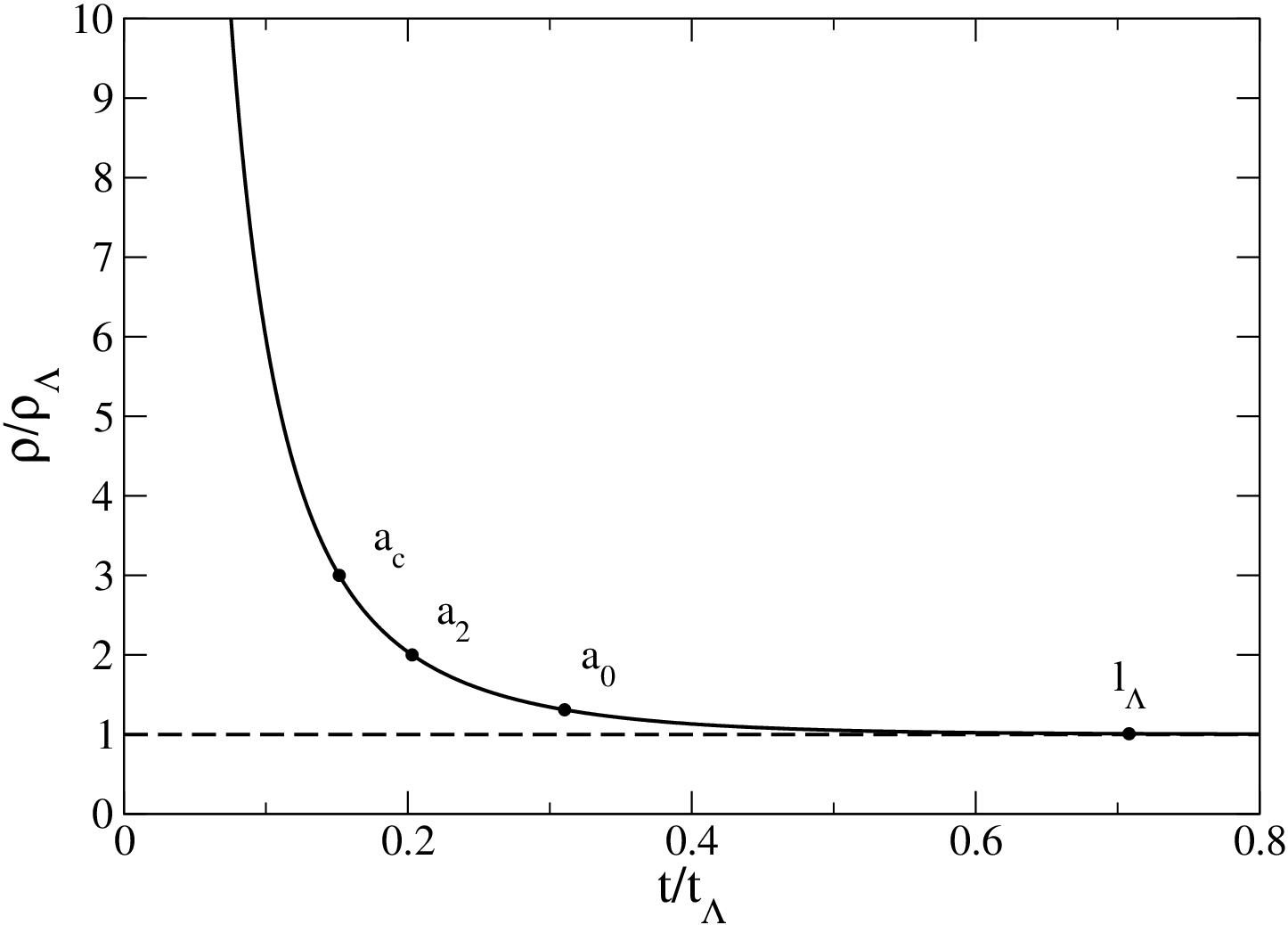}
\caption{Evolution of the density $\rho$ with the time $t$ in linear scales. General relativity (cosmological constant) limits the decay of the density to the cosmological value $\rho_{\Lambda}=7.02\, 10^{-24}\, {\rm g}/{\rm
m}^3$.}
\label{trhosansradiationLINLIN}
\end{center}
\end{figure}

Using the results of Sec. \ref{sec_dec}, we find that the universe is decelerating for $a<a_c$ ({\it i.e.}
$\rho>\rho_c$) and accelerating for $a>a_c$ ({\it i.e.}  $\rho<\rho_c$) where
\begin{equation}
\label{s19}
\rho_c=3\rho_{\Lambda}, \qquad a_c=(1/2)^{1/3}a_2.
\end{equation}
The time $t_c$ at which the universe starts accelerating is given by
\begin{equation}
\label{s20}
{t_c}=\frac{1}{\sqrt{6\pi}}{\rm argsinh}\left (\frac{1}{\sqrt{2}}\right )t_{\Lambda}.
\end{equation}
This corresponds to the time at which the curve $a(t)$ presents an inflexion point. Numerically, $\rho_c=3\,\rho_{\Lambda}=2.11\, 10^{-23} {\rm g}/{\rm m}^3$, $a_c=0.794 a_2=7.11\, 10^{25} {\rm m}$, and $t_c=0.152\, t_{\Lambda}=2.22\, 10^{17} {\rm s}$.

The evolution of the scale factor and  density  as a function of time are represented in Figs. \ref{tasansradiationLOGLOG}-\ref{trhosansradiationLINLIN} in logarithmic and linear scales.

\subsection{The present universe}
\label{sec_present}

The Hubble radius, the density and the Hubble time of the present universe are  $a_0=0.302 l_{\Lambda}=1.32\, 10^{26}{\rm m}$, $\rho_0=1.31\rho_{\Lambda}=9.20\, 10^{-24} {\rm g}/{\rm m}^3$, and $H_0^{-1}=0.302 t_{\Lambda}=4.41\, 10^{17} {\rm s}$. Using Eq. (\ref{s15}), the age of the universe is given by
\begin{equation}
\label{s21}
t_{0}=\frac{1}{\sqrt{6\pi}}{\rm argsinh}\left\lbrack \left (\frac{a_0}{a_2}\right )^{3/2}\right \rbrack t_{\Lambda}.
\end{equation}
Numerically, $t_{0}=0.310 t_{\Lambda}=4.54\, 10^{17}\, {\rm s}$. According to Eq. (\ref{s6b}), the present values of the deceleration parameter and of the equation of state parameter are
\begin{equation}
\label{s22}
q_0=\frac{1-3\Omega_{\Lambda,0}}{2}, \qquad w_0=-\Omega_{\Lambda,0}.
\end{equation}
Numerically,  $q_0=-0.645$ and $w_0=-0.763$.

It is striking to note that the present size of the universe
$a_0=1.32\, 10^{26}{\rm m}$ is of the order of the scale $a_2=8.95\,
10^{25}\, {\rm m}$ ($a_0=1.48 a_2$). Therefore, we live just at the
transition between the matter era and the dark energy era (see bullets
in
Figs. \ref{tasansradiationLOGLOG}-\ref{trhosansradiationLINLIN}). Another
way to state this result is to say that the present ratio
$\Omega_{\Lambda,0}/\Omega_{m,0}=3.22$ between dark energy and dark
matter is of order unity.  This coincidence is intriguing and often
referred to as the ``cosmic coincidence problem'' \cite{ccp}. Several
theories have been proposed to explain why
$\Omega_{\Lambda,0}/\Omega_{m,0}\sim 1$ \cite{ccptheories}. However,
this may be just a pure coincidence without deeper reason. Life (and
researchers inquiring about cosmology) may have emerged $\sim 14\,
{\rm Gyrs}$ after the Big Bang, precisely at the epoch where
$\Omega_{\Lambda}/\Omega_{m}\sim 1$. We leave this ``problem'' open.

\section{A model of peculiar cyclic universe}
\label{sec_cyclic}

The equation of state (\ref{b10}) with $n<0$ and $k>0$ leads to a model of non-inflationary universe exhibiting a future peculiarity. We take the same values of $\alpha$, $n$, $\rho_*$, and $a_2$ as in the previous section. We also take $k=\rho_{\Lambda}$. The basic  equations of the peculiar model are
\begin{equation}
\label{s23}
p=\rho_{\Lambda} c^2,
\end{equation}
\begin{equation}
\label{s24}
w=\frac{\rho_{\Lambda}}{\rho}, \qquad q=\frac{1}{2}\left (1+3\frac{\rho_{\Lambda}}{\rho}\right ),
\end{equation}
\begin{equation}
\label{s25}
\rho=\rho_{\Lambda}\left\lbrack \left (\frac{a_2}{a}\right )^{3}-1\right\rbrack.
\end{equation}
The temperature $T=T_*$ is constant. As the universe expands from $a=0$ to $a_2$, the density decreases from $+\infty$ to $0$, the parameter $w$ increases from $0$ to $+\infty$ and the deceleration parameter increases from $1/2$ to $+\infty$.

\begin{figure}[!h]
\begin{center}
\includegraphics[clip,scale=0.3]{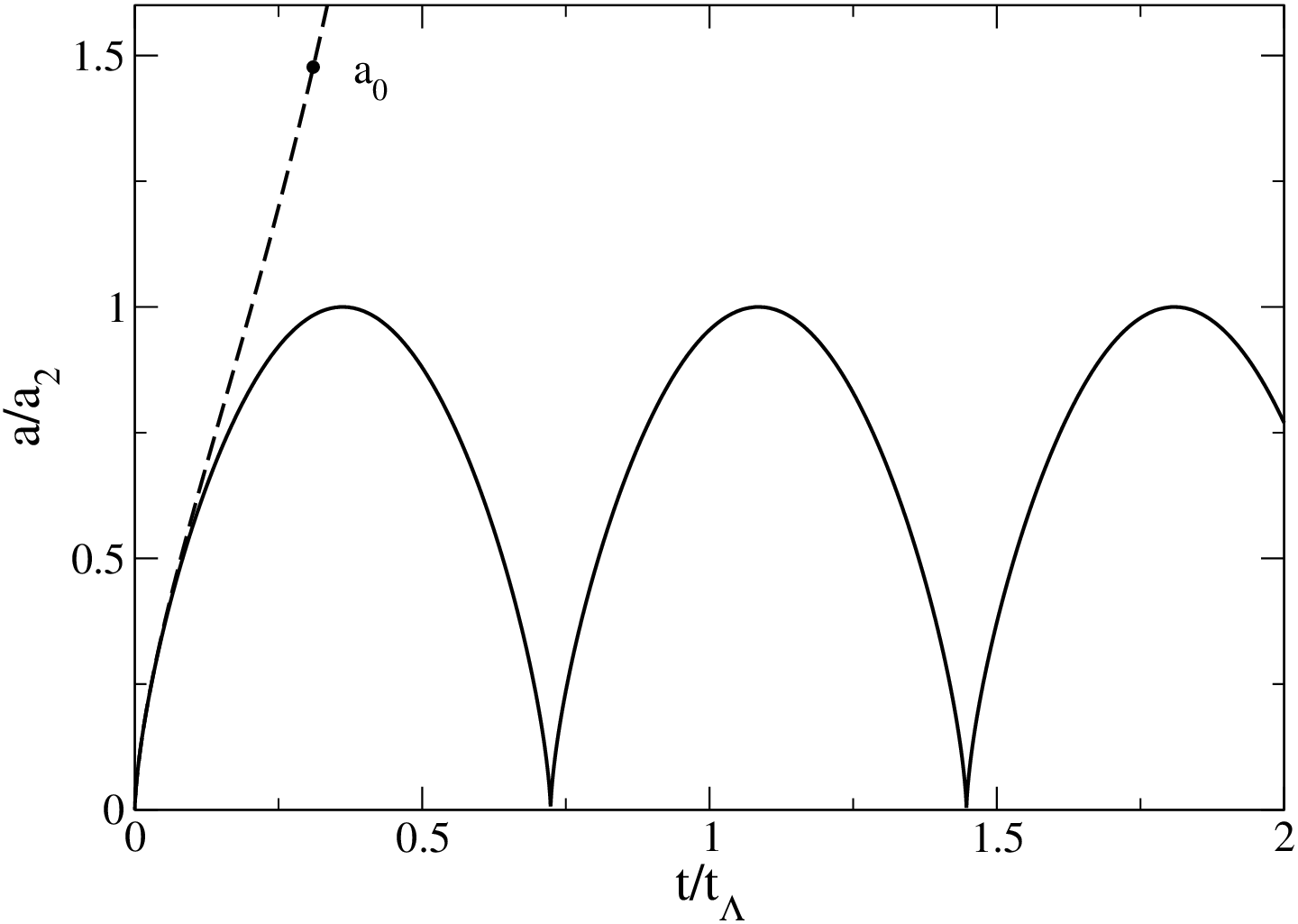}
\caption{Evolution of the scale factor $a$ with the time $t$. The dotted line corresponds to the real evolution of the universe according to the model of Sec. \ref{sec_solid}. The bullet corresponds to the value of the scale factor at the time $t_0=0.310t_{\Lambda}=4.54\, 10^{17}{\rm s}$ corresponding to the age of our universe. At this epoch, the cyclic universe has deviated relatively strongly from the real evolution.}
\label{cycliqueradius}
\end{center}
\end{figure}

\begin{figure}[!h]
\begin{center}
\includegraphics[clip,scale=0.3]{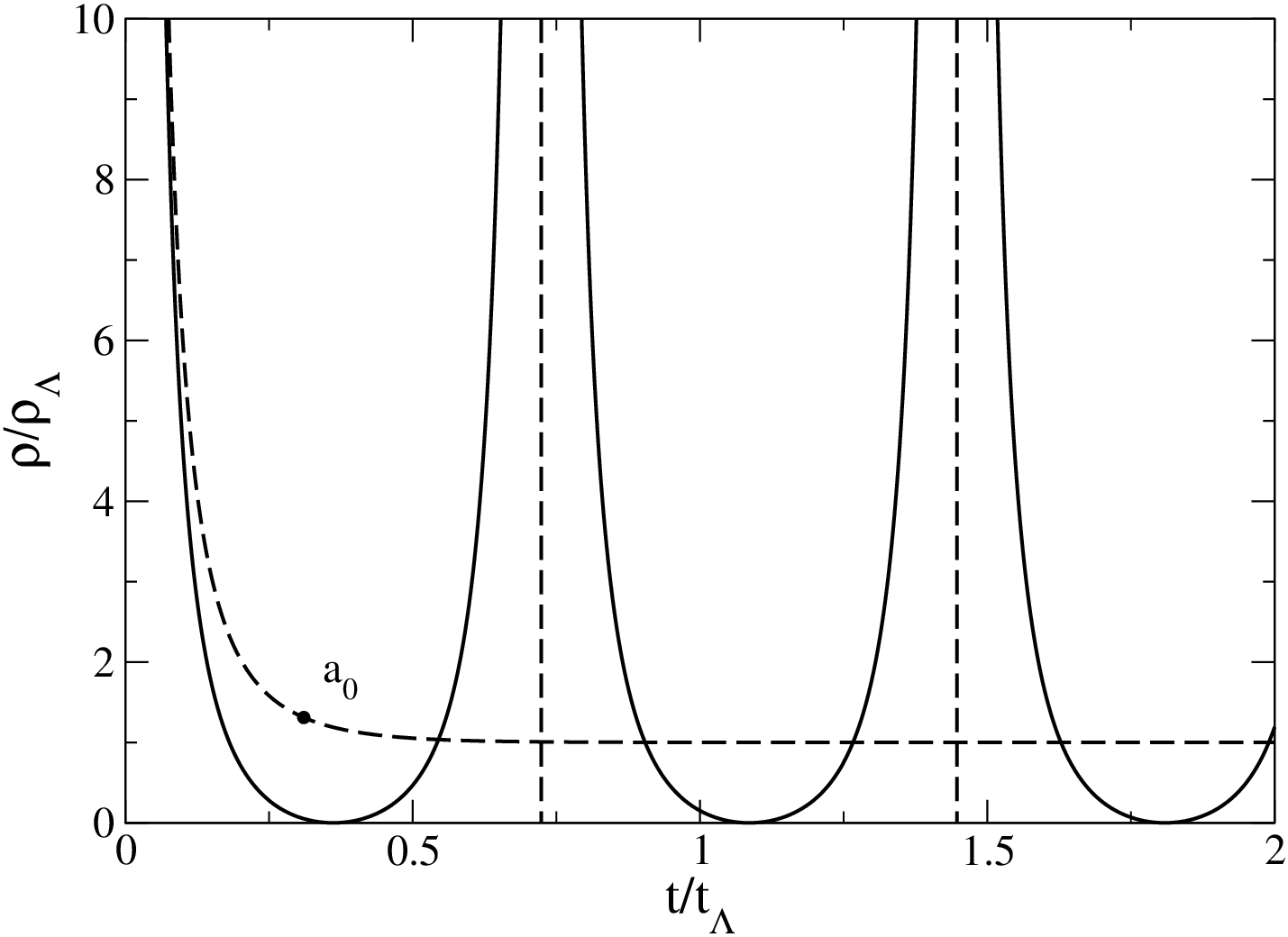}
\caption{Evolution of the density $\rho$ with the time $t$. The density vanishes periodically at the moment where the universe reaches its maximum size. On the other hand, it becomes infinite when the universe contracts to a point (Big Bang or Big Crunch).}
\label{cycliquedensity}
\end{center}
\end{figure}

For the equation of state (\ref{s23}), the general solution of the Friedmann equation (\ref{b9}) is
\begin{equation}
\label{s26}
\frac{a}{a_2}=\sin^{2/3}\left (\sqrt{6\pi}\frac{t}{t_{\Lambda}}\right ).
\end{equation}
The density evolves as
\begin{equation}
\label{s27}
\frac{\rho}{\rho_{\Lambda}}=\frac{1}{\tan^{2}\left (\sqrt{6\pi}\frac{t}{t_{\Lambda}}\right )}.
\end{equation}
This solution describes a peculiar cyclic universe. The universe reaches its maximum radius $a_2=0.204 l_{\Lambda}=8.95\, 10^{25}\, {\rm m}$ for the first time at
\begin{equation}
\label{s28}
t_2=\left (\frac{\pi}{24}\right )^{1/2}t_{\Lambda}.
\end{equation}
Numerically, $t_2=0.362t_{\Lambda}=5.29\, 10^{17}{\rm s}$. For $t\ll t_2$, we can make the approximation $p\simeq 0$. Therefore, this universe evolves just like the Einstein-de Sitter (EdS) universe describing the matter era (see Sec. \ref{sec_rp}). Its radius increases algebraically according to Eq. (\ref{s12}) while its density decreases algebraically according to Eq. (\ref{s14}).  When $t\rightarrow t_2$, the radius approaches its maximum value according to
\begin{equation}
\label{s29}
1-\frac{a}{a_2}\sim \frac{2\pi}{t_{\Lambda}^2}(t_{2}-t)^2.
\end{equation}
Correspondingly, the density decreases to zero as
\begin{equation}
\label{s30}
\frac{\rho}{\rho_{\Lambda}}\sim \frac{6\pi}{t_{\Lambda}^2}(t_{2}-t)^2.
\end{equation}
For $t>t_2$, the radius decreases with time while the density increases (this corresponds to the solution of Eq. (\ref{dark6}) with $\epsilon=-1$).  The radius vanishes at $t=2t_2$ (Big Crunch), then increases again. The density vanishes when $a=a_2$ and is infinite when $a=0$. This process continues periodically. Although there is no divergence at $t=t_2$, this model is peculiar because, when the radius reaches its maximum value $a_2$, the density vanishes ($\rho=0$), so the universe becomes empty. In a sense, the universe ``disappears'' at $t=t_2$, before re-appearing at $t>t_2$, and this in a periodic manner. According to this model, the universe would disappear for the first time at $t_2=0.362t_{\Lambda}=5.29\, 10^{17}{\rm s}$ that is not very far from the present age of the universe $t_0=0.310t_{\Lambda}=4.54\, 10^{17}{\rm s}$. Therefore, if we were living in this cyclic universe, the end of the world (in the sense of the whole universe) would take place in about $2.38$ billion years. We could fear to live in this universe. However, this universe is always decelerating in the phase of expansion. Therefore, the fact that our universe is accelerating (an observation that was made only recently \cite{novae}) shows that we do not live in this cyclic universe. In a sense, this is reassuring.

The evolution of the scale factor and density  as a function of time are represented in Figs. \ref{cycliqueradius} and \ref{cycliquedensity}.

{\it Remark:} The equation of state (\ref{s6})  corresponds to a  constant positive pressure (the velocity of sound $c_s=0$). This model turns out to be equivalent to the cold dark matter model with a negative cosmological constant (anti-$\Lambda$CDM model), although this was not obvious {\it a priori}. This is an example of cyclic universe corresponding to $k=0$, $\Lambda<0$ and $p=0$. Other types of cyclic universes  (corresponding to $k=\pm 1$, $\Lambda<0$ and $p=0$, or $k=+1$, $\Lambda=0$ and $p=0$) were studied by Friedmann \cite{friedmann}, Einstein \cite{einsteinzero}, and Lema\^itre \cite{lemaitre1933}, and were called ``phoenix-universes'' by Lema\^itre.

\section{A simple model for the whole evolution of the universe}
\label{sec_simple}

\subsection{The early universe}
\label{sec_early}

In Paper I, we have described the evolution of the early universe, corresponding to the  transition between the pre-radiation era and the radiation era, by a single equation of state of the form (\ref{b10}) with $\alpha=1/3$, $n=1$ and $k=-4/(3\rho_P)$, namely
\begin{equation}
\label{early1}
p=\frac{1}{3}\rho (1-4\rho/\rho_P)c^2.
\end{equation}
This equation of state leads to a non-singular early inflationary universe. The density is related to the scale factor by
\begin{equation}
\label{early2}
\rho=\frac{\rho_P}{(a/a_1)^4+1},
\end{equation}
where $a_1=2.61\, 10^{-6}{\rm m}$. This characteristic scale marks the transition between the pre-radiation era and the radiation era.

When $a\ll a_1$, the density tends to a maximum value
\begin{equation}
\label{early3}
\rho=\rho_{max}=\rho_P,
\end{equation}
identified with the Planck density $\rho_P=5.16\, 10^{99}{\rm g}/{\rm m}^3$. This leads to a phase of early inflation
\begin{equation}
\label{early4}
a\propto e^{(8\pi/3)^{1/2}t/t_{P}},
\end{equation}
whose timescale is the Planck time $t_P=5.39\, 10^{-44}{\rm s}$.

When $a\gg a_1$, we recover the equation $\rho_{rad}\sim\rho_P a_1^4/a^4$ corresponding to the pure radiation described by an equation of state $p=\rho c^2/3$.  The conservation of $\rho_{rad}a^4$ implies that $\rho_P a_1^4=\rho_{rad,0}a_0^4$. Writing $\rho_{rad,0}=\Omega_{rad,0}\rho_0$, we can rewrite Eq. (\ref{early2}) in the form
\begin{equation}
\label{early5}
\rho=\frac{\Omega_{rad,0}\rho_0}{(a/a_0)^4+ (a_1/a_0)^4},
\end{equation}
where $a_1/a_0=1.97\, 10^{-32}$. When $a\gg a_1$, it reduces to
\begin{equation}
\label{early6}
\rho_{rad}=\frac{\Omega_{rad,0}\rho_0}{(a/a_0)^4},
\end{equation}
which corresponds to the pure radiation. In the radiation era, the scale factor, the density, and the temperature evolve as  $a\propto t^{1/2}$, $\rho\propto t^{-2}$ and $T\propto t^{-1/2}$.

The Friedmann equation (\ref{b9}) corresponding to Eq. (\ref{early5})  can be written
\begin{equation}
\label{early7}
\frac{H}{H_0}=\sqrt{\frac{\Omega_{rad,0}}{(a/a_0)^4+(a_1/a_0)^4}}.
\end{equation}
It has the analytical solution given by Eq. (I-128). The transition between the pre-radiation era and the radiation era takes place at a typical time $t_1=23.3 t_P=1.25\, 10^{-42}{\rm s}$ corresponding to $a=a_1$. This time also corresponds to the inflexion point of the curve $a(t)$. The universe is accelerating for $t<t_1$ and decelerating for $t>t_1$.

{\it Remark:} The equation of state (\ref{early1}) smoothly interpolates between a negative pressure vacuum energy regime ($p=-\rho c^2$, $\rho=\rho_P$) and a relativistic radiation phase ($p=\rho c^2/3$, $\rho\propto a^{-4}$). This amounts to summing the {\it inverse} of the densities of these two components taken independently. Indeed, Eq. (\ref{early5}) can be rewritten as
\begin{equation}
\label{early8}
\frac{1}{\rho}=\frac{1}{\rho_{rad}}+\frac{1}{\rho_P}.
\end{equation}

\subsection{The late universe}
\label{sec_late}

In this paper, we have described the evolution of the late universe, corresponding to the  transition between the matter era and the dark energy era, by a single equation of state of the form (\ref{b10}) with $\alpha=0$, $n=-1$ and $k=-\rho_{\Lambda}$, namely
\begin{equation}
\label{late1}
p=-\rho_{\Lambda}c^2.
\end{equation}
This equation of state leads to a non-singular late inflationary universe. The density is related to the scale factor by
\begin{equation}
\label{late2}
\rho=\rho_{\Lambda}\left\lbrack \left (\frac{a_2}{a}\right )^{3}+ 1\right\rbrack,
\end{equation}
where $a_2=0.204 l_{\Lambda}=8.95\, 10^{25}{\rm m}$.  This characteristic scale marks the transition between the matter era and the dark energy era.

When $a\gg a_2$, the density tends to a minimum value
\begin{equation}
\label{late3}
\rho=\rho_{min}=\rho_{\Lambda},
\end{equation}
identified with the cosmological density $\rho_{\Lambda}=7.02\, 10^{-24}{\rm g}/{\rm m}^3$. This leads to a phase of late inflation
\begin{equation}
\label{late4}
a\propto e^{(8\pi/3)^{1/2}t/t_{\Lambda}},
\end{equation}
whose timescale is the cosmological time $t_{\Lambda}=1.46\, 10^{18}{\rm s}$.

When $a\ll a_2$, we recover the equation $\rho_{m}\sim \rho_{\Lambda} a_2^3/a^3$ corresponding to the pressureless  matter described by an equation of state $p=0$.  The conservation of $\rho_{m}a^3$ implies that $\rho_{\Lambda} a_2^3=\rho_{m,0}a_0^3$. Writing $\rho_{m,0}=\Omega_{m,0}\rho_0$ and $\rho_{\Lambda,0}=\Omega_{\Lambda,0}\rho_0$, we can rewrite Eq. (\ref{late2}) in the form
\begin{equation}
\label{late5}
\rho=\frac{\Omega_{m,0}\rho_0}{(a/a_0)^3}+\Omega_{\Lambda,0}\rho_0.
\end{equation}
When $a\ll a_2$, it reduces to
\begin{equation}
\label{late5b}
\rho_{m}=\frac{\Omega_{m,0}\rho_0}{(a/a_0)^3},
\end{equation}
which corresponds to the pure matter. In the matter era, the scale factor and the density evolve as  $a\propto t^{2/3}$ and $\rho\propto t^{-2}$.

The Friedmann equation (\ref{b9}) corresponding to Eq. (\ref{late5})  can be written as
\begin{equation}
\label{late8v}
\frac{H}{H_0}=\sqrt{\Omega_{m,0}\left\lbrack (a_0/a)^3+(a_0/a_2)^3\right\rbrack},
\end{equation}
or, equivalently, as
\begin{equation}
\label{late8}
\frac{H}{H_0}=\sqrt{\frac{\Omega_{m,0}}{(a/a_0)^3}+\Omega_{\Lambda,0}},
\end{equation}
with $\Omega_{m}+\Omega_{\Lambda}=1$. It has the analytical solution given by Eq. (\ref{s15}). This can be rewritten as
\begin{equation}
\label{late10}
\frac{a}{a_0}=\left (\frac{\Omega_{m,0}}{\Omega_{\Lambda,0}}\right )^{1/3}\sinh^{2/3}\left (\frac{3}{2}\sqrt{\Omega_{\Lambda,0}}H_0 t\right ).
\end{equation}
The evolution of the density is given by
\begin{equation}
\label{late11}
\frac{\rho}{\rho_0}=\frac{\Omega_{\Lambda,0}}{\tanh^2\left (\frac{3}{2}\sqrt{\Omega_{\Lambda,0}}H_0 t\right )}.
\end{equation}
The transition between the matter era and the dark energy era takes place at a typical time $t_2=0.203 t_{\Lambda}=2.97\, 10^{17}{\rm s}$ at which $a=a_2$. The universe is decelerating for $t<t_c$ and accelerating for $t>t_c$ where $t_c=0.152 t_{\Lambda}=2.22\, 10^{17}{\rm s}$. Using the results of Sec. \ref{sec_solid} and Appendix \ref{sec_pcscales}, we can write
\begin{equation}
\label{marre1}
\frac{a_2}{a_0}=\left (\frac{\Omega_{m,0}}{\Omega_{\Lambda,0}}\right )^{1/3},\qquad  \frac{a_c}{a_0}=\left (\frac{\Omega_{m,0}}{2\Omega_{\Lambda,0}}\right )^{1/3},
\end{equation}
\begin{equation}
\label{marre2}
t_{2}=\frac{1}{H_0}\frac{2}{3}\frac{1}{\sqrt{\Omega_{\Lambda,0}}}{\rm argsinh}(1),
\end{equation}
\begin{equation}
\label{marre2bg}
t_{c}=\frac{1}{H_0}\frac{2}{3}\frac{1}{\sqrt{\Omega_{\Lambda,0}}}{\rm argsinh}(1/\sqrt{2}).
\end{equation}
Numerically, $a_2=0.677a_0$ and $a_c=0.538 a_0$. The age of the universe is given by Eq. (\ref{s21}) which can be rewritten
\begin{equation}
\label{late15}
t_{0}=\frac{1}{H_0}\frac{2}{3}\frac{1}{\sqrt{\Omega_{\Lambda,0}}}{\rm argsinh}\left\lbrack \left (\frac{\Omega_{\Lambda,0}}{\Omega_{m,0}}\right )^{1/2}\right\rbrack.
\end{equation}
Numerically, $t_0=0.310 t_{\Lambda}=4.53\, 10^{17}\, {\rm s}$.

{\it Remark:} The equation of state (\ref{late1}) smoothly interpolates between a non-relativistic matter phase ($p=0$, $\rho\propto a^{-3}$) and a negative pressure dark energy regime ($p=-\rho c^2$, $\rho=\rho_{\Lambda}$). This amounts to summing the density of these two components taken independently. Indeed, Eq. (\ref{late5}) can be rewritten as
\begin{equation}
\label{late16}
{\rho}={\rho_{m}}+{\rho_{\Lambda}}.
\end{equation}
Therefore, the equation of state (\ref{late1}) is equivalent to the standard $\Lambda$CDM model \cite{bt}.

\subsection{Connection between the early and the late universe}
\label{sec_conn}

The previous sections reveal a deep ``symmetry'' between the early and the late evolution of the universe. In the early universe ($n=1$), we have to add the inverse of the densities, and in the late universe ($n=-1$) we have to add the densities themselves. This suggests that the universe must be described by {\it two} equations of state.

In order to obtain a single equation describing the whole evolution of the universe, we just have to add the contribution of radiation (including pre-radiation)
\begin{equation}
\label{pr1}
\rho_{rad}=\frac{\Omega_{rad,0}\rho_0}{(a/a_0)^4+(a_1/a_0)^4},
\end{equation}
baryonic matter
\begin{equation}
\label{pr4}
\rho_{B}=\frac{\Omega_{B,0}\rho_0}{(a/a_0)^3},
\end{equation}
dark matter
\begin{equation}
\label{pr5}
\rho_{DM}=\frac{\Omega_{DM,0}\rho_0}{(a/a_0)^3},
\end{equation}
and dark energy
\begin{equation}
\label{pr6}
\rho_{\Lambda}=\Omega_{\Lambda,0}\rho_0,
\end{equation}
in the Friedmann equation (\ref{b9}). The radiation dominates in the early universe, while baryonic matter, dark matter and dark energy dominate in the late universe. Now that the radiation phase has been regularized in order to avoid the initial singularity, it is necessary to state explicitly that matter is formed after radiation\footnote{In the standard model, the radiation density diverges as $a^{-4}$ when $a\rightarrow 0$ while the matter density diverges as $a^{-3}$ which is subdominant. Therefore, we can safely extrapolate the matter density to $a=0$ (its contribution is negligible anyway). In the present model, the radiation density (\ref{pr1}) does not diverge anymore. Therefore, it becomes necessary to state explicitly that matter is formed after radiation in order to avoid spurious divergences of the matter density when $a\rightarrow 0$.}. To that purpose, it is sufficient to add a Heaviside function $H(a-a_1)$ in the matter density. If we write $\Omega_{B,0}^*=\Omega_{B,0}H(a-a_1)$ and $\Omega_{DM,0}^*=\Omega_{DM,0}H(a-a_1)$, the Friedmann equation (\ref{b9}) can be written
\begin{equation}
\label{pr7}
\frac{H}{H_0}=\sqrt{\frac{\Omega_{rad,0}}{(a/a_0)^4+ (a_1/a_0)^4}+\frac{\Omega_{B,0}^*}{(a/a_0)^3}+\frac{\Omega_{DM,0}^*}{(a/a_0)^3}+\Omega_{\Lambda,0}},
\end{equation}
with $\Omega_{B}+\Omega_{rad}+\Omega_{DM}+\Omega_{\Lambda}=1$. For $a_1=0$, we recover the standard model of the universe \cite{bt}. It exhibits a singularity at $t=0$ (Big Bang). For $a_1\neq 0$, we obtain a regularized model without primordial singularity. The universe always existed in the past but, for $t<0$, it has a very small radius, smaller than the Planck length. At $t=0$, it undergoes an inflationary expansion in a very short lapse of time of the order of the Planck time and connects to the standard model. A nice feature of this model is its simplicity since it incorporates an ``inflation phase'' in a very simple and very natural manner. We just have to add a term $+(a_1/a_0)^4$ in the standard equation of $H/H_0$ given in \cite{bt}. Therefore,  the modification implied by Eq. (\ref{pr7}) to erase the initial  singularity is very natural. On the other hand, the equation of state (\ref{late1}) in the late universe is equivalent to adding the contribution of matter and dark energy individually, as in the standard model \cite{bt}. This does not bring any modification to the usual equation, which is a virtue of this description since the standard model works well at late times.

\begin{figure}[!h]
\begin{center}
\includegraphics[clip,scale=0.3]{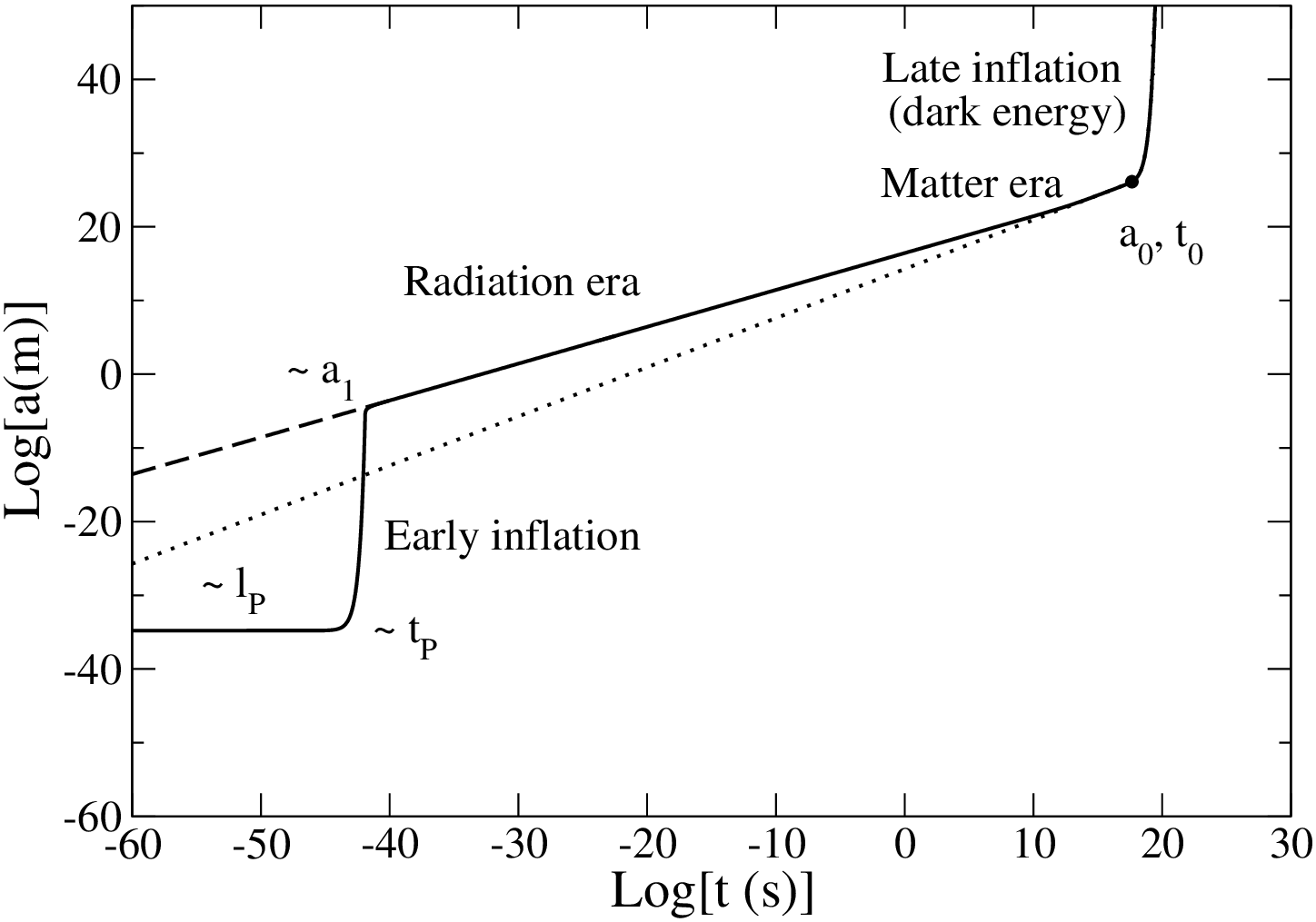}
\caption{Temporal evolution of the scale factor in logarithmic scales. The universe exists at all times in the past and in the future and  there is no singularity (aioniotic universe). The early universe undergoes a phase of inflation that takes it to the radiation era. This is followed by the matter era and the dark energy era responsible for the accelerated expansion of the universe. The universe exhibits two types of inflation: An early inflation corresponding to the Planck density $\rho_P$ due to quantum mechanics and a late inflation corresponding to the dark energy density $\rho_{\Lambda}$ due to the cosmological constant (general relativity). The evolution of the early and late universe is remarkably symmetric. We have represented in dashed line the standard model leading to a primordial singularity (Big Bang). The dotted lines corresponds to the model of Sec. \ref{sec_late} where the radiation is neglected. This analytical model provides a good description of the late universe. We have also represented the location of the present universe that is just at the transition between the matter era and the dark energy era.}
\label{taLOGLOGcomplet}
\end{center}
\end{figure}

\begin{figure}[!h]
\begin{center}
\includegraphics[clip,scale=0.3]{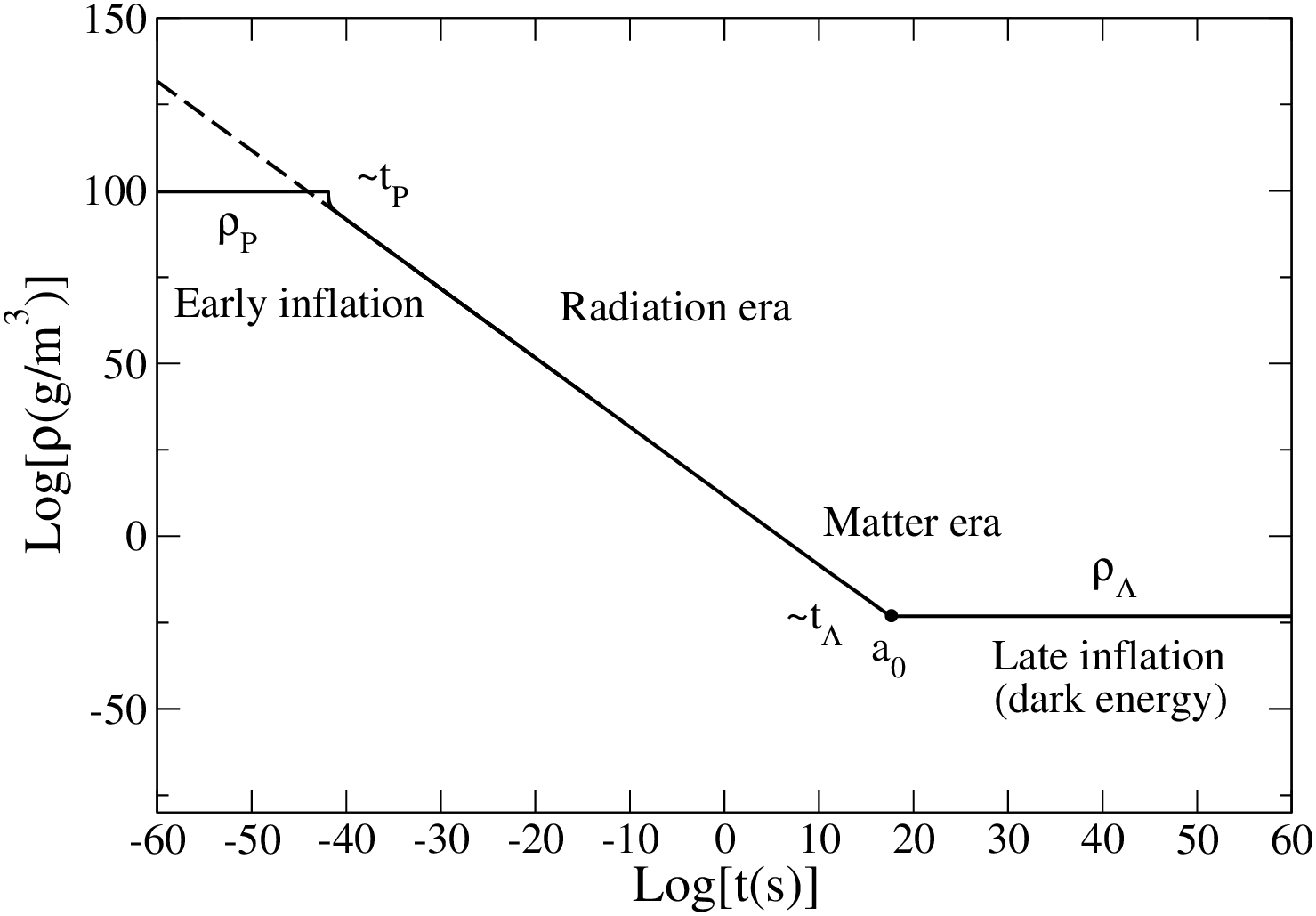}
\caption{Temporal evolution of the density in logarithmic scales. The density goes from a maximum value $\rho_{max}=\rho_P$ determined by the Planck constant (quantum mechanics) to a minimum value $\rho_{min}=\rho_{\Lambda}$ determined by the cosmological constant (general relativity). These two bounds are responsible for the early and late inflation of the universe. In between, the density decreases as $t^{-2}$.}
\label{trhoLOGLOGcomplet}
\end{center}
\end{figure}

\begin{figure}[!h]
\begin{center}
\includegraphics[clip,scale=0.3]{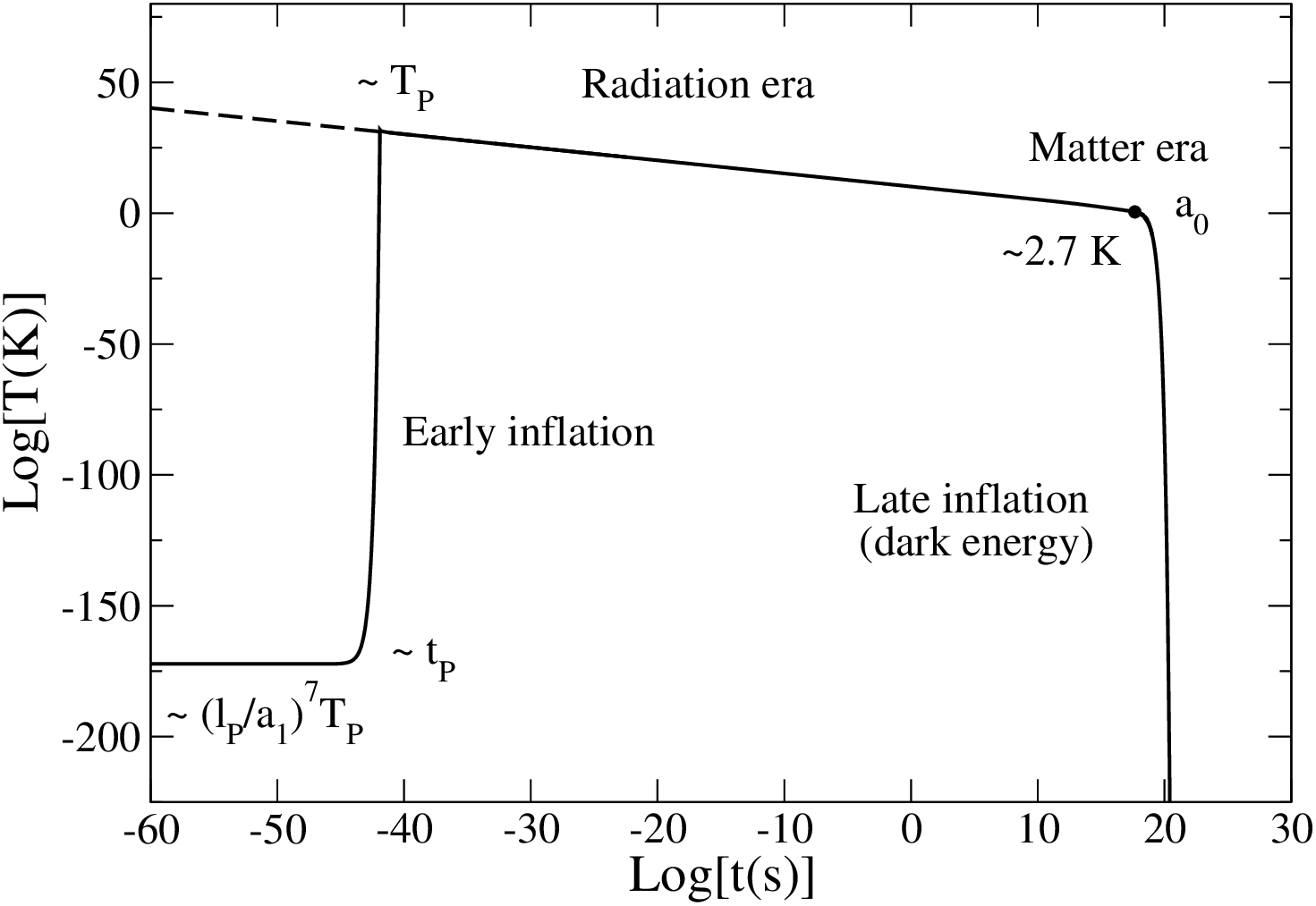}
\caption{Temporal evolution of the temperature of the radiation in logarithmic scales. The temperature increases considerably during the early inflation and decreases considerably during the late inflation. The present value of the temperature is $\sim 2.7 K$.}
\label{ttempLOGLOGcomplet}
\end{center}
\end{figure}

The evolution of the scale factor with time is obtained by solving the first order differential equation (\ref{pr7}). This yields
\begin{equation}
\label{pr8}
\int_{a_i/a_0}^{a/a_0} \frac{dx}{x\sqrt{\frac{\Omega_{rad,0}}{x^4+ (a_1/a_0)^4}+\frac{\Omega_{B,0}^*}{x^3}+\frac{\Omega_{DM,0}^*}{x^3}+\Omega_{\Lambda,0}}}=H_0 t,
\end{equation}
where $a_i=a_1=0$ in the standard model and $a_i=l_P$ in the regularized model. The age of the universe is
\begin{equation}
\label{pr9}
t_{0}=\frac{1}{H_0}\int_{0}^{1} \frac{dx}{x\sqrt{\frac{\Omega_{rad,0}}{x^4}+\frac{\Omega_{B,0}}{x^3}+\frac{\Omega_{DM,0}}{x^3}
+\Omega_{\Lambda,0}}}.
\end{equation}
Of course, for the determination of the age of the universe, we can neglect the pre-radiation era and take $a_1=0$ (strictly speaking, the age of the universe is infinite since it has no origin; however, we define the age of the universe from the time $t=0$ at which $a=l_P$). We obtain the standard result $t_{0}=1.03\, H_0^{-1}=4.53\, 10^{17}\, {\rm s}=14.4 \, {\rm Gyr}$ \footnote{The Hubble constant is usually written as $H_0=2.268\, h_7\, 10^{-18}{\rm s}^{-1}$ where the dimensionless parameter $h_7$ is about $10\%$ of unity \cite{bt}. For simplicity, we have taken $h_7=1$ in the numerical applications. The current value is $h_7=1.05\pm 0.05$. If we take $h_7=1.05$, the age of the universe is $t_0=13.7 \, {\rm Gyr}$.}.  Actually, we find the same result if we neglect radiation and use the analytical expression (\ref{late15}) instead of Eq. (\ref{pr9}). It is rather fortunate that the age of the universe almost coincides with the Hubble time $H_0^{-1}$.

In Figs. \ref{taLOGLOGcomplet}-\ref{ttempLOGLOGcomplet}, we have represented the evolution of the radius, density and temperature of the universe as a function of time. The universe exhibits two types of inflations: An early inflation due to the Planck density $\rho_P=5.16\, 10^{99}{\rm g}/{\rm m}^3$ and a late inflation due to the cosmological (dark energy) density $\rho_{\Lambda}=7.02\, 10^{-24}{\rm g}/{\rm m}^3$. The early inflation can be described by a polytropic equation of state with positive index $n=1$. The late inflation can be described by a polytropic equation of state with negative index $n=-1$. There exists a striking  ``symmetry'' between the early and the late evolution of the universe,  the cosmological constant in the late universe playing the same role as the Planck constant in the early universe. In particular, Fig. \ref{trhoLOGLOGcomplet} shows that the density varies between two bounds $\rho_{max}=\rho_P$ and $\rho_{min}=\rho_{\Lambda}$ that are fixed by fundamental constants (see Appendix \ref{sec_pcscales}). These values differ by a factor of the order $10^{122}$. The early universe is governed by quantum mechanics ($\hbar$) and the late universe by general relativity ($\Lambda$).

\begin{figure}[!h]
\begin{center}
\includegraphics[clip,scale=0.3]{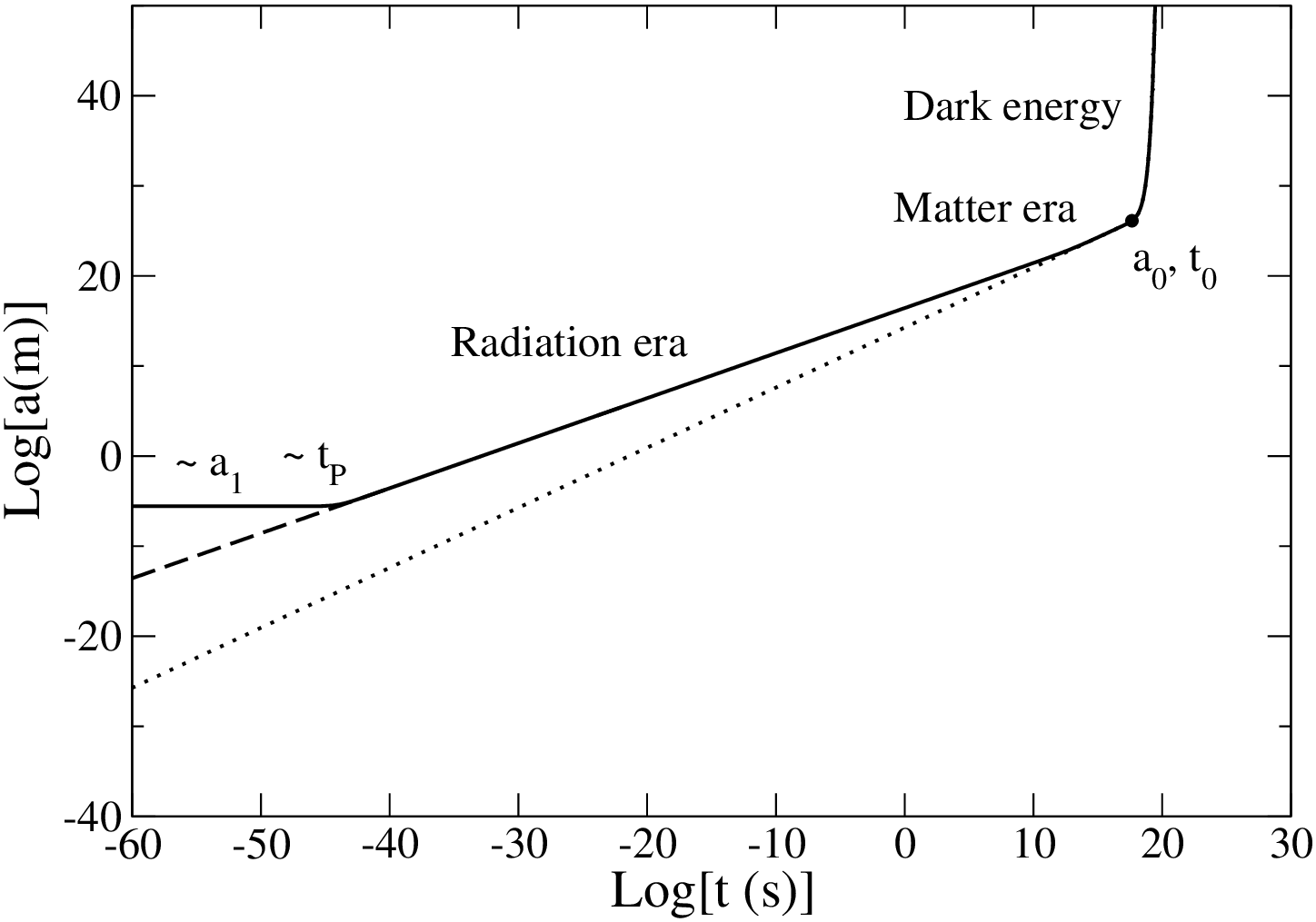}
\caption{Evolution of the scale factor in the singular model corresponding to $k<0$. The universe starts at $t=0$ from a singularity at which the radius is finite but the density and the temperature are infinite. This is followed by the radiation era, the matter era, and the dark energy era.}
\label{taLOGLOGcompletPOS}
\end{center}
\end{figure}

For the sake of completeness, we have also represented in Fig. \ref{taLOGLOGcompletPOS} the evolution of the scale factor in the model with the new type of initial singularity considered in Paper I. This model is obtained by replacing the sign $+$ by $-$ in Eq. (\ref{pr1}) and in the following formulae (we must then take $a_i=a_1$ in Eq. (\ref{pr8})). In that case, the universe emerges ``from nothing'' at $t=0$. It has an infinite density and temperature but a finite radius $a_1=2.61\, 10^{-6}{\rm m}$. This universe becomes physical only for  $t>t_i=0.154 t_P=8.32\, 10^{-45}{\rm s}$ when the velocity of sound becomes smaller than the speed of light.

{\it Remark:} An additional refinement can be introduced if dark matter is made of BECs instead of being pressureless. In that case, the term ${\Omega_{DM,0}^*}/{(a/a_0)^3}$ should be replaced by $\Omega_{DM,0}^*/\lbrack (a/a_0)^3\mp (a_{BEC}/a_0)^3\rbrack$ as proposed in \cite{harko,c4}. It is shown in these papers that the presence of BECs can substantially accelerate the formation of the large scale structures of the universe.

\subsection{The most natural universe}
\label{sec_aioniotic}

We may wonder whether the actual evolution of the universe can be predicted, without making any observation, from the principle of ``simplicity'' and ``aesthetics'' that was defended by great scientists such as Einstein, Eddington, and Chandrasekhar \cite{beauty}. We make the further assumption that the universe is non-singular and non-peculiar. Therefore, we try to construct the simplest non-singular model of universe.

The Friedmann equations (\ref{b2})-(\ref{b4}) governing the evolution of the universe require the value of the curvature of space $k$, the cosmological constant $\Lambda$, and the equation of state $p=p(\rho)$. Concerning the curvature of space, the most natural guess is $k=0$ (this leads to the simplest mathematical models). Actually, this value is consistent with observations. Even if an {\it apparently} flat universe could be the result of the primordial inflation (leading to $k/a^2\ll 1$) \cite{linde}, it may well be that $k=0$  exactly. This is what we assume. The cosmological constant can be eliminated from the basic equation since it can be taken into account in the equation of state. Concerning the equation of state, the simplest guess is to take $p=0$. This choice was made by early cosmologists. A pressureless universe describes the matter era (EdS model). However, this equation of state neglects relativistic effects which become important in the past when the universe was hot. Therefore, cosmologists considered a linear equation of state of the form $p=\alpha\rho c^2$. For $\alpha=1/3$, this equation of state describes the (relativistic) radiation era. However, this equation of state is not fully satisfactory since it leads to a primordial singularity (Big Bang). The early inflation  and the late inflation (dark energy) are usually described by a linear equation of state of the form $p=-\rho c^2$ leading to a constant density. However, this equation of state does not contain in itself the value of the constant density (a feature that we regard as a drawback). On the other hand, being linear, it cannot be ``added'' to another linear equation of state $p=\alpha\rho c^2$ to describe the transition between the pre-radiation era ($\alpha=-1$) and the radiation era ($\alpha=1/3$) in the early universe, or the transition  between the matter era ($\alpha=0$) and the dark energy era ($\alpha=-1$) in the late universe.

The next simplest choice is therefore to consider a mixed equation of state of the form $p=(\alpha \rho+k\rho^{1+1/n})c^2$ with a linear component and a polytropic component. It comes naturally from the polytropic model that the indices $n>0$ describe the early universe while the indices $n<0$ describe the late universe. This suggests that the early and the late universe must be described separately by two different equations of state. We must take $\alpha=1/3$ (radiation) in the equation of state describing the early universe ($n>0$) and $\alpha=0$ (matter) in the equation of state describing the late universe ($n<0$). By analyzing all the possibilities, we have found that only the models with $k<0$  can produce a non-singular (or non-peculiar) universe. The models with $k>0$ lead to past or future singularities/peculiarities. Therefore, the polytropic pressure must be {\it negative}. Then, it comes naturally from the polytropic model that the density has a maximum value $\rho_{max}$ in the early universe (for $n>0$) and a minimum value $\rho_{min}$ in the late universe (for $n<0$). It is natural to identify the upper bound with the Planck density $\rho_P$ (quantum mechanics) and the lower bound with the cosmological density $\rho_{\Lambda}$ (general relativity). This determines the constant $k$ which is $k=-4/(3\rho_P^{1/n})$ in the early universe and $k=-1/\rho_{\Lambda}^{1/n}$ in the late universe. Finally, according to the principle of simplicity, it is natural (but not quite compulsory) to select the index $n=+1$ (quadratic equation of state) in the early universe and the index $n=-1$ (constant equation of state) in the late universe. This leads to the model represented in Figs. \ref{taLOGLOGcomplet}-\ref{ttempLOGLOGcomplet}, featuring two symmetric phases of inflation. This evolution could have been predicted without making any observation of the universe, by only considering the simplest equation of state that does not yield any singularity or peculiarity. Interestingly, the observations of the real universe are remarkably consistent with that model.

In conclusion, the most natural non-singular model of universe
corresponds to a solution of the Einstein equations that exists
eternally in the past and in the future. This has been called {\it
aioniotic} universe in \cite{c4} since it has no beginning nor end. It
is the only model of the family of equations of state (\ref{b10}) that
has no singularity/peculiarity. For that reason, it may be selected by
Nature. Furthermore, this universe is very ``symmetric'' in its early
and late phases\footnote{We may argue that a model of late universe
with $\rho_{\Lambda}=0$ (corresponding to a vanishing cosmological
constant) is also non-singular and simpler.  However, if we add a
principle of ``symmetry'' ($\sim$ aesthetics), in addition to the
principle of ``simplicity'', we are left with the model discussed
previously.}. This symmetry can be seen in the values of the
polytropic indices $n=+1$ and $n=-1$ that characterize the early and
the late universe, respectively. All the formulae obtained in Paper I
and in the present paper are symmetric with respect to each other
(compare Secs. \ref{sec_early} and \ref{sec_late}). This symmetry is
also apparent in the two types of inflation that describe the early
and late universe: An early inflation due to the Planck density
$\rho_P$ (quantum mechanics) and a late inflation due to the
cosmological density $\rho_{\Lambda}$ (general relativity). They
correspond to the upper and lower bounds of the density shown in
Fig. \ref{trhoLOGLOGcomplet}, which is also strikingly
symmetric. Therefore, this universe is the simplest and most symmetric
non-singular cosmological solution of the Einstein equations.

Of course, some reservations should be made: The universe may not be ``simple''. In particular, the very early universe (before the Planck time) may not be described in terms of an equation of state $p(\rho)$, or even in terms of the Einstein equations, as we have assumed. The description of the very early universe may require the development of a theory of quantum gravity that does not exist for the moment. An interesting description of the early inflation has been given by \cite{monerat} in terms of a quantized model based on a simplified Wheeler-DeWitt equation. In that model, a quantum tunneling process explains the birth of the universe with a well defined size after tunneling. Therefore, other inflationary scenarios are possible in addition to the one given in Paper I in terms of the generalized equation of state (\ref{b10}). However, our aim was to explore all the consequences of this generalized equation of state, pushing it to its limits.

\subsection{A cosmological constant ``problem''?}
\label{sec_problem}

The cosmological constant $\Lambda$ is equivalent to a constant density $\rho_{\Lambda}=\Lambda/(8\pi G)$ called dark energy. Its value resulting from observations is $\rho_{\Lambda}=7.02\, 10^{-24}{\rm g}/{\rm m}^3$ (see Appendix \ref{sec_pcscales}). Since the late universe is very dilute, it is oftentimes argued that the cosmological density $\rho_{\Lambda}$  should correspond to the  vacuum energy density due to quantum fluctuations. However, according to particle physics and quantum field theory, the vacuum energy density is of the order of the Planck density $\rho_P=5.16\, 10^{99}{\rm g}/{\rm m}^3$ which is $10^{122}$ times larger than the cosmological density. This leads to the so-called cosmological constant problem \cite{weinbergcosmo}.

Actually, as illustrated in Fig. \ref{trhoLOGLOGcomplet}, the Planck density and the cosmological density represent fundamental upper and lower density bounds acting in the early and late universe, respectively. It is not surprising therefore that they are so different: $\rho_{\Lambda}\ll \rho_P$. Because of these bounds, the universe undergoes two phases of inflation. The inflation in the early universe is due to quantum mechanics and is related to the Planck density $\rho_P$. The inflation in the late universe is due to the cosmological constant and is related to the cosmological density  $\rho_{\Lambda}$. Quantum mechanics is negligible in the late universe. The cosmological constant should be interpreted as a new fundamental constant of physics. It applies to the very large universe (cosmophysics) exactly like the Planck constant applies to the very small universe (microphysics). Actually, there is a complete symmetry between the small and large universe where $\hbar$ and $\Lambda$ play symmetric roles. Therefore, we interpret  the cosmological constant as a fundamental constant of physics describing the cosmophysics (late universe) in the same sense that the Planck constant describes the microphysics (early universe).

If this interpretation is correct, the origin of the  dark energy density $\rho_{\Lambda}$ should not be sought in quantum mechanics, but in pure general relativity. In this sense, the cosmological constant ``problem'' may be a false problem. If $\Lambda$ is a fundamental constant of physics, independent from the others, its value should not cause problem. It is fixed by nature, just like the value of $G$, $c$, and $\hbar$. One can just expect that $\rho_P$ is ``very large'' and $\rho_{\Lambda}$ is ``very small''. Of course, the origin of the cosmological constant still needs to be understood by developing a theory of cosmophysics. In addition, it is extremely important to understand why $\rho_P$ and $\rho_{\Lambda}$ represent upper and lower bounds, and if these bounds are as fundamental as, for example, the bound on the velocity fixed by the speed of light. 

{\it Remark:} We may note that the appearance of maximum and minimum density bounds from the polytropic equation of state (\ref{b10}) is, in some sense, similar to the appearance of the maximum mass of relativistic white dwarf stars discovered by Chandrasekhar \cite{chandramass} also from a  polytropic equation of state.

\section{Scalar field models}
\label{sec_scalar}

The phase of inflation in the very early universe is usually described by a scalar field \cite{linde}. On the other hand, in  alternative theories to the cosmological constant, the present-day acceleration of the universe (dark energy) is also described by a scalar field called quintessence \cite{cst,quintessence}. A tachyon field \cite{cst,tachyon} has also been considered due to its connection with string theory \cite{string}.  Actually, to any ``fluid'' described by a barotropic equation of state $p=p(\rho)$, it is possible to attach a scalar field with a potential $V(\phi)$. In this section, we determine the potential of the scalar field (quintessence and tachyon field) corresponding to the equation of state (\ref{b10}) using the general methodology exposed in \cite{cst}. We treat polytropic indices $n$ and polytropic constants $k$ of arbitrary sign but we assume $\alpha+1+k\rho^{1/n}>0$ ({\it i.e.} $w\ge -1$). The case $\alpha+1+k\rho^{1/n}<0$ ({\it i.e.} $w\le -1$) leads to phantom (ghost) fields that are studied in Paper III.

\subsection{Quintessence}
\label{sec_quintessence}

Quintessence \cite{quintessence,cst} is described by an ordinary scalar field $\phi$ minimally coupled to gravity. The scalar field evolves according to the equation
\begin{equation}
\label{quintessence1}
\ddot \phi+3H\dot\phi+\frac{dV}{d\phi}=0,
\end{equation}
where $V(\phi)$ is the potential of the scalar field. The quintessence scalar field tends to run down the potential towards lower energies. The density and the pressure of the universe are related to the scalar field by
\begin{equation}
\label{quintessence2}
\rho c^2=\frac{1}{2}\dot\phi^2+V(\phi),\qquad p=\frac{1}{2}\dot\phi^2-V(\phi).
\end{equation}
When the kinetic energy dominates the potential energy, we obtain the equation of state $p=\rho c^2$ of stiff matter.  When the potential energy dominates the kinetic energy, we obtain the vacuum equation of state $p=-\rho c^2$.

From Eq. (\ref{quintessence2}), we get
\begin{equation}
\label{quintessence3}
\dot\phi^2=(1+w)\rho c^2,
\end{equation}
where we have written $p=w\rho c^2$. We assume $w\ge -1$ in order to have $\dot\phi^2\ge 0$.  Using $\dot\phi=(d\phi/da) H a$, and the Friedmann equation (\ref{b9}) valid for a flat universe, we get
\begin{equation}
\label{quintessence4}
\frac{d\phi}{da}=\left (\frac{3c^2}{8\pi G}\right )^{1/2}\frac{\sqrt{1+w}}{a}.
\end{equation}
For the equation of state (\ref{b10}), using Eqs. (\ref{ges2}) and (I-69), and setting $R=a/a_*$, we can rewrite Eq. (\ref{quintessence4}) in the form
\begin{equation}
\label{quintessence5}
\frac{d\phi}{dR}=\left (\frac{3c^2}{8\pi G}\right )^{1/2}\frac{\sqrt{\alpha+1}}{R}\frac{R^{3(1+\alpha)/2n}}{\sqrt{R^{3(1+\alpha)/n}\mp 1}}.
\end{equation}
With the change of variables
\begin{equation}
\label{quintessence6}
x=R^{3(\alpha+1)/2n},\qquad \psi=\left (\frac{8\pi G}{3c^2}\right )^{1/2}\frac{3\sqrt{\alpha+1}}{2n}\phi,
\end{equation}
we find that
\begin{equation}
\label{quintessence7}
\psi=\int \frac{dx}{\sqrt{x^2\mp 1}}.
\end{equation}
For $k>0$, we obtain
\begin{equation}
\label{quintessence8}
\psi={\rm Argcosh}(x),
\end{equation}
and for $k<0$, we get
\begin{equation}
\label{quintessence9}
\psi={\rm Argsinh}(x).
\end{equation}
On the other hand, according to Eq. (\ref{quintessence2}), we have
\begin{equation}
\label{quintessence10}
V=\frac{1}{2}(1-w)\rho c^2.
\end{equation}
For the equation of state (\ref{b10}), using Eqs. (\ref{ges2}) and (I-69), we obtain
\begin{equation}
\label{quintessence11}
V=\frac{1}{2}\rho_* c^2 \frac{(1-\alpha)x^2\mp 2}{(x^2\mp 1)^{n+1}}.
\end{equation}
For $k>0$, since $x=\cosh(\psi)$ the scalar field potential is explicitly given by
\begin{equation}
\label{quintessence12}
V(\psi)=\frac{1}{2}\rho_* c^2 \frac{(1-\alpha)\sinh^2\psi-(\alpha+1)}{\sinh^{2(n+1)}\psi}.
\end{equation}
For $k<0$, since $x=\sinh(\psi)$ the scalar field potential is explicitly given by
\begin{equation}
\label{quintessence14}
V(\psi)=\frac{1}{2}\rho_* c^2 \frac{(1-\alpha)\cosh^2\psi+\alpha+1}{\cosh^{2(n+1)}\psi}.
\end{equation}
In these models, $\psi\ge 0$. Let us consider particular cases.

(i) For $n=1$, $k=-4/(3\rho_P)$ and $\alpha=1/3$ (non-singular early universe), we obtain
\begin{equation}
\label{quintessence16}
V(\psi)=\frac{1}{3}\rho_{P} c^2 \frac{\cosh^2\psi+2}{\cosh^4\psi},
\end{equation}
and $R^{2}=\sinh(\psi)$.

(ii) For $n=1$, $k=4/(3\rho_P)$ and $\alpha=1/3$ (singular early universe), we obtain
\begin{equation}
\label{quintessence16b}
V(\psi)=\frac{1}{3}\rho_{P} c^2 \frac{\sinh^2\psi-2}{\sinh^4\psi},
\end{equation}
and $R^{2}=\cosh(\psi)$.

(iii) For $n=-1$, $k=-\rho_{\Lambda}$ and $\alpha=0$ (late universe, $\Lambda$CDM model), we obtain
\begin{equation}
\label{quintessence18}
V(\psi)=\frac{1}{2}\rho_{\Lambda} c^2 (\cosh^2\psi+1),
\end{equation}
and ${R^{-3/2}}=\sinh(\psi)$.

(iv) For $n=-1$, $k=\rho_{\Lambda}$ and $\alpha=0$ (late universe, anti-$\Lambda$CDM model), we obtain
\begin{equation}
\label{quintessence18b}
V(\psi)=\frac{1}{2}\rho_{\Lambda} c^2 (\sinh^2\psi-1),
\end{equation}
and ${R^{-3/2}}=\cosh(\psi)$.

(v) For $n=-1/2$, $k<0$ and $\alpha=0$ (Chaplygin gas), we obtain
\begin{equation}
\label{quintessence20}
V(\psi)=\frac{1}{2}\rho_{*} c^2 \left (\cosh\psi+\frac{1}{\cosh\psi}\right ),
\end{equation}
and ${R^{-3}}=\sinh(\psi)$. This returns the result of \cite{chaplygin}.

(vi) For $n=-1/2$, $k>0$ and $\alpha=0$ (anti-Chaplygin gas), we obtain
\begin{equation}
\label{quintessence20b}
V(\psi)=\frac{1}{2}\rho_{*} c^2 \left (\sinh\psi-\frac{1}{\sinh\psi}\right ),
\end{equation}
and ${R^{-3}}=\cosh(\psi)$.

The case $\alpha=-1$ and $k>0$ must be treated specifically (see Appendix A of Paper I and Appendix \ref{sec_eosgm} of the present paper). Repeating the preceding procedure, we find that the potential of the scalar field is
\begin{eqnarray}
\label{quintessence22}
V(\phi)=\rho_* c^2 \left (\frac{|n| c^2}{2\pi G}\right )^n \left (1-\frac{n^2c^2}{12\pi G\phi^2}\right )\frac{1}{\phi^{2n}},
\end{eqnarray}
\begin{equation}
\label{quintessence23}
\ln R=\frac{2\pi G}{nc^2}\phi^2.
\end{equation}

Finally, for a linear equation of state $p=\alpha\rho c^2$ with $\alpha>-1$, writing the relation between the density and the scale factor as  $\rho/\rho_*=(a_*/a)^{3(1+\alpha)}$, we obtain \cite{cst}:
\begin{equation}
\label{quintessence24}
V(\phi)=\frac{1}{2}\rho_* c^2 (1-\alpha) e^{-3\sqrt{\alpha+1}\left (\frac{8\pi G}{3c^2}\right )^{1/2}\phi},
\end{equation}
\begin{equation}
\label{quintessence25}
\phi=\left (\frac{3c^2}{8\pi G}\right )^{1/2}\sqrt{1+\alpha}\, \ln R,
\end{equation}
where $R=a/a_*$. Since $R\propto t^{2/\lbrack 3(1+\alpha)\rbrack}$, the scalar field evolves with time as $\phi= \lbrack c^2/6\pi G(1+\alpha)\rbrack^{1/2}\ln t$.  We note that $V(\phi)=0$ for $\alpha=1$ (stiff matter). On the other hand, there is no scalar field corresponding to the vacuum equation of state $p=-\rho c^2$ ({\it i.e.} $\alpha=-1$) used to describe inflation. By contrast, the equation of state (\ref{b10}) admits a scalar field representation in all cases, in particular for the models with $k<0$ exhibiting  a phase of inflation. This may be an advantage of the polytropic equation of state (\ref{b10}) with respect to the linear equation of state $p=-\rho c^2$.

\subsection{Tachyon field}
\label{sec_tachyon}

A tachyon field \cite{tachyon,cst} has an equation of state $p=w\rho c^2$ with $-1\le w\le 0$. This scalar field evolves according to the equation
\begin{equation}
\label{tachyon1}
\frac{\ddot \phi}{1-\dot\phi^2}+3H\dot\phi+\frac{1}{V}\frac{dV}{d\phi}=0.
\end{equation}
The density and the pressure are given by
\begin{equation}
\label{tachyon2}
\rho c^2=\frac{V(\phi)}{\sqrt{1-\dot\phi^2}},\qquad p=-V(\phi)\sqrt{1-\dot\phi^2}.
\end{equation}
From these equations, we obtain
\begin{equation}
\label{tachyon3}
\dot\phi^2=1+w,
\end{equation}
where we have written $p=w\rho c^2$. Using $\dot\phi=(d\phi/da) H a$, and the Friedmann equation (\ref{b9}), we get
\begin{equation}
\label{tachyon4}
\frac{d\phi}{da}=\left (\frac{3c^2}{8\pi G}\right )^{1/2}\frac{\sqrt{1+w}}{\sqrt{\rho c^2} a}.
\end{equation}
For the equation of state (\ref{b10}), using Eqs. (\ref{ges2}) and (I-69), we can rewrite Eq. (\ref{tachyon4}) in the form
\begin{eqnarray}
\label{tachyon5}
\frac{d\phi}{dR}=\frac{1}{\sqrt{\rho_*c^2}}\left (\frac{3c^2}{8\pi G}\right )^{1/2}\frac{\sqrt{\alpha+1}}{R}R^{3(1+\alpha)/2n}\nonumber\\
\times\left\lbrack R^{3(1+\alpha)/n}\mp 1\right\rbrack^{(n-1)/2}.
\end{eqnarray}
With the change of variables
\begin{equation}
\label{tachyon6}
x=R^{3(\alpha+1)/2n},\qquad \psi=\sqrt{\rho_*c^2}\left (\frac{8\pi G}{3c^2}\right )^{1/2}\frac{3\sqrt{1+\alpha}}{2n}\phi,
\end{equation}
we find that
\begin{equation}
\label{tachyon7}
\psi=\int (x^2\mp 1)^{(n-1)/2}\, {dx}.
\end{equation}
On the other hand, from Eq. (\ref{tachyon2}), we have
\begin{equation}
\label{tachyon8}
V^2=-w\rho^2 c^4.
\end{equation}
For the equation of state (\ref{b10}), using Eqs. (\ref{ges2}) and (I-69), we obtain
\begin{equation}
\label{tachyon9}
V^2=-\rho_*^2 c^4 \frac{\alpha x^2\pm 1}{(x^2\mp 1)^{2n+1}}.
\end{equation}
Therefore, the scalar field potential $V(\psi)$ is given in parametric form by Eqs. (\ref{tachyon7}) and (\ref{tachyon9}). Let us consider particular cases.

(i) For $n=1$, we find that $x=\psi$. Therefore, we obtain
\begin{equation}
\label{tachyon10}
V^2=-\rho_*^2 c^4 \frac{\alpha \psi^2\pm 1}{(\psi^2\mp 1)^{3}},
\end{equation}
and $R^{3(\alpha+1)/2}=\psi$.  In the case $k<0$ with $\alpha\le 0$, where $w$ is always between $-1$ and $0$, we obtain
\begin{equation}
\label{tachyon12}
V(\psi)=\rho_* c^2 \frac{\sqrt{1-\alpha\psi^2}}{(\psi^2+1)^{3/2}},
\end{equation}
with $\psi\ge 0$. We note that for $n=1$, $k=-4/(3\rho_P)$ and $\alpha=1/3$ (early universe), the parameter $w$ becomes positive when $R>R_w=3^{1/4}$ so that the tachyon field is not defined for all times. Similarly, in the case $k>0$, since $w\rightarrow +\infty$ when $R\rightarrow 1$, the tachyon field is not defined for all times.

(ii) For $n=-1$ and $k>0$, we find that $x=-1/\tanh\psi$. Therefore, we obtain
\begin{equation}
\label{tachyon13}
V^2=-\rho_*^2 c^4 \left (\frac{\alpha}{\tanh^2\psi}+1\right )\frac{1}{\sinh^2\psi},
\end{equation}
and $R^{3(\alpha+1)/2}=-\tanh\psi$,  with $\psi\le 0$. Since $w\rightarrow +\infty$ when $R\rightarrow 1$, the tachyon field is not defined for all times. In particular, for $n=-1$, $k=\rho_{\Lambda}$ and $\alpha=0$ (anti-$\Lambda$CDM model), the parameter $w$ is always positive so that the tachyon field is not defined.

(iii) For $n=-1$ and $k<0$, we find that $x=\tan\psi$. Therefore, we obtain
\begin{equation}
\label{tachyon15}
V^2=-\rho_*^2 c^4 (\alpha \tan^2\psi-1)\frac{1}{\cos^2\psi},
\end{equation}
and $R^{3(\alpha+1)/2}=1/\tan\psi$ with $0\le \psi\le\pi/2$. If $\alpha\le 0$, $w$ remains between $-1$ and $0$, so the tachyon field is well-defined. In particular, for $n=-1$, $k=-\rho_{\Lambda}$ and $\alpha=0$ (late universe, $\Lambda$CDM model), we find
\begin{equation}
\label{tachyon17}
V(\psi)=\frac{\rho_{\Lambda} c^2}{\cos\psi},
\end{equation}
and $R^{3/2}=1/\tan\psi$ with $0\le \psi\le\pi/2$.

(iv) For $n=-2$ and $k>0$, we find that $x=\psi/\sqrt{\psi^2-1}$. Therefore, we obtain
\begin{equation}
\label{tachyon20}
V^2=-\rho_*^2 c^4 \frac{(\alpha+1)\psi^2+1}{(\psi^2-1)^4},
\end{equation}
and $R^{3(\alpha+1)/4}=\sqrt{\psi^2-1}/\psi$ with $\psi\ge 1$. Since $w\rightarrow +\infty$ when $R\rightarrow 1$, the tachyon field is not defined for all times.

(v) For $n=-2$ and $k<0$, we find that $x=\psi/\sqrt{1-\psi^2}$. Therefore, we obtain
\begin{equation}
\label{tachyon22}
V^2=-\rho_*^2 c^4 \frac{(\alpha+1)\psi^2-1}{(1-\psi^2)^4},
\end{equation}
and $R^{3(\alpha+1)/4}=\sqrt{1-\psi^2}/\psi$,
with $0\le \psi\le 1$. If $\alpha\le 0$, $w$ is between $-1$ and $0$, so the tachyon field is well-defined. In particular, for $\alpha=0$, we have
\begin{equation}
\label{tachyon24}
V(\psi)=\frac{\rho_* c^2}{(1-\psi^2)^{3/2}},
\end{equation}
and $R^{3/4}=\sqrt{1-\psi^2}/\psi$ with $0\le \psi\le 1$.

(vi) For $n=-1/2$, $\alpha=0$, and $k<0$ (Chaplygin gas), we find that $V(\phi)=\rho_* c^2$ is constant. The tachyon field is not defined in the anti-Chaplygin gas ($n=-1/2$, $\alpha=0$, and $k>0$).

The case $\alpha=-1$ and $k>0$ must be treated specifically (see Appendix A of Paper I and Appendix \ref{sec_eosgm} of the present paper). Repeating the preceding procedure, we find that the potential of the scalar field is
\begin{eqnarray}
\label{tachyon29}
V(\phi)^2=\rho_*^2 c^4 \left \lbrack \frac{|n|}{2\pi G\rho_* (n+1)^2}\right \rbrack^{2n/(n+1)}\frac{1}{\phi^{4n/(n+1)}}\nonumber\\
\times\left \lbrace 1-\frac{|n|}{3}\left \lbrack \frac{|n|}{2\pi G\rho_* (n+1)^2}\right \rbrack^{1/(n+1)}\frac{1}{\phi^{2/(n+1)}}\right \rbrace,\nonumber\\
\end{eqnarray}
\begin{equation}
\label{tachyon30}
\ln R={\rm sgn}(n) \left \lbrack \frac{2\pi G \rho_* (n+1)^2}{|n|}\right \rbrack^{1/(n+1)}\phi^{2/(n+1)}.
\end{equation}
For $n=1$ and $t>t_i$, and for $n=-1$ and $t<t_f$, corresponding to the period over which the velocity of sound is smaller than the speed of light,  the parameter $w$ is between $-1$ and $0$ so that the tachyon field is perfectly well-defined.

Finally, for a linear equation of state $p=\alpha\rho c^2$ with $-1<\alpha\le 0$, writing the relation between the density and the scale factor as  $\rho/\rho_*=(a_*/a)^{3(1+\alpha)}$, we obtain \cite{cst}:
\begin{equation}
\label{tachyon27}
V(\phi)=\frac{\sqrt{-\alpha}}{1+\alpha}\frac{c^2}{6\pi G}\frac{1}{\phi^2},
\end{equation}
\begin{equation}
\label{tachyon28}
\phi=\frac{2}{3}\frac{1}{\sqrt{\rho_* c^2}} \left (\frac{3c^2}{8\pi G}\right )^{1/2}\frac{1}{\sqrt{1+\alpha}}R^{3(1+\alpha)/2},
\end{equation}
where $R=a/a_*$. Since $\rho=\rho_*/R^{3(1+\alpha)}=1/\lbrack 6\pi G(1+\alpha)^2t^2\rbrack$, the scalar field evolves with time as $\phi=\sqrt{1+\alpha} \, t$. We note that $V(\phi)=0$ for $\alpha=0$. On the other hand, there is no tachyon field corresponding to the vacuum equation of state $p=-\rho c^2$ ({\it i.e.} $\alpha=-1$) used to describe inflation. By contrast, the equation of state (\ref{b10}) admits a tachyon field representation in the models with $k<0$ and $\alpha\le 0$  exhibiting  a phase of inflation. This may be an advantage of the polytropic equation of state (\ref{b10}) with respect to the linear equation of state $p=-\rho c^2$.

\section{Conclusion}

In this paper, and in the previous one, we have carried out an exhaustive study of the generalized equation of state (\ref{b10}), considering all the possible cases. We have obtained the following results (we assume here $\alpha\neq -1$):

(i) For $n>0$ and $k<0$, the universe undergoes an early inflation. It starts from $t=-\infty$ with a vanishing radius and a finite density $\rho_{max}$. Its radius increases with time while its density decreases. The universe exists at any time in the past and is non-singular.

(ii) For $n>0$ and $k>0$, the universe undergoes a new type of primordial singularity. It starts at $t=0$ with a finite radius and an infinite density. For $t>0$, its radius increases while its density decreases.

(iii) For $n<0$ and $k<0$, the universe undergoes a late inflation. Its radius increases to $+\infty$ as  $t\rightarrow +\infty$ while its density decreases to a constant value $\rho_{min}$. The universe exists at any time in the future and is non-singular.

(iv) For $n<0$ and $k>0$, the universe undergoes a future peculiarity. Its radius increases to a maximum value while its density decreases to zero (the universe ``disappears''). For $n\le -2$, this peculiarity is reached in infinite time. For $n>-2$, this peculiarity is reached in a finite time (and there is a future singularity when $n>-1$ due to the divergence of the pressure). Then, the radius decreases to zero while the density increases to $+\infty$. These phases of expansion and contraction continue periodically (cyclic universe).

The generalized equation of state (\ref{b10}) can be viewed as a ``mixture'' of a linear equation of state describing a classical universe filled with radiation ($\alpha=1/3$) or matter ($\alpha=0$),
and a polytropic equation of state whose origin remains to be understood (in Paper I, we have mentioned the connection with Bose-Einstein condensates, but other possibilities may be contemplated).  Positive indices describe the early universe (the polytropic component dominates the linear component when the density is high) and negative indices describe the late universe (the polytropic component dominates the linear component when the density is low). A positive polytropic pressure ($k>0$) leads to singular or peculiar models, while a negative polytropic pressure ($k<0$) leads to non-singular models that exist for all times.

For $k<0$, the generalized equation of state (\ref{b10}) implies the existence of upper and lower bounds on the density. It is natural to identify the upper bound with the Planck density $\rho_{max}=\rho_P$ (quantum mechanics) and the lower bound with the cosmological density $\rho_{min}=\rho_{\Lambda}$ (general relativity). By taking $n=1$ in the early universe and $n=-1$ in the late universe, we have obtained a non-singular model that is consistent with the known properties of our universe. This model improves the standard $\Lambda$CDM model by removing the primordial singularity. An attractive feature of this model is its simplicity and the  ``symmetry'' that it reveals between the past and the future. Furthermore, this model admits a scalar field representation based on a quintessence field (in the early and late universe) or a  tachyon field (in  the late universe).

By using only the principle of ``simplicity'', and rejecting singular or peculiar solutions, we have obtained in a natural manner a simple model of universe. It corresponds to an ``aioniotic'' universe that exists eternally in past and future and whose early and late evolutions present some striking symmetry. The remarkable point is that this model, which is obtained in a purely theoretical manner independent of observations, turns out to be fully consistent with the known structure of the real universe. A theoretical challenge would now be to {\it justify} a quadratic equation of state $p=-4\rho^2c^2/(3\rho_P)$ ($n=1$) in the early universe and a constant pressure $p=-\rho_{\Lambda}c^2$ ($n=-1$) in the late universe. It may also be interesting to study an equation of state of the form $p/c^2=-(\alpha+1)\rho^2/\rho_P+\alpha\rho-(\alpha+1)\rho_{\Lambda}$ exhibiting two phases of inflation \cite{prep}. For $\alpha=1/3$, it provides a unification of pre-radiation, radiation, and dark energy (see Appendix \ref{sec_unification}).

Our study suggests that the Planck density $\rho_P$ and the cosmological density $\rho_{\Lambda}$  represent fundamental upper and lower bounds. These bounds are responsible for a phase of early and late inflation. It is oftentimes argued that the dark matter density $\rho_{\Lambda}$ should be identified with the vacuum energy. Since the vacuum energy is of the order of the Planck density $\rho_P$, which is $10^{122}$ times larger than the cosmological density, this leads to the so-called cosmological constant problem \cite{weinbergcosmo}. Actually, if the Planck density and the cosmological density represent fundamental upper and lower bounds, it is not surprising that they differ by about ${122}$ orders of magnitudes. Therefore, the origin of the cosmological density should not be sought in quantum mechanics but in a new theory of cosmophysics based on general relativity.

We have also studied singular models of universe corresponding to a positive polytropic pressure ($k<0$). Although these models do not appear to be selected by nature, they are interesting on a mathematical point of view. For $n>0$, we have obtained a model exhibiting a primordial singularity at $t=0$ at which the radius has a finite  value while the density and the temperature are infinite. For $n<0$, we have obtained models exhibiting future peculiarities in which the density of the universe vanishes while its radius reaches a maximum value. This peculiarity may occur in infinite time ($n\le -2$), or periodically in time ($n>-2$). In particular, we have obtained a simple analytical solution of a cyclic universe (anti-$\Lambda$CDM model) in which the density disappears and re-appears periodically. Since this solution is always decelerating (in the phase of expansion), this peculiar model is not compatible with the known properties of our universe.

There remains a type of solutions that we have not described so far: The case where the  density increases as the universe expands ($w<-1$). These {\it a priori} un-natural solutions which violate the null dominant energy condition  are associated with phantom scalar fields \cite{caldwell,caldwellprl}. They are studied in Paper III in which we construct models of ``phantom universes''. Actually, there is a rich recent literature \cite{cst,ghosts} on these solutions since observations do not exclude the possibility that we live in a phantom universe \cite{observations}. These models usually predict a future singularity at which the radius and the density of the universe become infinite in a finite time. This would lead to the death of the universe in a Big Rip or a Cosmic Doomsday \cite{caldwellprl}.

The main feature of the generalized equation of state (\ref{b10}) is its polytropic component $p=k\rho^{\gamma}c^2$. A polytropic equation of state occurs in many situations of astrophysical interest \cite{bt}. For example, a polytropic equation of state with index $\gamma=5/3$ describes adiabatic processes in main sequence stars \cite{chandra}. Compact stars such as white dwarfs, neutron stars and BEC stars are also described by a polytropic equation of state. Classical  white dwarf stars correspond to a polytropic  index $n=3/2$ and relativistic white dwarf stars correspond to a polytropic  index $n=3$ \cite{chandra,st}. Newtonian and semi-relativistic BEC stars with a self-interaction correspond to a polytropic index $n=1$ \cite{ch}. From a mathematical point of view, polytropic stars represent a particular class of steady states of the Euler-Poisson system.  Similarly, stellar polytropes  represent a particular class of steady states of the  Vlasov-Poisson system \cite{bt}.  Polytropic distributions also appear in statistical physics, in relation with generalized thermodynamics \cite{tsallisbook} and nonlinear Fokker-Planck equations \cite{frank}. In previous works, we have studied polytropic equations of state in several situations of astrophysical \cite{aaantonov,cs}, physical \cite{cc}, and biological \cite{nfp} interest. It was therefore natural to consider the application of polytropic equations of state in cosmology also.

\appendix

\section{Equation of state $p=(-\rho+k\rho^{\gamma})c^2$ with $n<0$ and $k>0$}
\label{sec_eosgm}

In this Appendix, we specifically study the equation of state (\ref{b10}) with $\alpha=-1$, $n<0$, and $k>0$, namely
\begin{eqnarray}
\label{eosgm0}
p=(-\rho+k\rho^{\gamma})c^2.
\end{eqnarray}
It generalizes  the equation of state $p=-\rho c^2$ of the vacuum  energy. For the equation of state (\ref{eosgm0}), the continuity equation (\ref{b7}) can be integrated into
\begin{eqnarray}
\label{eosgm1}
\rho=\frac{\rho_*}{\ln(a_*/a)^n},
\end{eqnarray}
where $\rho_*=(|n|/3k)^n$ and $a_*$ is a constant of integration.
The density is defined for $a\le a_*$. When $a\rightarrow 0$, $\rho\rightarrow +\infty$ and $p\rightarrow -\infty$. When $a\rightarrow a_*$,  $\rho\rightarrow 0$. In the same limit, $p\rightarrow 0$ for $n<-1$, $p$ tends to a finite value for $n=-1$, and $p \rightarrow +\infty$ for $n>-1$. 

\begin{figure}[!h]
\begin{center}
\includegraphics[clip,scale=0.3]{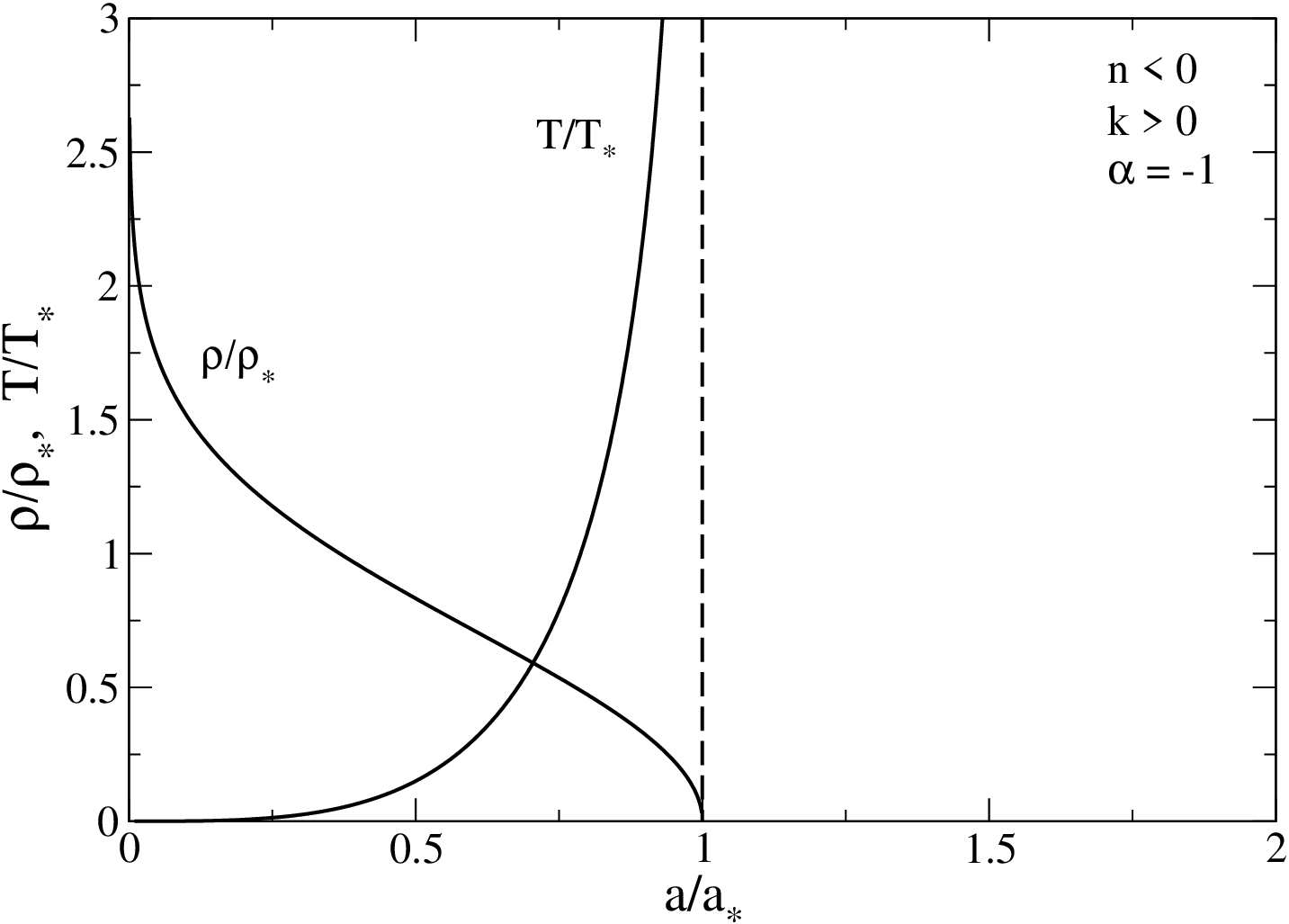}
\caption{Evolution of the density and temperature as a function of the scale factor. We have taken $n=-1/2$. }
\label{annexedensity}
\end{center}
\end{figure}

The thermodynamical equation (\ref{t2}) can be integrated into
\begin{eqnarray}
\label{eosgm2}
T=T_* \left (\frac{\rho}{\rho_*}\right )^{(n+1)/n} e^{-3(\rho_*/\rho)^{1/n}},
\end{eqnarray}
where $T_*$ is a constant of integration. Combined with Eq. (\ref{eosgm1}), we obtain
\begin{eqnarray}
\label{eosgm3}
T= \frac{T_*}{\ln(a_*/a)^{n+1}}\left (\frac{a}{a_*}\right )^3.
\end{eqnarray}
When $a\rightarrow 0$, $T\rightarrow 0$. When $a\rightarrow a_*$, $T\rightarrow 0$ for $n<-1$ and $T\rightarrow +\infty$ for $n>-1$. For $n<-1$, the temperature reaches its maximum at
\begin{equation}
\label{eosgm4a}
\frac{\rho_{e}}{\rho_*}=\left (\frac{3}{|n+1|}\right )^n,\qquad \frac{a_{e}}{a_*}=e^{(n+1)/3},
\end{equation}
\begin{equation}
\label{eosgm4b} \frac{T_e}{T_*}=\left (\frac{3}{|n+1|}\right )^{n+1}e^{n+1}.
\end{equation}
The evolution of the density and temperature  as a function of the scale factor is plotted in Fig. \ref{annexedensity}.

The equation of state can be written as $p=w\rho c^2$ with
\begin{equation}
\label{eosgm5}
w=-1+\frac{|n|}{3}\left (\frac{\rho}{\rho_*}\right )^{1/n}.
\end{equation}
The pressure vanishes ($w=0$) at
\begin{equation}
\label{eosgm6}
\frac{\rho_{w}}{\rho_*}=\left (\frac{3}{|n|}\right )^n,\qquad \frac{a_{w}}{a_*}=e^{n/3}.
\end{equation}
When $a\rightarrow 0$, $w\rightarrow -1$; when  $a\rightarrow a_*$, $w\rightarrow +\infty$. The pressure is negative when $a<a_w$ and positive when $a_w<a<a_*$.

The deceleration parameter is given by Eqs. (I-77) and (I-78). Together with Eq. (\ref{eosgm5}), we obtain
\begin{equation}
\label{eosgm7}
q=-1+\frac{|n|}{2}\left (\frac{\rho}{\rho_*}\right )^{1/n}.
\end{equation}
The curve $a(t)$ presents an inflexion point ($\ddot a=q=0$) at
\begin{equation}
\label{eosgm8}
\frac{\rho_{c}}{\rho_*}=\left (\frac{2}{|n|}\right )^n,\qquad \frac{a_{c}}{a_*}=e^{n/2}.
\end{equation}
When $a\rightarrow 0$, $q\rightarrow -1$; when $a\rightarrow a_*$, $q\rightarrow +\infty$. The universe is accelerating when $a<a_c$ and decelerating when $a_c<a<a_*$.

Finally, the velocity of sound is given by
\begin{equation}
\label{eosgm9}
\frac{c_s^2}{c^2}=-1- \frac{n+1}{3}\left (\frac{\rho}{\rho_*}\right )^{1/n}.
\end{equation}
We have to distinguish two cases. For $n>-1$, the velocity of sound is always imaginary. When $a\rightarrow 0$, $(c_s/c)^2\rightarrow -1$; when $a\rightarrow a_*$, $(c_s/c)^2\rightarrow -\infty$.
For $n<-1$, the velocity of sound vanishes at the point (\ref{eosgm4a})-(\ref{eosgm4b}) where the temperature is maximum. At that point, the pressure is maximum with value
\begin{equation}
\label{eosgm9b}
\frac{p_e}{\rho_* c^2}=\frac{3^n}{|n+1|^{n+1}}.
\end{equation}
When $a\rightarrow 0$, $(c_s/c)^2\rightarrow -1$; when $a\rightarrow a_*$, $(c_s/c)^2\rightarrow +\infty$. The velocity of sound is imaginary when $a<a_e$ and real when $a_e<a<a_*$. On the other hand, the velocity of sound is equal to the speed of light at
\begin{equation}
\label{eosgm10b}
\frac{\rho_{s}}{\rho_*}=\left (\frac{6}{|n+1|}\right )^n,\qquad \frac{a_{s}}{a_*}=e^{(n+1)/6}.
\end{equation}
The velocity of sound is smaller than the speed of light when $a<a_s$ and larger when $a_s<a<a_*$.

\begin{figure}[!h]
\begin{center}
\includegraphics[clip,scale=0.3]{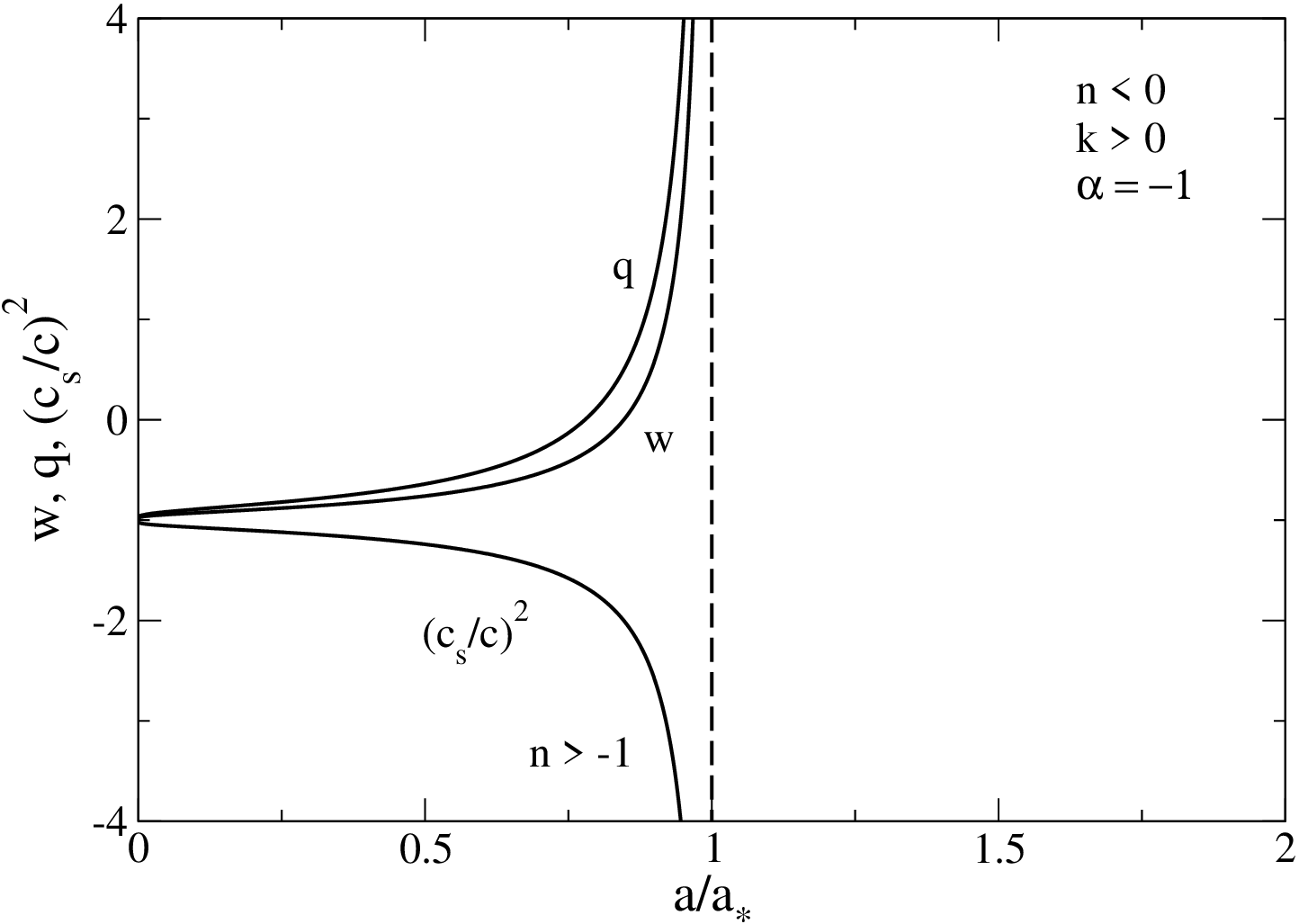}
\caption{Evolution of $w$, $q$, and $(c_s/c)^2$ as a function of the scale factor $a$. We have taken $n=-1/2$. }
\label{annexew}
\end{center}
\end{figure}

The evolution of $w$, $q$, and $(c_s/c)^2$ as a function of the scale factor $a$ is represented in Fig. \ref{annexew}.

\begin{figure}[!h]
\begin{center}
\includegraphics[clip,scale=0.3]{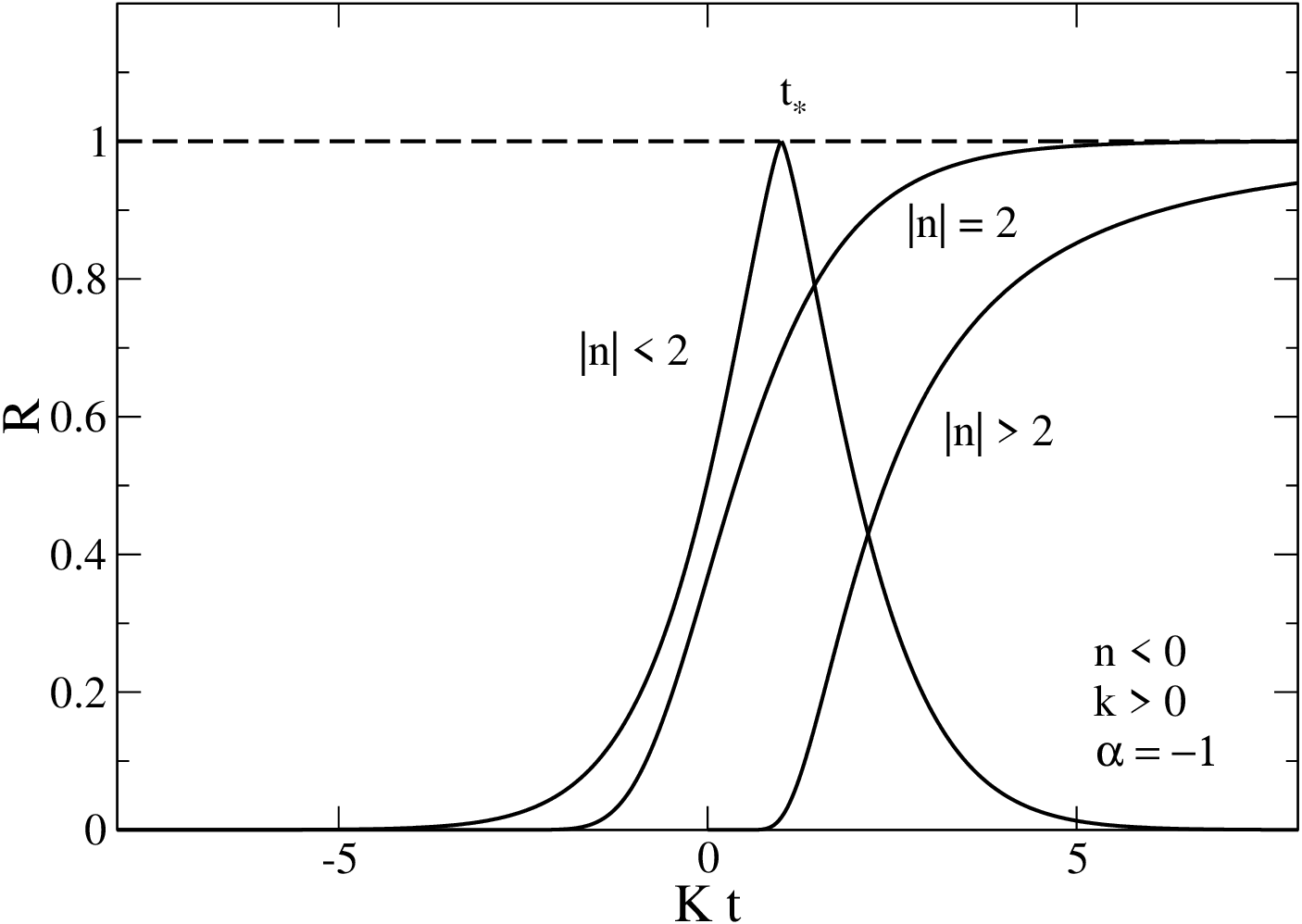}
\caption{Evolution of the scale factor with time for $n<-2$, $n=-2$, and $n>-2$ (specifically $n=-3$, $n=-2$ and $n=-1/2$). }
\label{annexeKposNneg}
\end{center}
\end{figure}

Setting $R=a/a_*$, the Friedmann equation (\ref{b9}) can be written
\begin{eqnarray}
\label{eosgm12}
\dot R=\frac{\epsilon K R}{(-\ln R)^{n/2}},
\end{eqnarray}
where $K=(8\pi G\rho_*/3)^{1/2}$ and $\epsilon=\pm 1$ as explained previously.  We must distinguish three cases.

For $n<-2$,
\begin{eqnarray}
\label{eosgm14}
R(t)=e^{-\left (\frac{|n|-2}{2} Kt\right )^{-2/(|n|-2)}},
\end{eqnarray}
\begin{eqnarray}
\label{eosgm14marre}
\frac{\rho(t)}{\rho_*}=\left (\frac{|n|-2}{2} Kt\right )^{-2|n|/(|n|-2)}.
\end{eqnarray}
The universe starts at $t=0$ with a vanishing radius, an infinite density, an infinite pressure, and a vanishing temperature. The universe expands and its density decreases. It reaches its maximum radius $R=1$ in infinite time, while the density and the pressure tend to zero algebraically rapidly.

For $n=-2$,
\begin{eqnarray}
\label{eosgm15}
R(t)=e^{-Ae^{-Kt}},
\end{eqnarray}
\begin{eqnarray}
\label{eosgm15marre}
\frac{\rho(t)}{\rho_*}=A^{2} e^{-2Kt}.
\end{eqnarray}
The universe starts from $t=-\infty$ with a vanishing radius, an infinite density, an infinite pressure, and a vanishing temperature. The universe expands and its density decreases. It reaches its maximum radius $R=1$ in infinite time, while its density and pressure tend to zero exponentially rapidly.

For $n>-2$,
\begin{eqnarray}
\label{eosgm13}
R(t)=e^{-\left \lbrack \frac{2-|n|}{2} K(t_{*}-t)\right \rbrack^{2/(2-|n|)}},
\end{eqnarray}
\begin{eqnarray}
\label{eosgm13marre}
\frac{\rho(t)}{\rho_*}=\left \lbrack \frac{2-|n|}{2} K(t_{*}-t)\right \rbrack^{2|n|/(2-|n|)}.
\end{eqnarray}
The universe starts from $t=-\infty$ with a vanishing radius, an infinite density, an infinite pressure, and a vanishing temperature. The universe expands and its density decreases. When $t=t_*$, the universe reaches its maximum radius $R=1$ while its density vanishes (future peculiarity). The universe ``disappears''. When $t>t_*$, the universe contracts (this corresponds to the solution of Eq. (\ref{eosgm12}) with $\epsilon=-1$) and its density increases. When $t\rightarrow +\infty$, the universe has a vanishing radius and an infinite density. At $t=t_*$, the pressure vanishes for $n<-1$, is constant for $n=-1$, and is infinite for $n>-1$. In this last case, there is a future singularity.

For $n<-1$, the velocity of sound becomes larger than the speed of light when $t>t_s$ corresponding to Eq. (\ref{eosgm10b}). This marks the end of the physical universe. The solutions (\ref{eosgm14})-(\ref{eosgm13marre}) may not have sense for $t>t_f=t_s$.  The evolution of the scale factor with time is plotted in Fig. \ref{annexeKposNneg}.

\section{The Planck scales and the cosmological scales}
\label{sec_pcscales}

It is expected that quantum mechanics plays a fundamental role in the early universe where the scale factor is small. From the Planck constant $\hbar$, the speed of light $c$, and the constant of gravity $G$, we can construct characteristic scales, known as the Planck scales. These are the Planck density
\begin{eqnarray}
\label{pc1}
\rho_P=\frac{c^5}{G^2\hbar}=5.16\, 10^{99}\, {\rm g/m^3},
\end{eqnarray}
the Planck time
\begin{eqnarray}
\label{pc2}
t_P=\left (\frac{\hbar G}{c^5}\right )^{1/2}=5.39\, 10^{-44}\, {\rm s},
\end{eqnarray}
the Planck length
\begin{eqnarray}
\label{pc3}
l_P=\left (\frac{G\hbar}{c^3}\right )^{1/2}=1.62\, 10^{-35}\, {\rm m},
\end{eqnarray}
the Planck mass
\begin{eqnarray}
\label{pc5}
M_P=\left (\frac{\hbar c}{G}\right )^{1/2}=2.17\, 10^{-5}\, {\rm g},
\end{eqnarray}
and the Planck temperature
\begin{eqnarray}
\label{pc4}
T_P=\frac{M_P c^2}{k_B}=1.42\, 10^{32}\, {\rm K}.
\end{eqnarray}

We have the obvious relations $l_P=c t_P$, $\rho_P=M_P/l_P^3$, $t_P=(G\rho_P)^{-1/2}$. We have given in Paper I an interpretation of these different scales. In particular, the Planck density is expected to represent a fundamental upper bound for the density leading to a phase of early inflation.

Similarly, we expect that the cosmological constant $\Lambda$ plays a fundamental role in the late universe where the scale factor is large. We shall regard the cosmological constant as a fundamental constant of nature that describes the cosmophysics (hence the late universe) exactly like the Planck constant describes the microphysics (hence the early universe). If this interpretation is correct, we should not seek the meaning of the dark energy in the vacuum energy, or in quantum mechanics. It would have another origin purely connected to general relativity. In this point of view, the cosmological constant problem is a false problem (see Sec. \ref{sec_problem}). However, the true nature of the cosmological constant remains to be found. It would be necessary to develop the counterpart  of quantum mechanics to cosmophysics. For the moment, there is no such theory. However, using dimensional analysis, we can easily introduce cosmological scales that are the counterpart of the Planck scales.

From observations, we know the value of the dark energy density $\rho_{\Lambda}$. It is usually written as $\rho_{\Lambda}=\Omega_{\Lambda,0}\rho_0$ where $\Omega_{\Lambda,0}=0.763$ is the present fraction of dark energy and $\rho_0=9.20\, 10^{-24}{\rm g}/{\rm m}^3$ is the present density of the universe determined by the Hubble constant $H_0=2.27\, 10^{-18}{\rm s}^{-1}$ through the Friedmann equation (\ref{b9}). This yields $\rho_{\Lambda}=7.02\, 10^{-24}{\rm g}/{\rm m}^3$.
From now on, $\rho_{\Lambda}$ will be called the cosmological density (the word ``cosmological'' in the late universe replacing the name ``Planck'' in the early universe). From the cosmological density $\rho_{\Lambda}$, the speed of light $c$, and the constant of gravity $G$, we can construct a cosmological time $t_{\Lambda}=1/(G\rho_{\Lambda})^{1/2}$, a cosmological length $l_{\Lambda}=c t_{\Lambda}$, and a cosmological mass $M_{\Lambda}=\rho_{\Lambda}l_{\Lambda}^3$. We have $\rho_{\Lambda}=\Omega_{\Lambda,0}\rho_0$, $t_{\Lambda}=(8\pi/3\Omega_{\Lambda,0})^{1/2}H_0^{-1}$, and $l_{\Lambda}=(8\pi/3\Omega_{\Lambda,0})^{1/2}a_0$.

The dark energy density is related to the cosmological constant by $\rho_{\Lambda}=\Lambda/(8\pi G)$.  From now on, we will consider that $\Lambda=1.18\, 10^{-35}\, {\rm s}^{-2}$ is a fundamental constant of physics. Therefore, the corresponding scales $\rho_{\Lambda}$, $t_{\Lambda}$, $l_{\Lambda}$... are fundamental scales determined from $\Lambda$ in the same sense that the Planck  scales are determined from $\hbar$. These cosmological scales can be written as the Planck scales  by introducing the notation
\begin{eqnarray}
\label{pc6}
\hbar_{\Lambda}=\frac{8\pi c^5}{\Lambda G}.
\end{eqnarray}
Numerically, $\hbar_{\Lambda}=7.35\, 10^{122}\hbar=7.75\, 10^{88} {\rm J\,  s}$. This constant might play a fundamental role in the theory of cosmophysics, just like the Planck constant $\hbar$ plays a fundamental role in the theory of microphysics. This is, however, a pure speculation that is simply motivated by the apparent ``symmetry'' between the early universe and the late universe (in any case, $\hbar_{\Lambda}$ can be introduced as a convenient notation). The theory of cosmophysics would then lead to characteristic scales that are the cosmological density
\begin{eqnarray}
\label{pc7}
\rho_{\Lambda}=\frac{c^5}{G^2\hbar_{\Lambda}}=\frac{\Lambda}{8\pi G}=7.02\, 10^{-24}\, {\rm g}/{\rm
m}^3,
\end{eqnarray}
the cosmological time
\begin{eqnarray}
\label{pc8}
t_{\Lambda}=\left (\frac{\hbar_{\Lambda} G}{c^5}\right )^{1/2}=\left (\frac{8\pi}{\Lambda}\right )^{1/2}=1.46\, 10^{18}\, {\rm s},
\end{eqnarray}
the cosmological length
\begin{eqnarray}
\label{pc9}
l_{\Lambda}=\left (\frac{G\hbar_{\Lambda}}{c^3}\right )^{1/2}=\left (\frac{8\pi c^2}{\Lambda}\right )^{1/2}=4.38\, 10^{26}\, {\rm m},
\end{eqnarray}
the cosmological mass
\begin{eqnarray}
\label{pc11}
M_{\Lambda}=\left (\frac{\hbar_{\Lambda} c}{G}\right )^{1/2}=\frac{c^3}{(8\pi
\Lambda G^2)^{1/2}}=5.90\, 10^{56}\, {\rm g},\qquad
\end{eqnarray}
and the cosmological temperature
\begin{eqnarray}
\label{pc10}
T_{\Lambda}=\frac{M_{\Lambda} c^2}{k_B}=3.84\, 10^{93}{\rm K}.
\end{eqnarray}
We have given in this paper an interpretation of these different scales (except for the temperature whose nature is unknown to us). In particular, the cosmological density is expected to represent a fundamental lower bound for the density leading to a phase of late inflation. Actually, these scales were first introduced by Einstein \cite{einsteincosmo} in his static model with a cosmological constant. Although they now have a different meaning (since the universe is not static), the scaling is of course the same. These scales, coming from the fundamental constant $\hbar_{\Lambda}$ (a redefinition of $\Lambda$) should ultimately emerge from a theory of cosmophysics that remains to be constructed. The apparent symmetry between the early and the late universe, may be a source of inspiration in the construction of such a theory.

\section{Unification of pre-radiation, radiation, and dark energy}
\label{sec_unification}

We have seen in Paper I that the transition between the pre-radiation
era and the radiation era can be described by an equation of state
$p/c^2=-4\rho^2/(3\rho_P)+\rho/3$. On the other hand, the transition
between a linear equation of state era and the dark energy era can be
described by an equation of state $p/c^2=\alpha \rho
-(\alpha+1)\rho_{\Lambda}$ (see Secs. \ref{sec_ges} and
\ref{sec_dark}).  In Sec. \ref{sec_solid}, we have taken $\alpha=0$ to
describe the transition between the matter era and the dark energy
era. Alternatively, if we take $\alpha=1/3$, we describe the transition
between the radiation era and the dark energy era. Finally, the
quadratic equation of state
\begin{eqnarray}
\label{unif1}
p=-\frac{4\rho^2}{3\rho_P}c^2+\frac{1}{3}\rho c^2-\frac{4}{3}\rho_{\Lambda}c^2,
\end{eqnarray}
describes, in a unified manner, the pre-radiation era, the radiation
era, and the dark energy era. A nice feature of this equation of state
is that both the Planck density (early universe) and the cosmological
density (late universe) explicitly appear. Therefore, this equation of
state reproduces both the early inflation and the late inflation. This
may suggest that the pre-radiation, radiation, and dark energy are the
manifestation of a {\it unique} form of ``generalized
radiation''. When $\rho\rightarrow
\rho_P$, we get $p\sim -\rho_Pc^2$ (vacuum energy); when $\rho\rightarrow
\rho_{\Lambda}$, we get $p\sim -\rho_{\Lambda}c^2$ (dark energy); when
$\rho_{\Lambda}\ll\rho\ll\rho_{P}$, we get $p\sim \rho c^2/3$ (radiation).

For the equation of state (\ref{unif1}), the continuity equation (\ref{b7}) can be integrated into
\begin{eqnarray}
\label{unif2}
\rho=\frac{\rho_P}{1+(a/a_1)^4}+\rho_{\Lambda},
\end{eqnarray}
where $a_1=2.61\, 10^{-6}{\rm m}$ (see Paper I). To obtain this
expression, we have used the fact that $\rho_P\gg \rho_{\Lambda}$ so
that $p/c^2+\rho\simeq
(4/3\rho_P)(\rho-\rho_{\Lambda})(\rho_P-\rho)$. When
$a\rightarrow 0$, $\rho\rightarrow
\rho_{P}$ (early inflation); when $a\rightarrow +\infty$,
$\rho\rightarrow
\rho_{\Lambda}$ (late inflation); when
$\rho_{\Lambda}\ll\rho\ll\rho_{P}$, $\rho\sim
\rho_P (a/a_1)^{-4}$ (power-law evolution). We note that
Eq. (\ref{unif2}) can be rewritten as
\begin{eqnarray}
\label{unif2b}
\rho=\frac{\Omega_{rad}\rho_0}{(a/a_0)^4+(a_1/a_0)^4}+\Omega_{\Lambda,0}\rho_0.
\end{eqnarray}
It leads to Eq. (\ref{pr7}) with
$\Omega_m=\Omega_B+\Omega_{DM}=0$. It is therefore necessary to add
``by hand'' the density of matter (\ref{pr4}) and (\ref{pr5}) in the
Friedmann equation (\ref{b9}). This suggests that matter on the one
hand, and pre-radiation $+$ radiation $+$ dark energy on the other
hand should be treated as two different ``species''. This is at
variance with usual models, including those considered in the main
part of this paper, that try to unify matter and dark energy. 

For the equation of state (\ref{unif1}), the thermodynamical equation (\ref{t2}) can be integrated into
\begin{eqnarray}
\label{unif3}
T=T_P \left (\frac{15}{\pi^2}\right )^{1/4}\left (\frac{\rho}{\rho_P}-\frac{\rho_{\Lambda}}{\rho_{P}}\right )^{1/4}\left (1-\frac{\rho}{\rho_P}\right )^{7/4}.\qquad
\end{eqnarray}
where we have determined the constant of integration in order to
recover the Stefan-Bolzmann law in the radiation era
$\rho_{\Lambda}\ll\rho\ll \rho_{P}$ (see Paper I). Combined with Eq. (\ref{unif2}), we get
\begin{eqnarray}
\label{unif4}
T=T_P \left (\frac{15}{\pi^2}\right )^{1/4}\frac{(a/a_1)^7}{\left\lbrack (a/a_1)^4+1\right \rbrack^2}.
\end{eqnarray}
This is exactly the same expression as in the pre-radiation $+$ radiation era (see Paper I) but it is also valid in the dark energy era. This may be a confirmation that dark energy is of the same nature as pre-radiation and radiation. When $a\ll a_1$, $T\sim T_P (15/\pi^2)^{1/4}(a/a_1)^7$ (pre-radiation); when $a\gg a_1$, $T\sim T_P (15/\pi^2)^{1/4}(a_1/a)$ (radiation and dark energy).

Simple analytical expressions of the Friedmann equation (\ref{b9})
with the density-radius relation (\ref{unif2}) can be given in
particular limits. In the pre-radiation era, the scale factor
increases exponentially as in Eq. (I-119). In the pre-radiation $+$
radiation era, the evolution of the scale factor is given by
Eq. (I-128); in the pure radiation era, the scale factor increases
algebraically as in Eq.  (I-124); in the radiation $+$ dark energy
era, the evolution of the scale factor is given by Eq. (\ref{dark14})
with $\alpha=1/4$, $a_*=(\rho_P/\rho_{\Lambda})^{1/4}a_1$, and
$K=(8\pi/3)^{1/2}t_{\Lambda}^{-1}$; in the dark energy era, the scale
factor increases exponentially as in Eq. (\ref{s10}). Actually, it is possible to solve the Friedmann equation (\ref{b9}) with the density-radius relation (\ref{unif2}) exactly. Introducing $R=a/a_1$ and $\lambda=\rho_{\Lambda}/\rho_{P}\ll 1$, we obtain 
\begin{eqnarray}
\label{unif5}
\int \frac{\sqrt{1+R^4}}{R\sqrt{1+\lambda R^4}}\, dR=\left (\frac{8\pi}{3}\right )^{1/2}t/t_P+C,
\end{eqnarray}
which can be integrated into
\begin{eqnarray}
\label{unif6}
\frac{1}{\sqrt{\lambda}}\ln\left \lbrack 1+2\lambda R^4+2\sqrt{\lambda(1+R^4+\lambda R^8)}\right\rbrack\qquad\nonumber\\
-\ln\left\lbrack \frac{2+R^4+2\sqrt{1+R^4+\lambda R^8}}{R^4}\right\rbrack=4\left (\frac{8\pi}{3}\right )^{1/2}\frac{t}{t_P}+C,\nonumber\\
\end{eqnarray}
where the constant $C$ is determined such that $a=l_P$ at $t=0$ (see Paper I).

\end{document}